\documentclass[print]{elsarticle}

\usepackage{caption}
\usepackage{subcaption}
\usepackage{xcolor}
\usepackage{lineno,hyperref}
\usepackage{dblfloatfix}  
\usepackage{graphicx}  
\usepackage{enumitem}
\usepackage{relsize}
\usepackage{amsmath}
\usepackage{amssymb}
\usepackage{multirow}
\usepackage{bbm}

\newcommand{\bftab}{\fontseries{b}\selectfont}

\journal{}

\begin{document}

\begin{titlepage}
        \vspace*{1cm}
            
        \LARGE
        \noindent \textbf{Connecting User and Item Perspectives in Popularity Debiasing for Collaborative Recommendation}
            
        \vspace{2cm}
        \LARGE
        
        \noindent Journal Item
            
        \vspace{2cm}
 
        \hrule
        
        \vspace{0.5cm}
 
        \small
        
        \noindent \textbf{How to cite:}
        
        \vspace{2mm} 
        
        \noindent Boratto, L., Fenu, G., \& Marras, M. (2021). Connecting user and item perspectives in popularity debiasing for collaborative recommendation. Information Processing \& Management, 58(1), 102387. \url{https://doi.org/10.1016/j.ipm.2020.102387}
        
        \vspace{0.5cm}
        
        \hrule
        
        \vspace{0.5cm}
            
        \noindent Version: Accepted Manuscript
        
\end{titlepage}

\begin{frontmatter}

\title{Connecting User and Item Perspectives in Popularity Debiasing for Collaborative Recommendation}

\author[eurecataddress]{Ludovico Boratto\corref{equalcontribution}}
\ead{ludovico.boratto@acm.org}
\author[unicaaddress]{Gianni Fenu}
\ead{fenu@unica.it}
\author[unicaaddress]{Mirko Marras\corref{equalcontribution}}
\cortext[equalcontribution]{The authors equally contributed to this work.}
\ead{mirko.marras@unica.it}
\address[eurecataddress]{Data Science and Big Data Analytics, \\Eurecat - Centre Tecnol\`ogic de Catalunya, Barcelona, Spain.}
\address[unicaaddress]{Department of Mathematics and Computer Science, \\University of Cagliari, Cagliari, Italy}

\begin{abstract}
Recommender systems learn from historical users' feedback that is often non-uniformly distributed across items. As a consequence, these systems may end up suggesting popular items more than niche items progressively, even when the latter would be of interest for users. This can hamper several core qualities of the recommended lists (e.g., novelty, coverage, diversity), impacting on the future success of the underlying platform itself. In this paper, we formalize two novel metrics that quantify how much a recommender system equally treats items along the popularity tail. The first one encourages equal probability of being recommended across items, while the second one encourages true positive rates for items to be equal. We characterize the recommendations of representative algorithms by means of the proposed metrics, and we show that the item probability of being recommended and the item true positive rate are biased against the item popularity. To promote a more equal treatment of items along the popularity tail, we propose an in-processing approach aimed at minimizing the biased correlation between user-item relevance and item popularity. Extensive experiments show that, with small losses in accuracy, our popularity-mitigation approach leads to important gains in beyond-accuracy recommendation quality.
\end{abstract}


\end{frontmatter}


\section{Introduction}
Recommender systems are bridging users and relevant products, services and peers on the Web. By leveraging past behavioural data, such automated systems aim to learn patterns behind users' preferences and predict the future interests of users~\cite{DBLP:reference/sp/RicciRS15}. Notable examples are integrated into platforms that cover different contexts, including e-commerce (e.g., Amazon, eBay), entertainment (e.g., YouTube, Netflix), and education (e.g., edX, Udemy). The success of these platforms strongly depends on the effectiveness of their recommender system.  

\begin{figure}[!t] 
\centering
\minipage{1.0\textwidth}
    \includegraphics[width=1.0\linewidth]{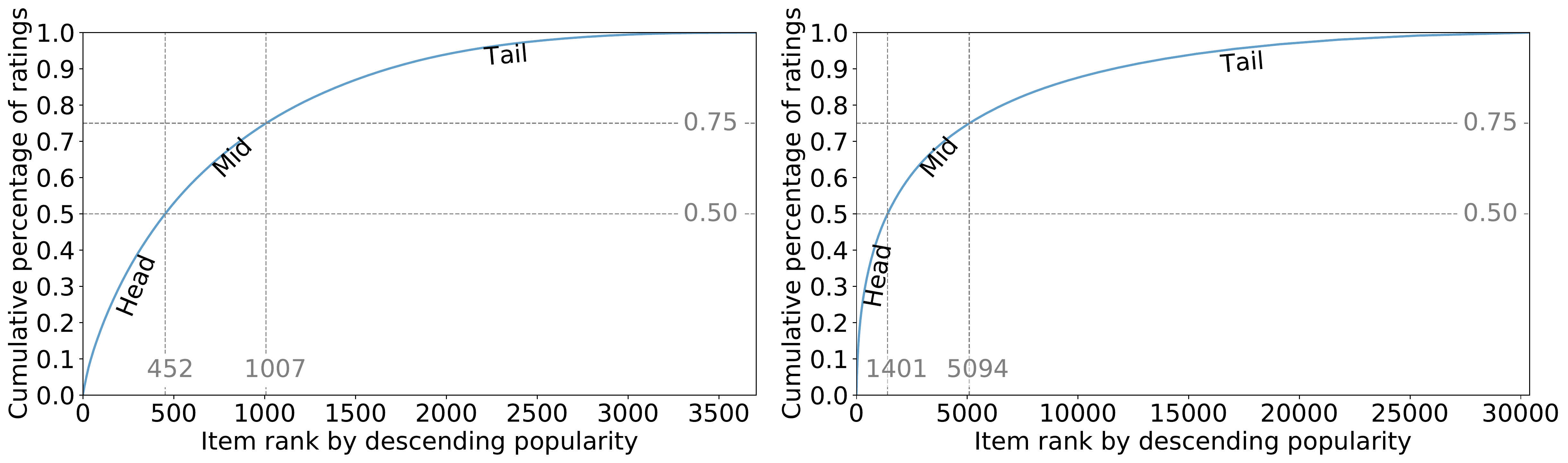}
	\caption{\textbf{Imbalanced users' feedback distributions}. Cumulative percentage of users' feedback for movies in Movielens1M \cite{DBLP:journals/tiis/HarperK16} (left) and online courses in COCO \cite{DBLP:conf/worldcist/DessiFMR18} (right). Each curve is divided in head, mid and tail based on $50\%$ and $75\%$ percentiles. Head items gather 50\% of ratings. Section~\ref{sec:data} provides a detailed descriptions of the mentioned datasets.}
    \label{fig:pop-dist}
\endminipage\hfill
\end{figure}

The increasing adoption of recommender systems in online platforms has spurred investigations on issues of bias in their internal mechanisms. One aspect that has received attention is the recommender systems' tendency of emphasizing a ``\emph{rich-get-richer}'' effect in favor of popular items~\cite{DBLP:journals/jasis/NikolovLFM19}. Such a phenomenon leads to a loop where recommender systems trained on data non-uniformly distributed across items tend to suggest popular items more than niche items, even when the latter would be of interest. Thus, popular items gain higher visibility and become more likely to be selected. The awareness of this type of bias might even lead providers to bribe users, so that they rate or increase the ratings given to their items, thus allowing these items to get more visibility~\cite{RamosBC20,DBLP:conf/icdm/SaudeRCK17}. Under this repeated loop, the training data becomes imbalanced towards a tiny set of items more and more (Figure \ref{fig:pop-dist})\footnote{Please note that all the figures in this manuscript are best seen in color.}, as a result of a strong popularity bias. 

Recommender systems that suggest the most popular items have been proved to achieve competitive accuracy, compared with advanced recommendation techniques~\cite{DBLP:journals/umuai/JannachLKJ15}. However, the literature has acknowledged that other aspects beyond the accuracy of the recommender system, such as whether the suggested items are novel and cover well the catalog, can make a positive impact on the overall recommendation quality~\cite{DBLP:journals/kbs/BobadillaOHG13}. Unfortunately, a bias against item popularity may emphasize the occurrence of filter bubbles, thus hampering users' interest and several beyond-accuracy aspects~\cite{DBLP:journals/corr/NematzadehCMF17,DBLP:conf/sigir/CanamaresC18,DBLP:conf/cikm/MehrotraMBL018}. Trading such aspects for item popularity might likely not be accepted by the users who finally receive the recommendations. Therefore, mitigating the impact of a popularity bias can help to meet a better trade-off between accuracy and beyond-accuracy objectives, with a clear benefit for the recommendation quality on the whole~\cite{DBLP:journals/tiis/KaminskasB17}.

Existing frameworks and techniques for mitigating a popularity bias \cite{DBLP:conf/recsys/KamishimaAAS14,hou2018balancing,DBLP:conf/flairs/AbdollahpouriBM19,DBLP:conf/recsys/AbdollahpouriBM17} are often based on evaluation metrics that do not account for users' preferences, being assessed only on the level of popularity of items in a recommended list. However, controlling popularity cannot be a final objective concept, as it strongly depends on users' preferences and on how data has been collected. It follows that popularity metrics and bias mitigation need to account for the users' tastes and the visibility given to items thanks to recommendations, creating a bridge between these perspectives by means of beyond-accuracy objectives.

In this paper, we tackle this key challenge with a new popularity-mitigation framework. Two novel metrics quantify how much a recommender equally treats items along the popularity tail. The first metric encourages similar probabilities of being recommended among items, that is an important aspect when platform owners are interested in equally suggesting items (e.g., loan platforms). The second metric takes into account the ground-truth users' preferences, and encourages true positive rates of items to be equal. This goal is primary in contexts where the platform owners aim to preserve the imbalance across items in data, while avoiding any further distortion on recommendations caused by bias.

Then, we empirically prove that two widely-adopted classes of recommendation algorithms, based on point-wise and pair-wise optimization respectively, are highly biased against item popularity with respect to the two proposed metrics. To limit this negative effect, we propose an approach characterized by ($i$) a new user-item observation sampling that balances the training examples where the observed item is more or less popular than the unobserved item, and ($ii$) a new regularization term that minimizes the biased correlation between user-item relevance and item popularity. Extensive experiments show that the proposed approach provides a less biased treatment of items over the popularity tail. Further, with a small loss in accuracy, it also leads to important gains in novelty and catalog coverage, known to provide benefits to the overlying platform.

This paper is organized as follows: Section~\ref{related-work} presents related works on item popularity in recommendation. Section~\ref{problem-formulation} formalizes problem and metrics. Section~\ref{exploratory-analysis} describes the exploratory analysis. Then, Sections~\ref{popularity-reg} and~\ref{emp-eval} describe and validate our mitigation approach, respectively. Section~\ref{conclusions} concludes the paper. 

\section{Related Work}
\label{related-work}
This research relates to the literature from the recommender system and machine learning communities.

\subsection{Item Popularity in Recommendation}
Treating biases against item popularity has traditionally required a \emph{multi-objective setting} that investigates any accuracy loss resulting from taking popularity into consideration. Therefore, the final goal has been to strike a balance between accuracy (e.g., precision, recall) and pre-conceptualized metrics associated with the popularity of the recommended items (e.g., average recommended item popularity, average percentage and coverage of tail items)~\cite{DBLP:conf/recsys/AbdollahpouriBM17,DBLP:conf/flairs/AbdollahpouriBM19}. These metrics (i) tend to monitor popularity distribution skews for a complete system rather than for individual items, (ii) often require to define head/tail memberships for all items, which is indeed arbitrary and depends on the characteristics of the data, (iii) or assess biases against popularity without considering the ground truth of the users' interests. Our study in this paper addresses this gap with two metrics that monitor the popularity of each item individually: the first one enforces ranking probabilities for items to be the same, and the second one encourages true positive rates of items to be the same.

Under this setting, \emph{pre-processing} countermeasures consist of re-arranging training data in order to reduce the impact of users' feedback imbalances across items. For instance, Park and Tuzhilin \cite{DBLP:conf/recsys/ParkT08} divide the item set into head and tail based on the item popularity and cluster the ratings associated with these sets separately. Tail recommendations leverage ratings from the corresponding cluster, while the head ones use ratings from individual items. In~\cite{DBLP:journals/umuai/JannachLKJ15}, Jannach et al. sample training triplets including a user, an observed item, and an unobserved item, where the observed item is less popular than the unobserved item. The work of Chen et al. in~\cite{DBLP:conf/airs/ChenZLM18} picks up unobserved items following a probability distribution based on item popularity. Differently, our data sampling balances cases where the observed item is more or less popular than the unobserved item in the current training triplet, in order to let models learn from training examples with different relative popularity and to better fit with our regularization. In case of algorithms which receive user-item pairs rather than triplets, this balancing is performed for the pairs involving a user and an unobserved item. 

\emph{In-processing} countermeasures extend an existing algorithm in order to simultaneously consider both relevance and popularity of an item, doing a joint optimization or using one criterion as a constraint for the other. For instance, Oh et al.~\cite{DBLP:conf/icdm/OhPYSP11} consider the user's tendency of interacting with items based on their popularity in order to recommend items along the popularity tail. Kamishima et al.~\cite{DBLP:conf/recsys/KamishimaAAS14} propose to enhance the statistical independence between a recommendation and its popularity. Similarly, Abdollahpouri et al.~\cite{DBLP:conf/recsys/AbdollahpouriBM17} arrange the \textit{RankALS} algorithm in order to recommend lists that balance accuracy and head/tail recommendations. Moreover, Hou et al.~\cite{hou2018balancing} find common neighbours between two items with a given popularity and get a balanced common-neighbour similarity index, after pruning the popular neighbours. Our regularization in this paper differs in expressiveness and algorithmically, generalizing to several learning-to-rank approaches, such as those based on point- and pair-wise optimization. 

\emph{Post-processing} countermeasures re-rank a recommended list according to certain item-popularity constraints. In this view, Abdollahpouri et al.~\cite{DBLP:conf/flairs/AbdollahpouriBM19,abdollahpouri2018popularity} present two approaches for controlling the popularity of the recommended items. The first one consists of re-arranging the \textit{xQuAD} algorithm in order to balance the trade-off between accuracy and mid-tail item coverage. The second one requires to multiply the relevance score for a given item with a weight inversely proportional to its popularity, with items re-ranked according to the weighted scores. In~\cite{DBLP:journals/umuai/JannachLKJ15}, Jannach et al. introduce user-specific weights in order to balance accuracy and popularity. Post-processing countermeasures can provide elegant solutions, but often require to know the head/tail membership of items, may be sensitive to the original relevance, and lead to a decrease in algorithm efficiency.

\subsection{Bias Mitigation in Machine Learning}
The literature has primarily focused on mitigating biases inside classification models, with objectives mostly targeting fairness~\cite{DBLP:conf/kdd/SinghJ18}. Our study in this paper fundamentally reformulates the statistical parity and the equal opportunity notions~\cite{DBLP:conf/nips/HardtPNS16,DBLP:journals/corr/abs-1901-04730} for the item popularity bias problem, given that they were conceived to monitor differences in accuracy across classes in fairness-aware classification. Compared to previous works, our metrics monitor bias on individual items rather than groups and do not use any intrinsic notion of popularity group membership for an item. Furthermore, our metrics align with the probability of being recommended in a top-k list rather than being classified to a given label.

Many approaches have been proposed to address bias and fairness issues in machine learning. Notable examples of in-processing approaches primarily target fairness among classes and fall into two main categories: constraint-based optimization~\cite{DBLP:conf/icml/AgarwalBD0W18,DBLP:conf/nips/GohCGF16} and adversarial learning~\cite{DBLP:conf/aies/BeutelCDQWLKBC19,DBLP:conf/icdm/KamishimaAS11,DBLP:conf/kdd/BeutelCDQWWHZHC19}. Our approach builds on and reformulates the former class of strategies to fit with the popularity debiasing task. Differently from \cite{DBLP:conf/kdd/BeutelCDQWWHZHC19,DBLP:conf/aies/BeutelCDQWLKBC19}, we relax the assumption of knowing the group membership of input samples, targeting individual items regardless of their popularity head/tail membership. In our setting, we are concerned with the biased correlation between relevance and popularity rather than between predicted labels and group membership, which differently drives optimization and enables different data sampling strategies. Furthermore, as we tackle a popularity debiasing task rather than an unfairness mitigation task, our study leads to consider model facets so far under-explored.    

\section{Preliminaries}
\label{problem-formulation}
In this section, we present the main concepts underlying our study, including the recommender system and the new metrics that monitor a popularity bias.

\subsection{Recommender System Formalization}
Given a set of $M$ users $U=\{u_1, u_2,..., u_M\}$ and a set of $N$ items $I=\{i_1, i_2, ..., i_N\}$, we assume that users have expressed their interest for a subset of items in $I$. The collected feedback from observed user-item interactions can be abstracted to a set of ($u$, $i$) pairs implicitly obtained from user activity or ($u$, $i$, $v$) triplets explicitly computed, with $v \in V \subseteq \mathbb{R}$. Elements in $V$ are either ratings or frequencies (e.g., play counts). We denote the user-item matrix $R \in \mathbb{R}^{M*N}$ by $R(u,i) = 1$ for implicit feedback or $R(u,i) = v$ for explicit feedback or frequencies to indicate the (level of) preference of $u$ for $i$, $R(u,i)=0$ otherwise. 

Given this matrix, the recommender system aims to predict unobserved user-item relevance scores, and deliver a set of ranked items to a user. To this end, we assume that a function estimates relevance scores of unobserved entries in $R$ for a given user, and the recommender system uses them for ranking the items. Formally, this operation can be abstracted as learning $\widetilde{R}(u,i) = f(u,i|\theta)$, where $\widetilde{R}(u,i)$ denotes the predicted relevance score, $\theta$ denotes model parameters, and $f$ denotes the function that maps model parameters to the predicted score. 

We assume that each user/item is represented inside a model through a $D$-sized numerical vector. More precisely, the model includes a user-vector matrix $W$ and an item-vector matrix $X$. We assume that the function $f$ is parametrized by $\theta$ and depends on the recommendation algorithm under consideration. The higher the $\widetilde{R}(u,i)$ is, the higher the relevance of $i$ for $u$. To rank items, they are sorted by decreasing relevance for a user, and the top-$k$ items are recommended.

\subsection{Popularity Bias Metric Formalization}
As we deal with biases against item popularity, we define what we consider as a popularity bias and how we monitor such a bias throughout the paper.

\vspace{2mm}

\noindent \textbf{Item Statistical Parity (ISP)}. Recommender system may be inherently influenced by the item popularity. More popular items remain more popular since they are more likely to appear at the top of the recommended list. This phenomenon inadvertently leads to a few mainstream items being recommended and to an impedance for items of the tail-end popularity spectrum to become visible to users. In this scenario, we assume that the recommender system should equally cover items along the popularity tail. For instance, in some cases, this notion may be useful  in platforms which manage recommendations of individuals (e.g., people recommendations) or particularly delicate items (e.g., loans).  

To measure this property, we reformulate the statistical parity principle introduced in fairness-aware machine learning \cite{DBLP:journals/corr/abs-1901-04730}, for the popularity debiasing task. This task requires to control the item statistical parity, i.e., equalizing the outcomes across individual items. We operationalize this concept for items by computing the ratio between the number of users each item is recommended to and the number of users who can receive that item in the recommended list. We then encourage that such a ratio is equal across items. Considering that only top-k items are recommended, we assume that the outcome for an item is the probability of being ranked in the top-k list. We define such probability as:

\begin{equation}
p(i|k,\theta) = \frac{\sum_{u \in U} \varphi\left(u,i|k,\theta\right)} {\sum_{u \in U} 1 - min\left(R(u,i), 1\right)} 
\label{eq:pik}
\end{equation}
\vspace{1mm}

\noindent where $\varphi\left(u,i|k,\theta\right)=1$ if item $i$ is being ranked in top-k for user $u$ by model $\theta$, $0$ otherwise. In other words, the numerator counts the users who receive item $i$ in top-k, while the denominator counts the users who have never interacted with item $i$ and, thus, may receive $i$ as a recommendation. Finally, we compute the inverse of the Gini index in order to measure the equality of $p(i|k,\theta)$ across items. The Gini index is a scale-independent and bounded measure of inequality that ranges in $[0,1]$. Higher values represent higher inequality. The item statistical parity of a recommender system $\theta$ for a cutoff $k$ is computed as follows: 

\begin{equation}
ISP(k,\theta) = 1 - Gini\left(\{p(i|k,\theta) \, | \, \forall \, i \in I \}\right)
\label{eq:isp}
\end{equation}

\noindent If there is a perfectly equal distribution of recommendations across items, then $ISP=1$ and the statistical parity is met. $ISP$ decreases and gets closer to $0$ when the distribution of the recommendations is more unequal. For example, this situation occurs if most of the items never appeared in the recommended lists. In the extreme, where the same $n << N$ items appeared in all the recommendations, $ISP$ is very close to 0. Thus, $ISP$ will lie between 0 and 1, and the greater it is, the more equally the recommendations are distributed.

In some cases, the platform owners would not equalize the recommendations along the entire popularity tail. High statistical parity could lead to situations in which even items of very low interest get recommended the same amount of times with respect to more-of-interest items. This motivated us to complement our measurement of popularity bias with another representative metric. 

\vspace{2mm}

\noindent \textbf{Item Equal Opportunity (IEO)}. Instead of equalizing the recommendations themselves, we can equalize some statistics of the algorithm’s effectiveness (e.g., true positive rate across the items). In many applications, platform owners may care more about preserving and retaining a certain degree of item popularity, while checking that no further distortions are emphasized by algorithmic bias on recommendation distributions. For instance, guaranteeing a certain degree of popularity in recommendations resulted in higher users' acceptance in touristic contexts \cite{DBLP:conf/www/CremonesiGPQ14}. Under this view, a less biased algorithm tends to recommend each item proportionally to its representation in the ground-truth user preference. 

We operationalize this concept of equal opportunity across items by encouraging the true positive rates of different items to be the same. Its formulation builds upon the original concept proposed in the fairness-aware machine-learning domain \cite{DBLP:conf/nips/HardtPNS16}. Specifically, we define the true positive rate as the probability of being ranked within the top-k of a user, given the ground truth that the item is relevant for that user in the test set. This concept is denoted by $p(i|k,\theta,y=1)$, where $y=1$ defines that items are relevant for users in the ground truth:

\begin{equation}
p(i|k,\theta,y=1) = \frac{\sum_{u \in U} \varphi\left(u,i|k,\theta\right) \times R_{test}(u,i)} {\sum_{u \in U} R_{test}(u,i)} 
\label{eq:pik}
\end{equation}

\noindent where $R_{test}(u,i)=1$ if item $i$ is relevant for user $u$ in the ground-truth test set\footnote{While in this work we focus on an offline evaluation setting and the test set represents our ground truth, in case of an online evaluation (e.g., A/B testing), $R_{test}(u,i)=1$ if the user accepted the recommendation.}. The numerator counts the users who consider item $i$ as relevant in the test set and receive the item $i$ in their top-k. The denominator counts the users who consider item $i$ as relevant in the test set. Finally, we compute the item equal opportunity of a recommender system $\theta$ for a cutoff $k$ as the inverse of the Gini index across these probabilities: 

\begin{equation}
IEO(k,\theta) = 1 - Gini\left(\{p(i|k,\theta,y=1) \, | \, \forall \, i \in I \}\right)
\label{eq:ieo}
\end{equation}

\noindent If there is a perfect equality of being recommended when items are known to be of interest, then $IEO=1$. Conversely, $IEO$ decreases and gets closer to $0$ when the probability of being recommended is high for only few items of interest in the test set. This case occurs when most of the niche items never appear in the recommended lists, even if they are of interest (i.e., algorithmic bias emphasized the popularity phenomenon). Thus, $IEO$ ranges between 0 and 1, and the greater it is, the less the popularity bias is emphasized.

It should be noted that, while we provide a discussion on metrics for monitoring a popularity bias, it is ultimately up to the stakeholders (e.g., platform owners) to select the metrics and the trade-offs most suitable for their objectives.

\section{Exploratory Analysis on Popularity Bias}
\label{exploratory-analysis}
In this section, we show that representative algorithms from two of the most widely-adopted learning-to-rank optimization families~\cite{DBLP:journals/csur/ZhangYST19}, namely point-wise and pair-wise, produce biased recommendations with respect to ISP and IEO. 

\subsection{Datasets} \label{sec:data}
Since there is no standard benchmark framework for assessing a popularity bias in recommendation, our study in this paper considers two public datasets with different item distribution skews (Fig. \ref{fig:pop-dist}). This analysis treats ratings as a positive feedback to indicate that users are interested in the items they rated. Being oriented to learning-to-rank contexts, our analysis and the proposed debiasing approach can be applied to rating or frequency matrices as well. 

\begin{itemize}[leftmargin=*]
\item MovieLens1M (ML1M)~\cite{DBLP:journals/tiis/HarperK16} contains 998,131 ratings applied to 3,705 movies by 6,040 users of the online service MovieLens. The sparsity of the user-item matrix is 0.95. Each user rated at least 20 movies. 
\item COCO600k (COCO)~\cite{DBLP:conf/worldcist/DessiFMR18} includes 617,588 ratings provided by 37,704 learners to 30,399 courses of an online platform. The sparsity of the user-item matrix is 0.99. Each learner rated at least 10 courses.
\end{itemize}

\subsection{Recommendation Algorithms and Protocols} \label{sec:algo}
Our exploratory analysis covers four algorithms and investigates the recommendations they generate. Two of them are non-personalized algorithms with an opposite behavior with respect to item popularity: \emph{Random} is insensitive to popularity and uniformly recommends items, while \emph{MostPop} ignores tail items, and suggests the same few popular items to everyone\footnote{Even though comparing an algorithm against Random and MostPop has been previously studied~\cite{DBLP:journals/umuai/JannachLKJ15,DBLP:conf/ecir/BorattoFM19}, there is no evidence on how the new metrics monitor their outcomes.}. The other two algorithms \emph{NeuMF} \cite{DBLP:conf/sigir/0001C17} and \emph{BPR} \cite{DBLP:conf/wsdm/RendleF14} belong to the point-wise and the pair-wise families, respectively. They were chosen due to their performance and wide adoption as a basis of several state-of-the-art recommender systems \cite{DBLP:conf/ijcai/ZhangCZLS18,DBLP:conf/aaai/DengHWLY19,DBLP:conf/ijcai/XueDZHC17}. Please note that our methodology makes it easy to run this analysis on other algorithms.  

Point-wise approaches are generally required to optimize model parameters $\theta$ by minimizing the margin between the relevance $f(u,i|\theta)$ predicted for an item $i$ and the true binary relevance $y=R(u,i)$, given interactions in $R$:

\begin{equation}
\underset{\theta}{\operatorname{min}} \mathop{\sum}_{u\in U, i \in I} \mathcal{L}((u,i),y|\theta) = \mathop{\sum}_{u\in U, i \in I} - y \, ln\big(\widetilde{f}(u,i|\theta)\big) - (1-y) \, ln\big(1-\widetilde{f}(u,i|\theta)\big) + \xi\left\lVert \theta \right\rVert^2 
\label{eq:point-opt}
\end{equation}
\vspace{1mm}

\noindent where $\widetilde{f}(u,i|\theta)$ is the value obtained by applying the Sigmoid function $\sigma(\cdot)$ to the predicted relevance $f(u,i|\theta)$, $y \in \{0,1\}$ is the ground-truth relevance of $i$ for $u$, $\xi$ is the weight for l2 regularization, and $\theta$ represents the model parameters.

Conversely, pair-wise approaches estimate parameters $\theta$ by maximizing the margin between the relevance $f(u,i|\theta)$ predicted for an observed item $i$ and the relevance $f(u,j|\theta)$ predicted for an unobserved item $j$, given interactions in $R$:

\begin{equation}
\underset{\theta}{\operatorname{min}} \mathop{\sum}_{u\in U, i \in I^{+}_{u}, j \in I^{-}_{u}}  \mathcal{L}((u,i,j)|\theta) = - \mathop{\sum}_{u\in U, i \in I^{+}_{u}, j \in I^{-}_{u}} ln \, \sigma\big(f(u,i|\theta) - f(u,j|\theta)\big) + \xi\left\lVert \theta \right\rVert^2 
\label{eq:point-pair}
\end{equation}

\noindent where $I^{+}_{u}$ and $I^{-}_{u}$ are the items observed and not observed by $u$, $f(u,i|\theta)$ and $f(u,j|\theta)$ are the predicted relevances for the observed and unobserved item respectively, $\xi$ is the weight for l2 regularization, and $\theta$ marks model parameters. 

We performed a temporal train-test split with the most recent 20\% of ratings per user in the test set and the remaining 80\% ones in the training set. User and item matrices are initialized with values uniformly distributed in the range [0, 1]. Each model is served with training batches of $256$ examples. For NeuMF, for each user $u$, we created $t=4$ unobserved-item examples $((u,j),0)$ for each observed item $((u,i),1)$. For BPR, we created $t=4$ triplets $(u,i,j)$ per observed item $i$; the unobserved item $j$ is randomly selected. Such parameters were tuned in order to find a balance between training effectiveness and efficiency\footnote{In our study, we are more interested in better understanding beyond-accuracy characteristics of algorithms, so the further accuracy improvements that can probably be achieved through hyper-parameter tuning would not substantially affect the outcomes of our analyses.}. 

\subsection{Ranking Accuracy and Beyond-Accuracy Observations} \label{sec:exp}
\begin{figure}[!t] 
\begin{subfigure}[t]{0.49\linewidth}
    \centering
    \includegraphics[width=1.0\linewidth]{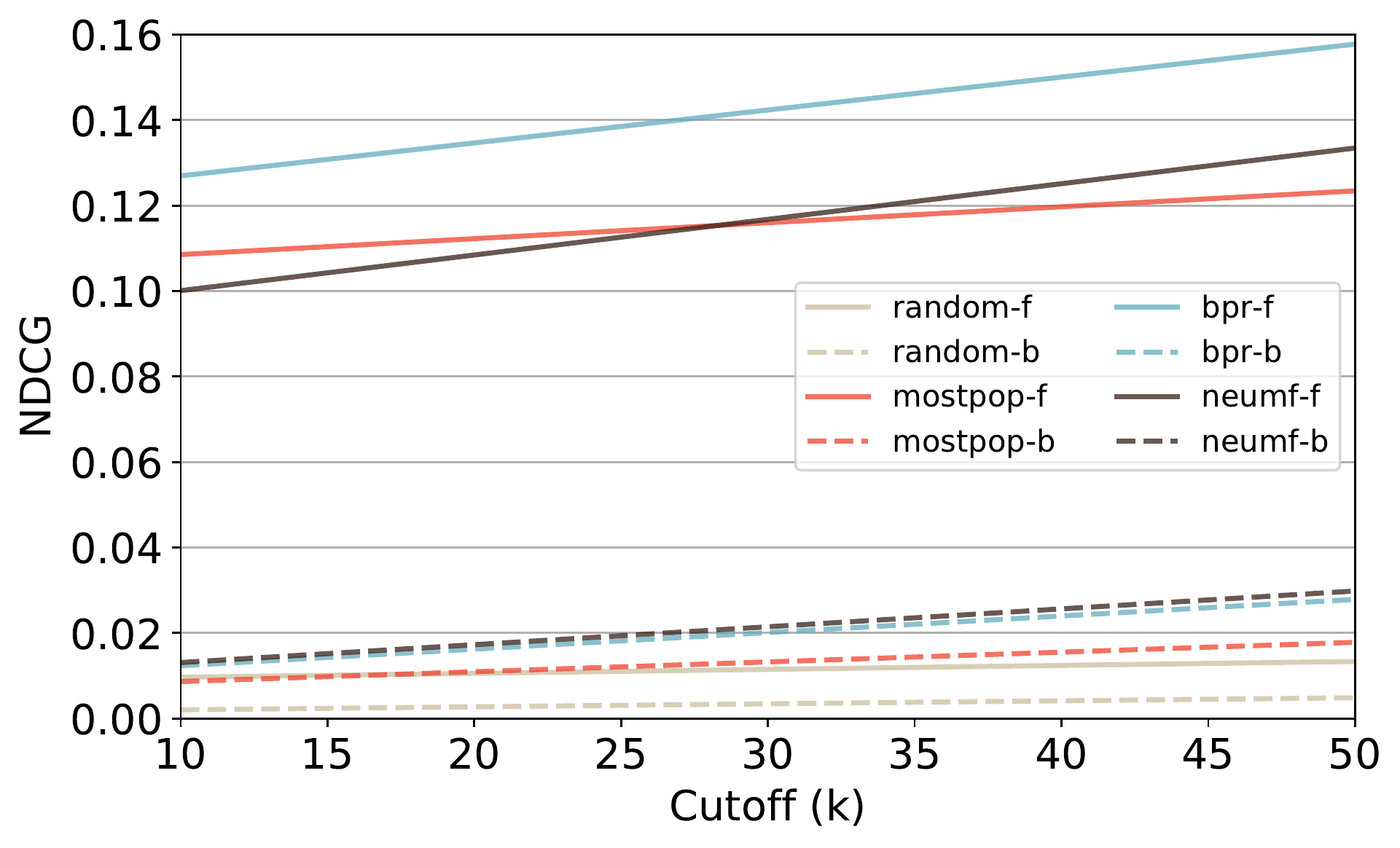}
    \caption{NDCG on ML-1M.}
\end{subfigure}
\begin{subfigure}[t]{0.49\linewidth}
    \centering
    \includegraphics[width=1.0\linewidth]{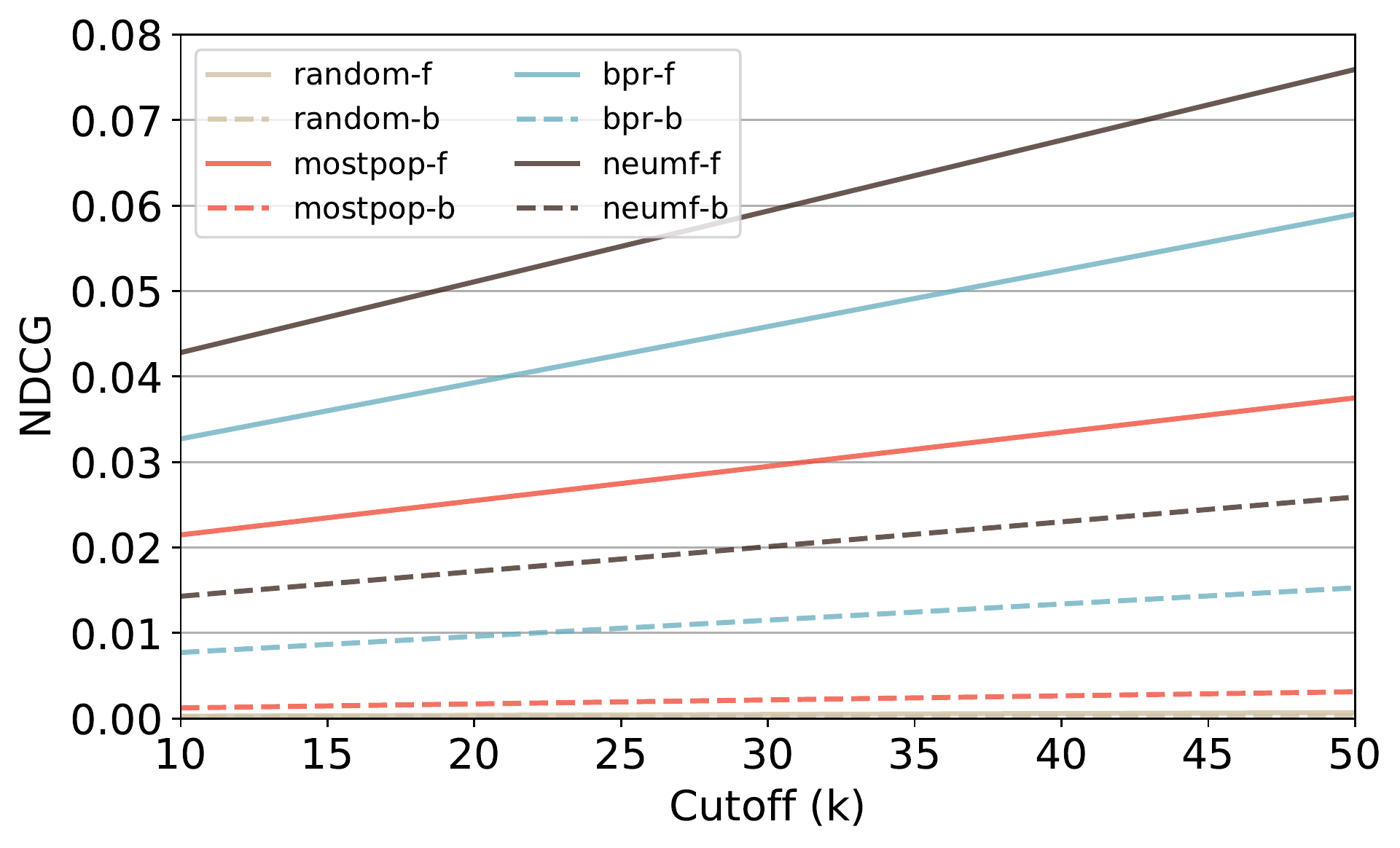}
    \caption{NDCG on COCO.}
\end{subfigure}
\begin{subfigure}[t]{0.49\linewidth}
    \centering
    \includegraphics[width=1.0\linewidth]{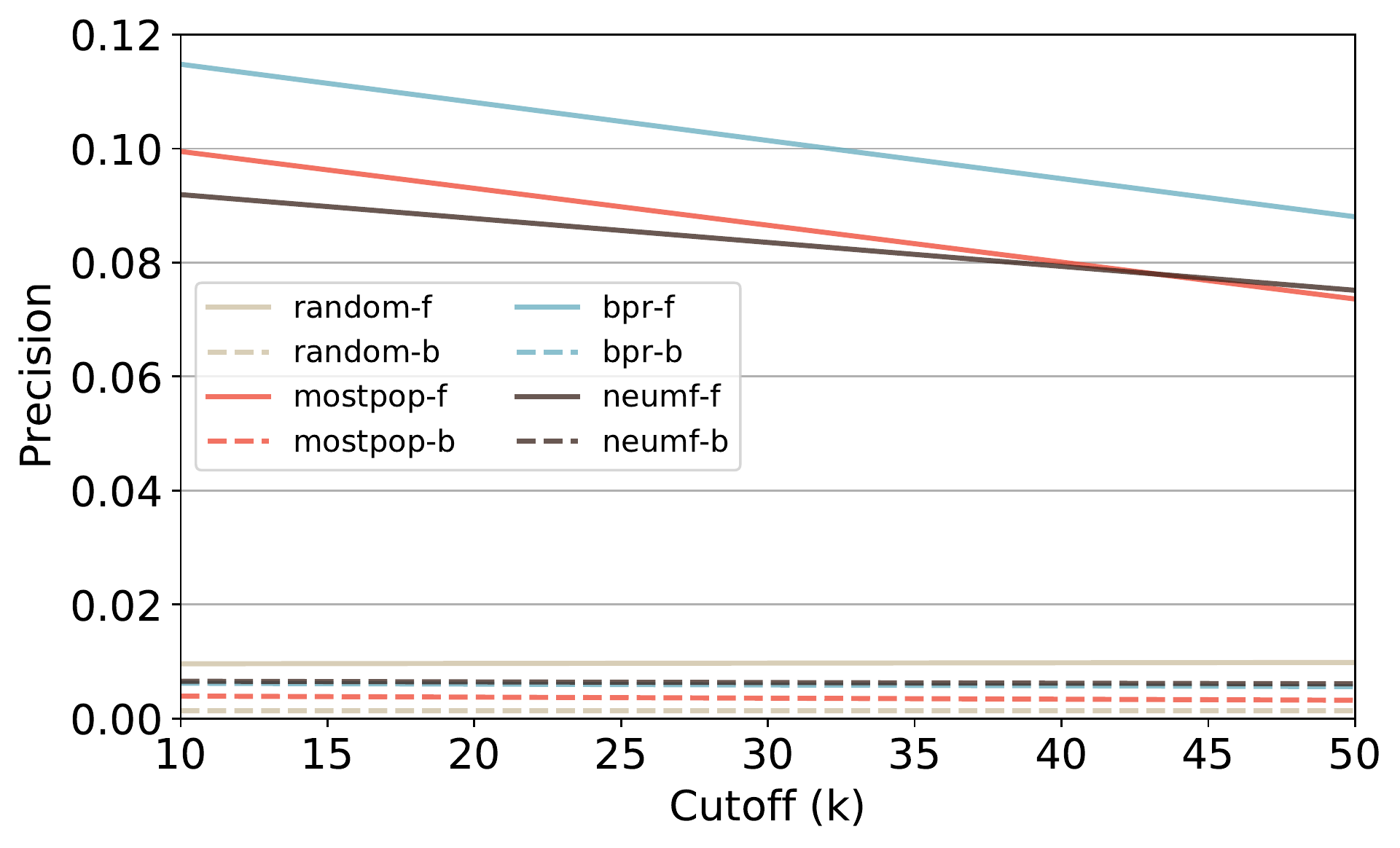}
    \caption{Precision on ML-1M.}
\end{subfigure}
\begin{subfigure}[t]{0.49\linewidth}
    \centering
    \includegraphics[width=1.0\linewidth]{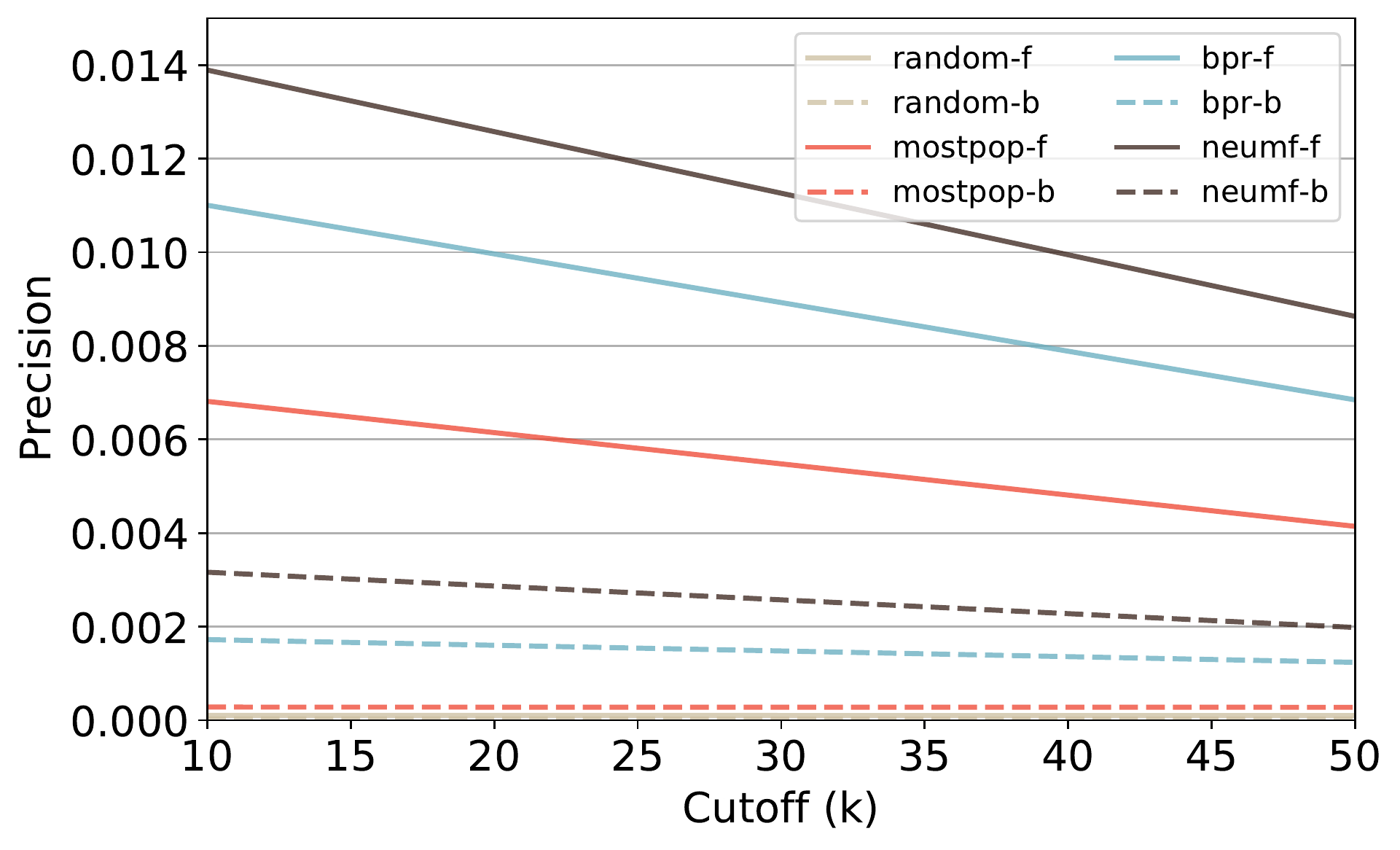}
    \caption{Precision on COCO.}
\end{subfigure}
\begin{subfigure}[t]{0.49\linewidth}
    \centering
    \includegraphics[width=1.0\linewidth]{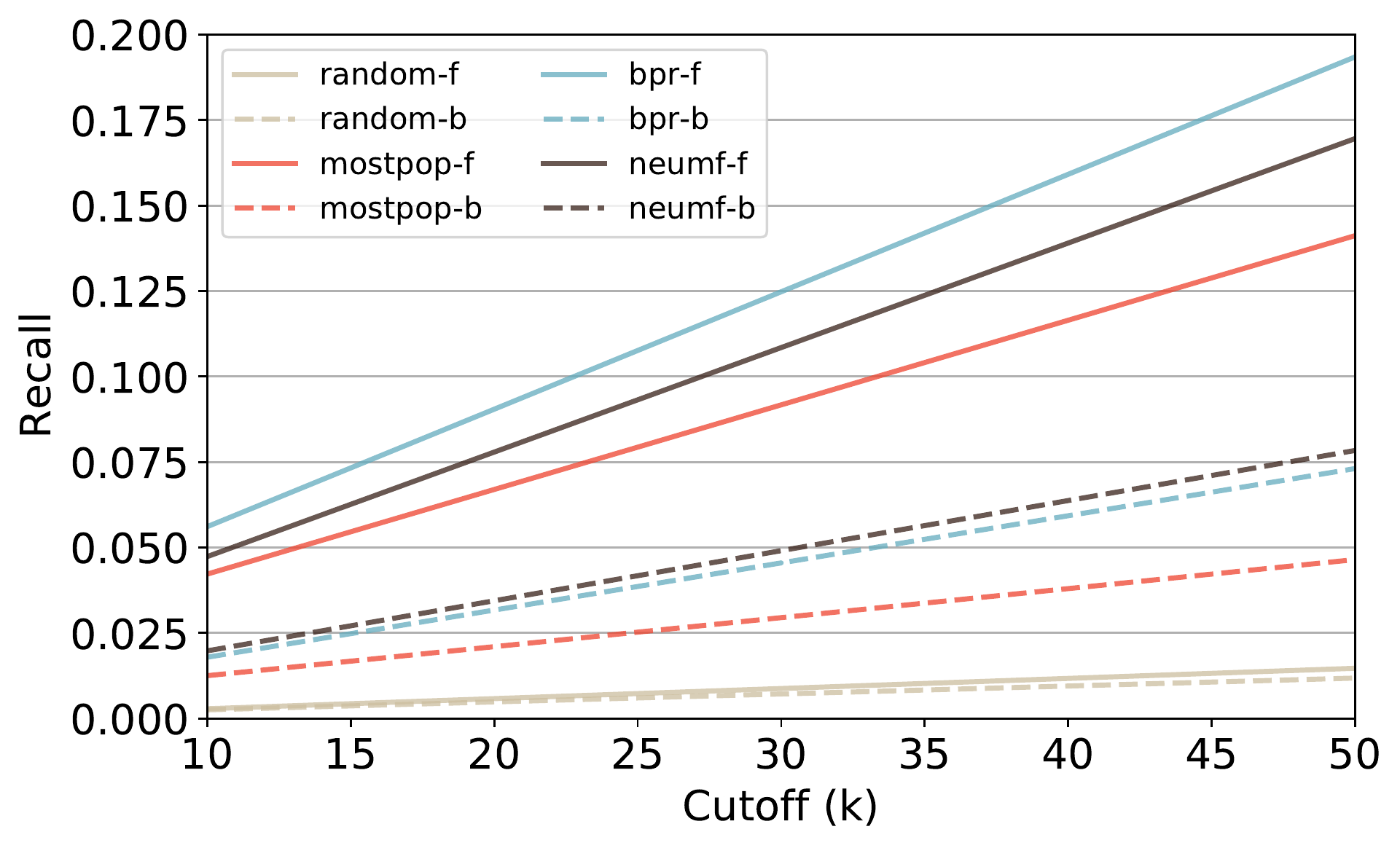}
    \caption{Recall on ML-1M.}
\end{subfigure}
\begin{subfigure}[t]{0.49\linewidth}
    \centering
    \includegraphics[width=1.0\linewidth]{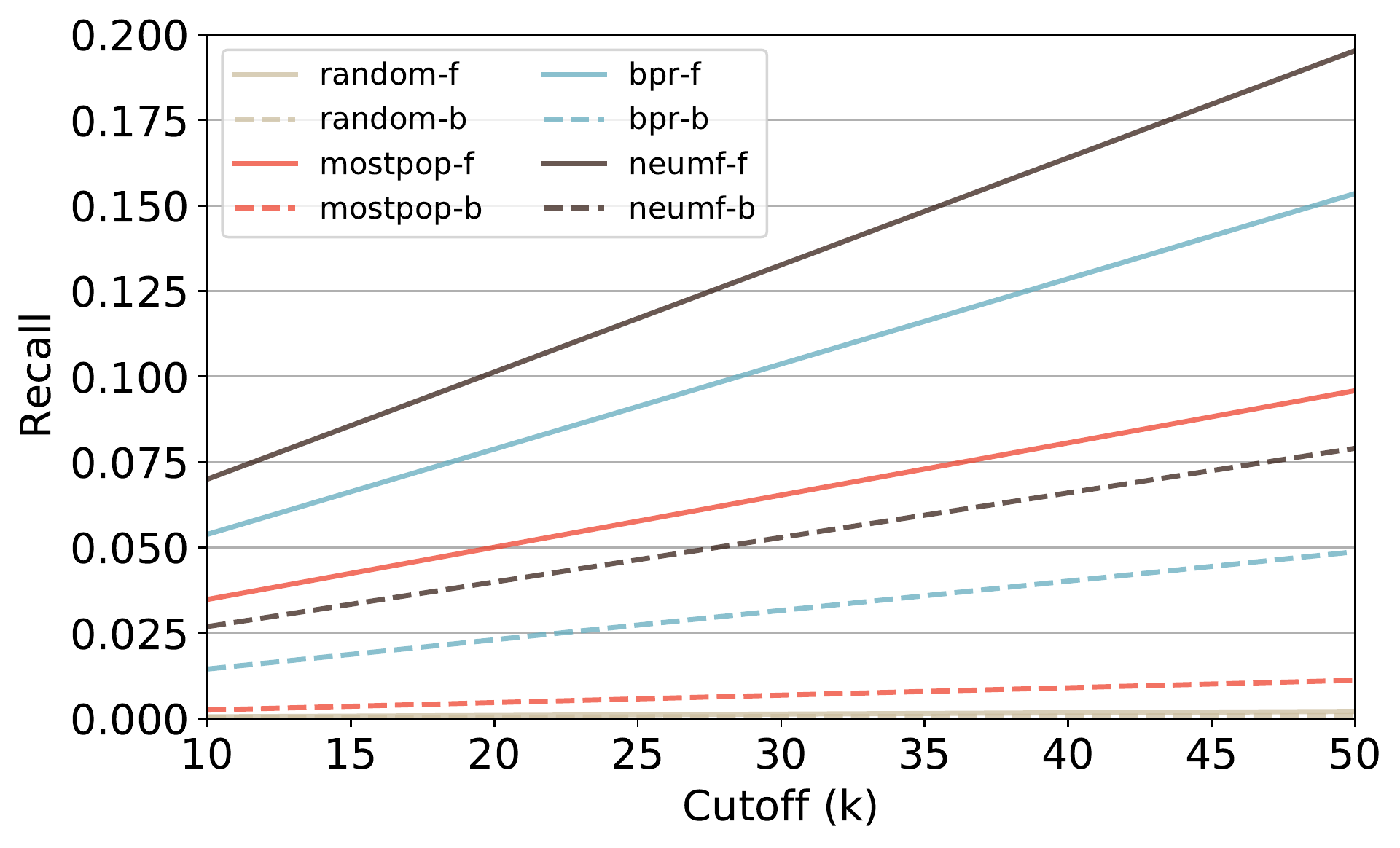}
    \caption{Recall on COCO.}
\end{subfigure}
\caption{\textbf{Ranking Accuracy}. Normalized  Discounted  Cumulative  Gain  (NDCG), Precision, and Recall achieved by the four considered recommendation algorithms. Straight lines (-f post-fix) indicate results on the full test set, while dashed lines (-b post-fix) represent results under a subset of the test set where all items have the same amount of test observations.}
\label{fig:exp-ndcg}
\end{figure}

First, we evaluated the ranking accuracy, considering Normalized Discounted Cumulative Gain (NDCG), Precision, and Recall as support metrics. The higher they are, the better the ranking. \figurename~\ref{fig:exp-ndcg} shows the ranking accuracy achieved by the four algorithms on ML1M and COCO. To test statistical significance, we performed paired two-tailed Students t-tests with a significance level of $0.05$.

The ranking accuracy achieved by MostPop (straight red line) on the full test set seemed to be highly competitive, especially in ML1M. This observation might reveal that the user-item feedback associated with the test set is unbalanced towards popular items, and it can bias evaluation metrics in favor of popular items. We thus examined an alternative experimental configuration, which considers a subset of the original test set where all items have the same amount of test observations~\cite{DBLP:journals/ir/BelloginCC17}. The accuracy scores decreased under the latter setup for BPR (dashed blue line), NeuMF (dashed brown line), and MostPop (dashed red line) over cutoffs. This fact confirms that all algorithms tend to be highly accurate because they mostly suggest popular items. 

\begin{figure}[!t] 
\begin{subfigure}[t]{0.49\linewidth}
    \centering
    \includegraphics[width=1.0\linewidth]{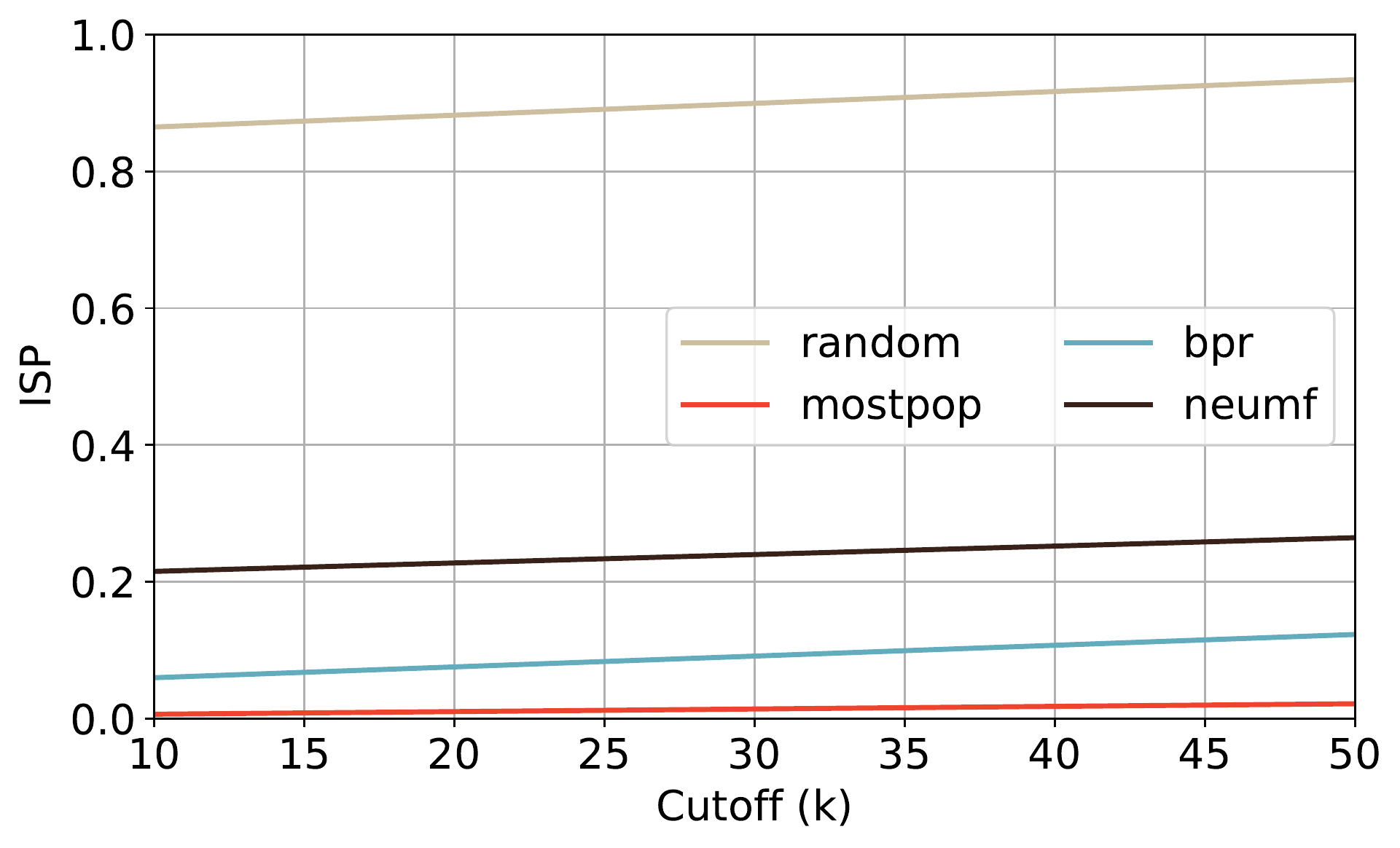}
    \caption{Item Statistical Parity on ML-1M.}
\end{subfigure}
\begin{subfigure}[t]{0.49\linewidth}
    \centering
    \includegraphics[width=1.0\linewidth]{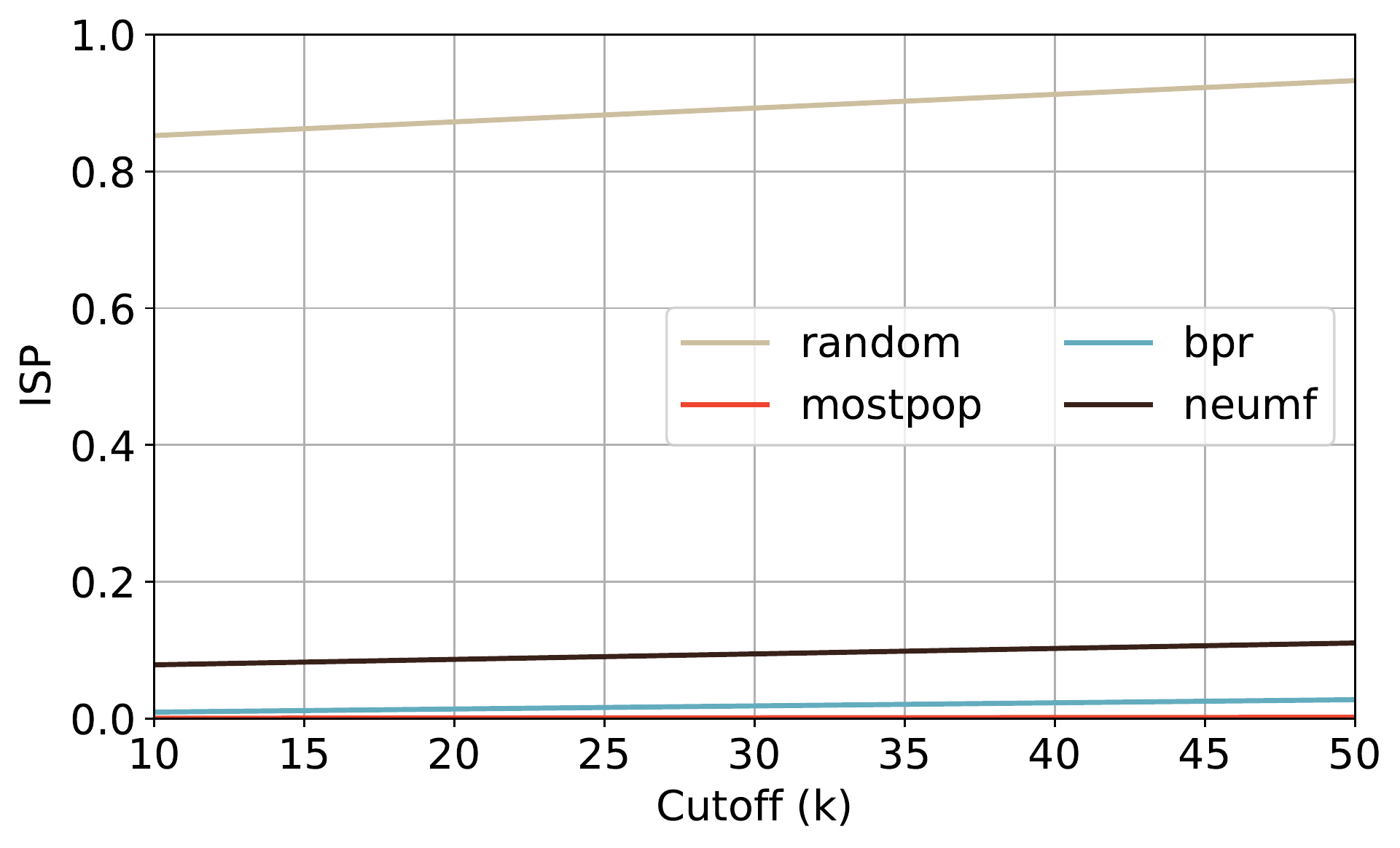}
    \caption{Item Statistical Parity on COCO.}
\end{subfigure}
\begin{subfigure}[t]{0.49\linewidth}
    \centering
    \includegraphics[width=1.0\linewidth]{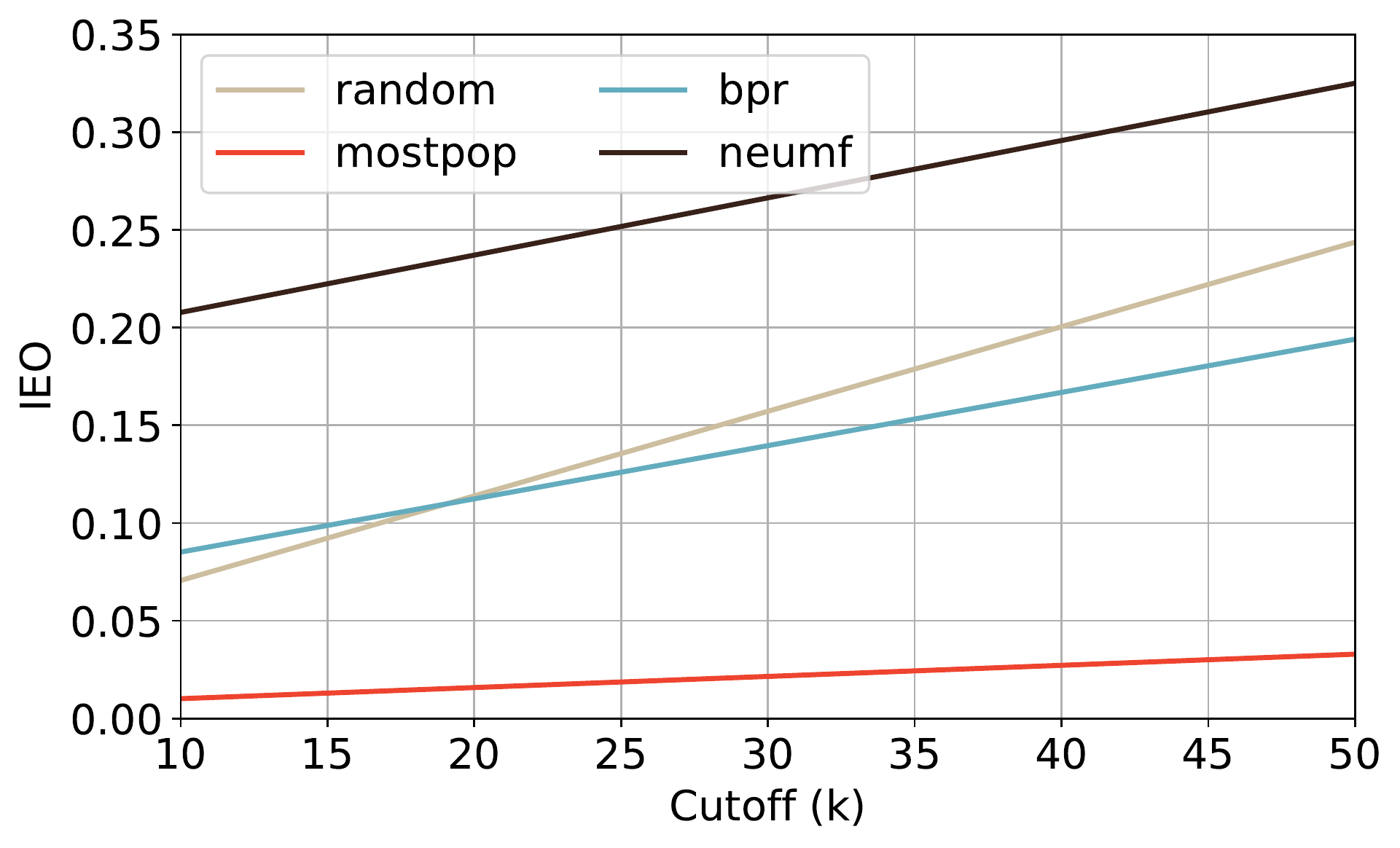}
    \caption{Item Equal Opportunity on ML-1M.}
\end{subfigure}
\begin{subfigure}[t]{0.49\linewidth}
    \centering
    \includegraphics[width=1.0\linewidth]{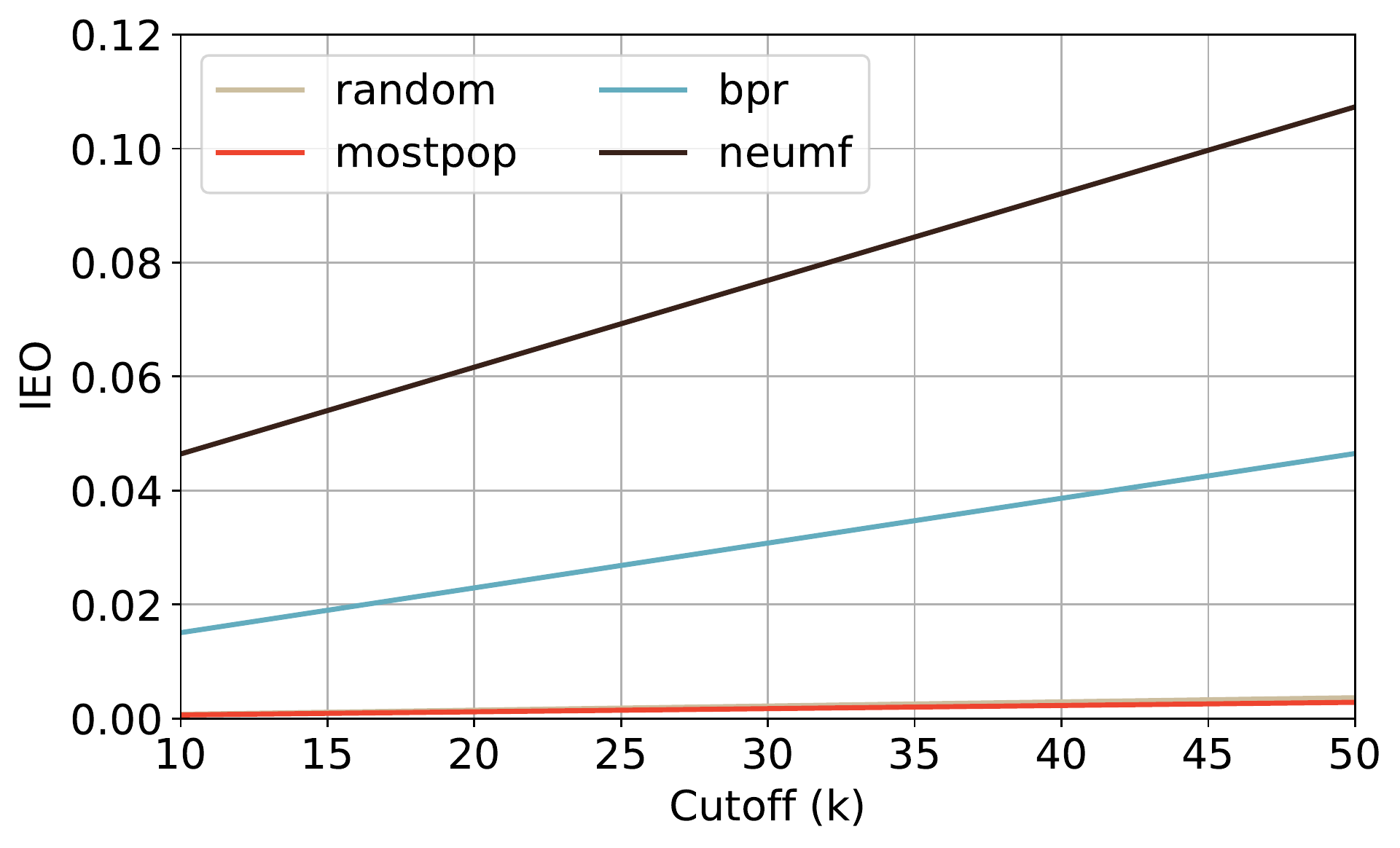}
    \caption{Item Equal Opportunity on COCO.}
\end{subfigure}
\caption{\textbf{Popularity Bias Monitoring}. Item Statistical Parity (ISP) that encourages the same probability of being recommended for items, and Item Equal Opportunity (IEO) that favors true positive rates of items to be equal. The higher the value is, the less the bias.}
\label{fig:exp-popularity}
\end{figure} 

The fact that MostPop performed similarly to BPR and NeuMF may imply that a recommender system optimized for ranking accuracy would not by default result in recommending sets with low item popularity estimates. We conjecture that optimizing for accuracy, without explicitly considering popularity bias, favors the latter. Motivated by the patterns observed in ranking accuracy, we analyzed the metrics introduced in Section \ref{problem-formulation} for monitoring popularity bias, namely Item Statistical Parity (ISP) and Item Equal Opportunity (IEO). From Figure~\ref{fig:exp-popularity}, we can observe that BPR, NeuMF, and MostPop did not achieve a good level of item statistical parity (top two plots). NeuMF and BPR statistical parity was significantly lower than the Random statistical parity, which was maximized by design. Moreover, the results on equal opportunity (bottom two plots) showed that the algorithmic bias emphasized the distortions in popularity across items. These low IEO values may uncover situations where ($i$) the true positive rates for popular items is high (i.e., the recommender's error for them is low) and ($ii$) the true positive rate for the other items is low or even zero. 
 
\vspace{2mm} \noindent \colorbox{gray!15}{\parbox{0.98\textwidth}{\textbf{Observation 1}. \textit{Point- and pair-wise optimization procedures reinforce disparate statistical parity and unequal opportunities over items. Such inequalities are stronger for pair-wise optimization, under sparsed datasets, at low cutoffs.}}} \vspace{2mm}  

\begin{figure}[!t] 
\begin{subfigure}[t]{0.49\linewidth}
    \centering
    \includegraphics[width=1.0\linewidth]{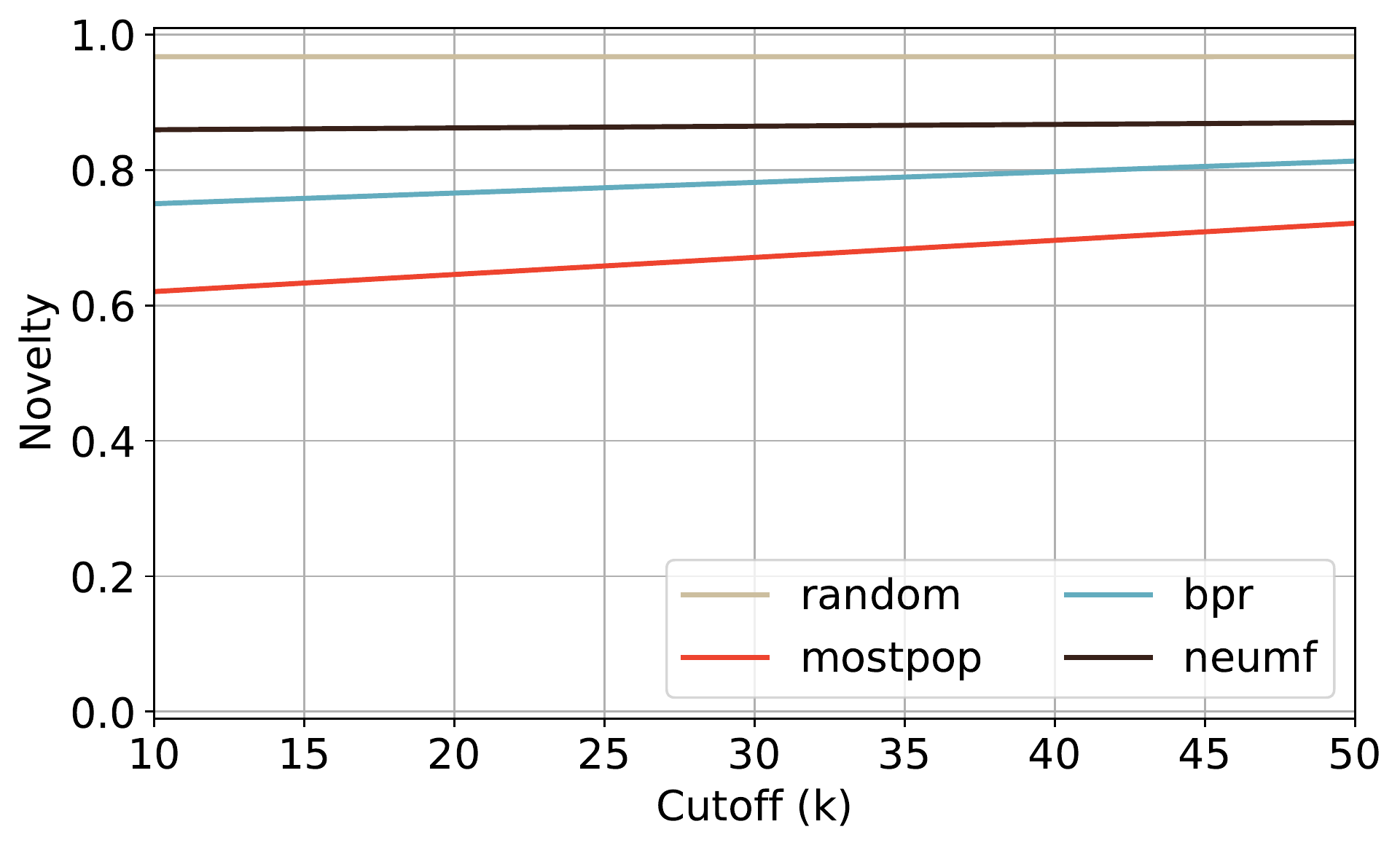}
    \caption{Novelty on ML-1M.}
\end{subfigure}
\begin{subfigure}[t]{0.49\linewidth}
    \centering
    \includegraphics[width=1.0\linewidth]{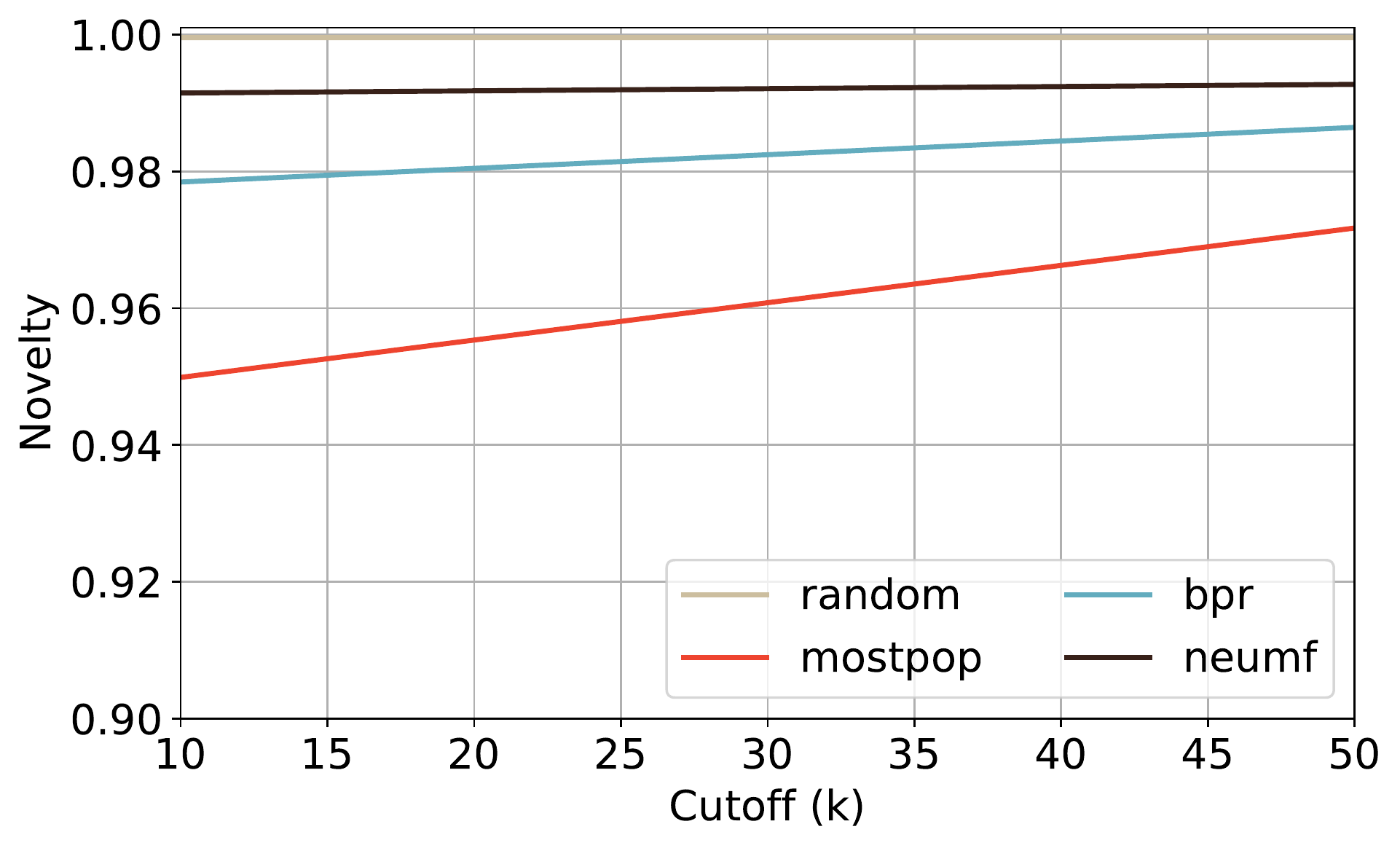}
    \caption{Novelty on COCO.}
\end{subfigure}
\begin{subfigure}[t]{0.49\linewidth}
    \centering
    \includegraphics[width=1.0\linewidth]{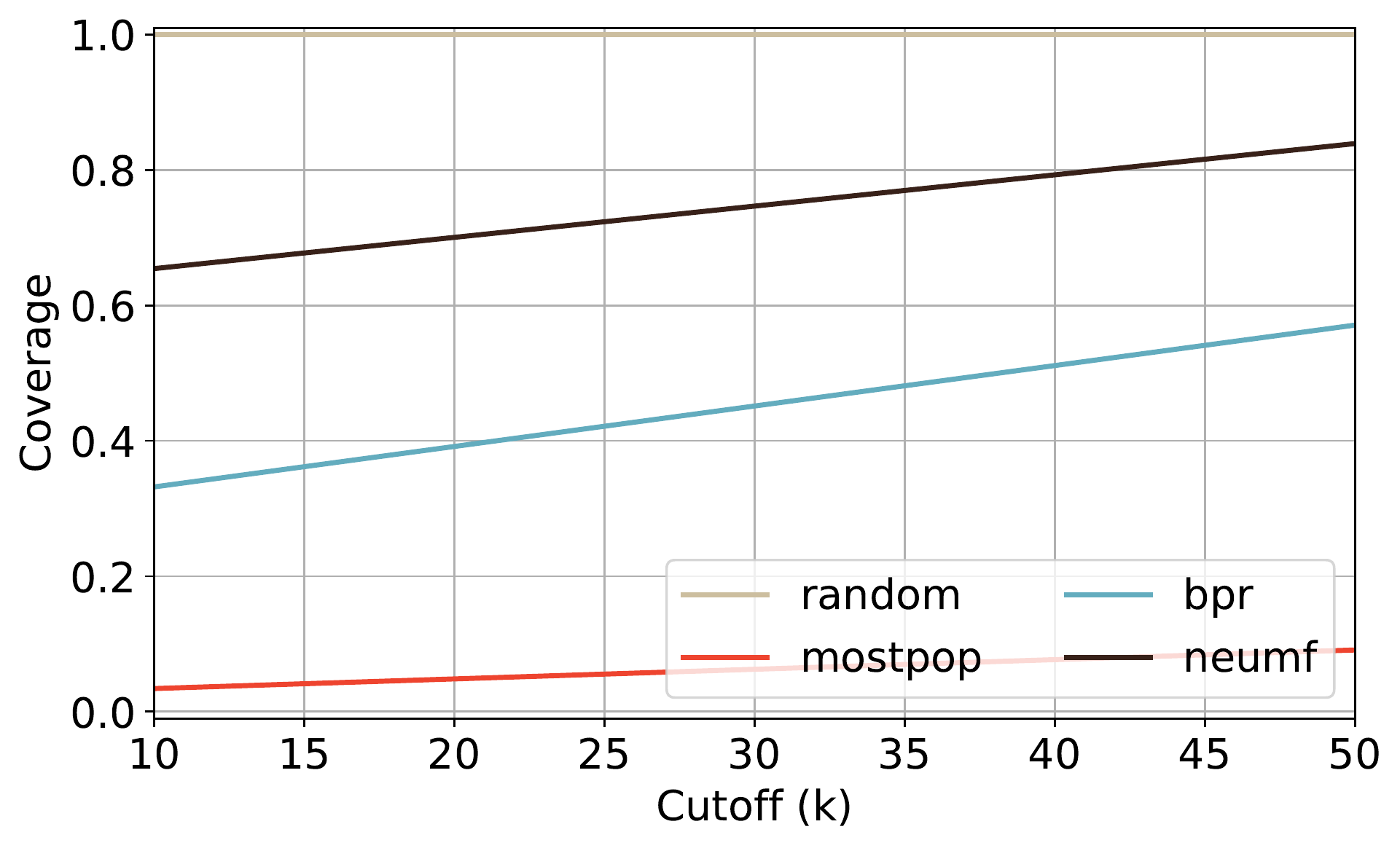}
    \caption{Catalog Coverage on ML-1M.}
\end{subfigure}
\begin{subfigure}[t]{0.49\linewidth}
    \centering
    \includegraphics[width=1.0\linewidth]{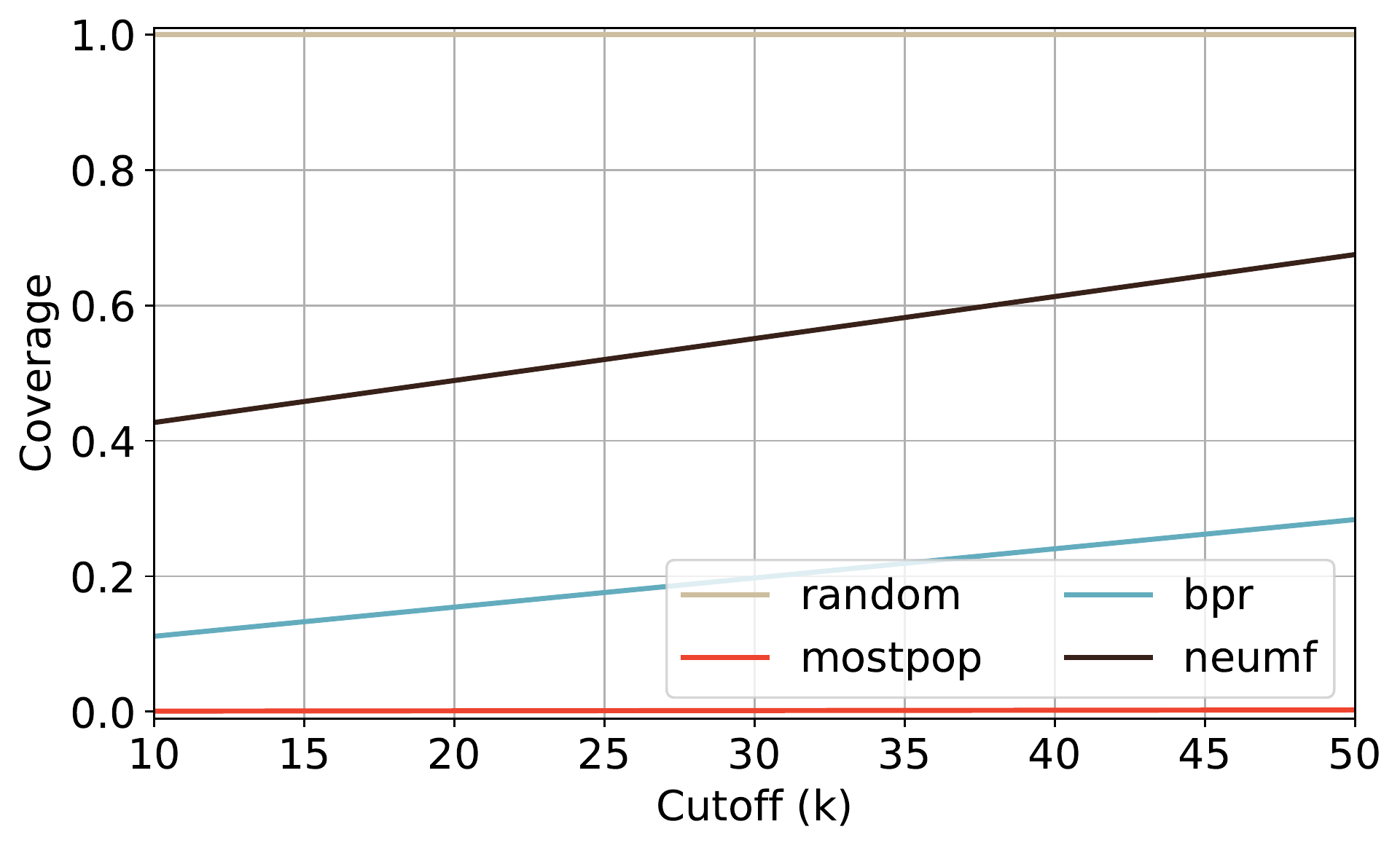}
    \caption{Catalog Coverage on COCO.}
\end{subfigure}
\caption{\textbf{Beyond-Accuracy Objectives}. Novelty computed as the inverse of the average popularity of a ranked item, and Coverage calculated as the ratio of items appearing at least once in a top-k list. The higher the score is, the better the beyond-accuracy goal is met.}
\label{fig:exp-beyond}
\end{figure} 

Low ISP and IEO reveal a high degree of bias against popularity according to our definitions, which may hamper the quality of recommendations (i.e., an algorithm might fail to learn users' preferences for niche items, even if they are known to be of interest). Due to social dynamics, information cascades, and highly subjective notions, it would not be feasible to devise a single definition of quality. Therefore, this study simultaneously looks at both ranking accuracy and beyond-accuracy metrics as a proxy of the overall quality of a recommended list~\cite{DBLP:journals/tiis/KaminskasB17}. These properties are particularly important in real-world systems, since users are likely to consider only a small set of top-k recommendations. It is thus crucial to ensure that this set is as much interesting and engaging as possible\footnote{While we bring forth the discussion about such metrics, it is ultimately up to the stakeholders to select the metrics and the trade-offs most suitable for their applicative context.}.

Figure~\ref{fig:exp-beyond} depicts item novelty and catalog coverage, which are two widely-known beyond-accuracy goals in recommendation, on ML1M and COCO~\cite{DBLP:journals/tiis/KaminskasB17}. Specifically, item novelty is computed as the inverse of the relative number of users who interacted with an item in the training set with respect to the total number of users (i.e., the less users interacted with an item, the higher is its novelty). Coverage is computed as the percentage of items in the catalog that are recommended at least in one top-k list. Both metrics range between 0 and 1. The higher the novelty/coverage is, the better the beyond-accuracy goal is met. As expected, while comparing metrics across the considered models, we observed that BPR and NeuMF recommend more novel items than MostPop and that the former models better cover the item catalog. However, the difference between the scores computed for MostPop, BPR, and NeuMF was often small, and all the novelty and coverage estimates were still far from the optimal value of 1, especially at small cutoffs\footnote{ML1M and COCO lead to different novelty and coverage scores. This difference across datasets reflects their characteristics, with COCO having a higher sparsity and a much larger number of users and items. Having a lot of users means that an item can be easily novel for someone, while having a lot of items implies that covering the whole catalog may be harder.}. Hence, we argue that the bias against item popularity might have a key role in the recommendation process carried out by BPR and NeuMF and that this factor can negatively influence the overall recommendation quality. This consideration motivated us to deepen some aspects connected to the internal mechanics of the recommendation algorithms. 
 
\vspace{2mm} \noindent \colorbox{gray!15}{\parbox{0.98\textwidth}{\textbf{Observation 2}. \textit{The higher the item statistical parity and equal opportunity are, the newer and catalog-wider the recommendations. This relation does not inversely affect the ranking accuracy, if a balanced test set is considered.}}} \vspace{2mm}

\subsection{Internal Mechanics Analysis}
\label{sec:diagnosis}
Motivated by our findings, we next explored the internal mechanics of the considered recommendation algorithms to better understand how disparate statistical parity and unequal opportunity across items are internally emphasized.

Each algorithm required to train a model by optimizing a certain objective function. This process aims to improve the model's ability of predicting a high relevance for items known to be of interest for users in the training set. The fact that MostPop NDCG is close to the NDCG of BPR and NeuMF, and that a low estimate of IEO was achieved by the considered models, suggested to further investigate such an algorithm's ability on the basis of the observed-item popularity. Therefore, we analyzed the effectiveness of each model in terms of pair-wise accuracy while predicting the user's relevance for head and mid observed items\footnote{The pair-wise accuracy can be considered as a proxy of the Area Under Curve (AUC) metric, which is associated to the probability that the model will rank a randomly chosen observed item higher than a randomly chosen unobserved item, for a given user.}. We randomly sampled four sets of user-feedback triplets $(u,i,j)$. Each triplet in the first set included an observed head item as $i$ and an unobserved head item as $j$. Triplets in the second set relied on observed head items as $i$ and unobserved mid items as $j$. The third set of triplets included observed mid items as $i$ and unobserved head items as $j$. The fourth set had observed mid items as $i$ and unobserved mid items as $j$. Head and mid popularity thresholds were set up according with the percentiles reported in Figure~\ref{fig:pop-dist}. For each model and set of triplets, we computed the model's accuracy on predicting a relevance for the observed item higher than the relevance for the unobserved item. 

\begin{table}[!t]
   
\resizebox{\textwidth}{!}{
\begin{small}
\begin{tabular}{cc|cc|cc}
   
\hline
\multirow{2}{*}{\textbf{Observed Item}} & \multirow{2}{*}{\textbf{Unobserved Item}} & \multicolumn{2}{c}{\textbf{ML-1M}} & \multicolumn{2}{c}{\textbf{COCO}} \\
 &  & \textbf{BPR} & \textbf{NeuMF} & \textbf{BPR} & \textbf{NeuMF} \\
\hline
Head & Any & 0.92 & 0.93 & 0.90 & 0.96 \\
Mid & Any & 0.72 & 0.81 & 0.76 & 0.94 \\
\hline
Head & Head & 0.86 & 0.89 & 0.84 & 0.93 \\
Head & Mid & 0.98 & 0.97 & 0.97 & 0.99 \\
Mid & Head & 0.53 & 0.71 & 0.61 & 0.89 \\
Mid & Mid & 0.89 & 0.91 & 0.91 & 0.98 \\
\hline 
\end{tabular} \end{small}}
\caption{\textbf{Pair-wise Accuracy}. Pair-wise accuracy over observation triplets spanning diverse parts of the popularity tail. The more the cases where the relevance for the observed item is higher than the relevance for the unobserved item are, the higher the pair-wise accuracy.}
\label{tab:pair-acc}
\end{table}

From Table~\ref{tab:pair-acc}, we observed that the pair-wise accuracy achieved by BPR and NeuMF strongly depends on the popularity of the considered items $i$ and $j$. Specifically, the models failed more frequently in giving higher relevance to observed mid items, especially when they were compared against unobserved head items. Conversely, the models ended up performing significantly better when observed head items were compared against unobserved mid items. 
 
\vspace{2mm} \noindent \colorbox{gray!15}{\parbox{0.98\textwidth}{\textbf{Observation 3}. \textit{Observed mid items, even when of interest, are more likely to receive less relevance with respect to head items. This effect is stronger when the users' feedback matrix is less sparse.}}} \vspace{2mm}

\begin{figure}[!t] 
\begin{subfigure}[t]{0.49\linewidth}
    \centering
    \includegraphics[width=1.0\linewidth]{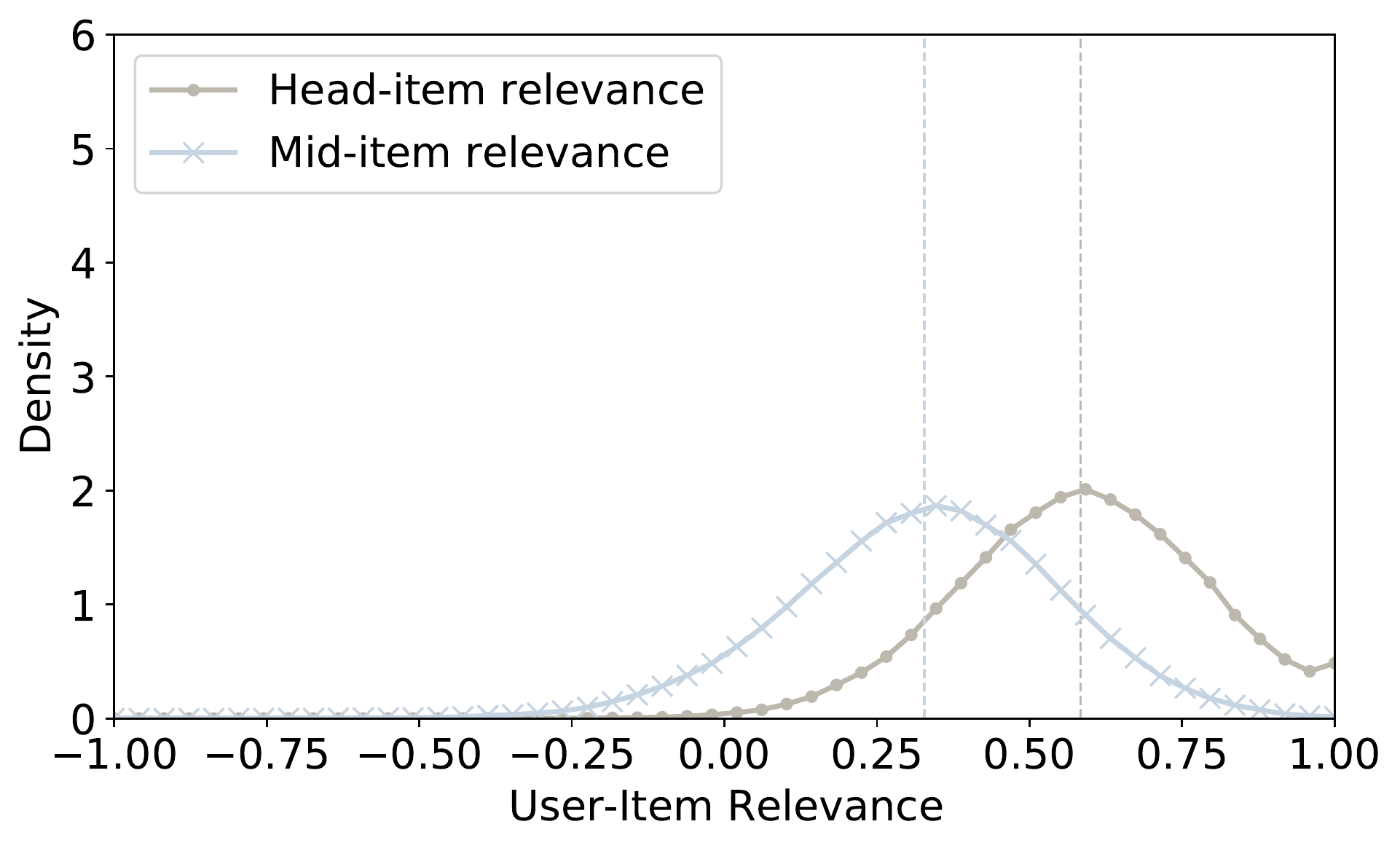}
    \caption{BPR on on ML-1M.}
\end{subfigure}
\begin{subfigure}[t]{0.49\linewidth}
    \centering
    \includegraphics[width=1.0\linewidth]{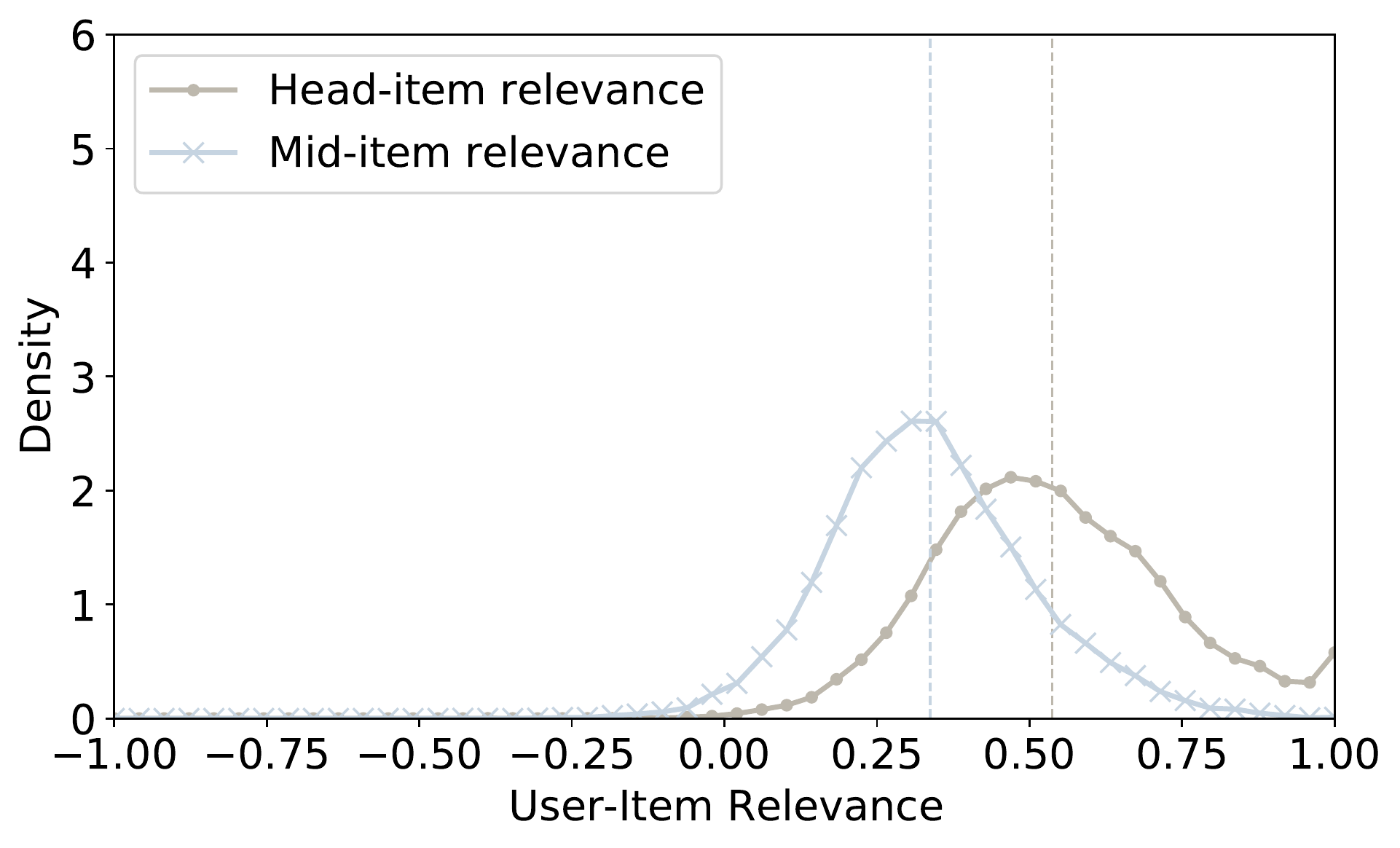}
    \caption{BPR on COCO.}
\end{subfigure}
\begin{subfigure}[t]{0.49\linewidth}
    \centering
    \includegraphics[width=1.0\linewidth]{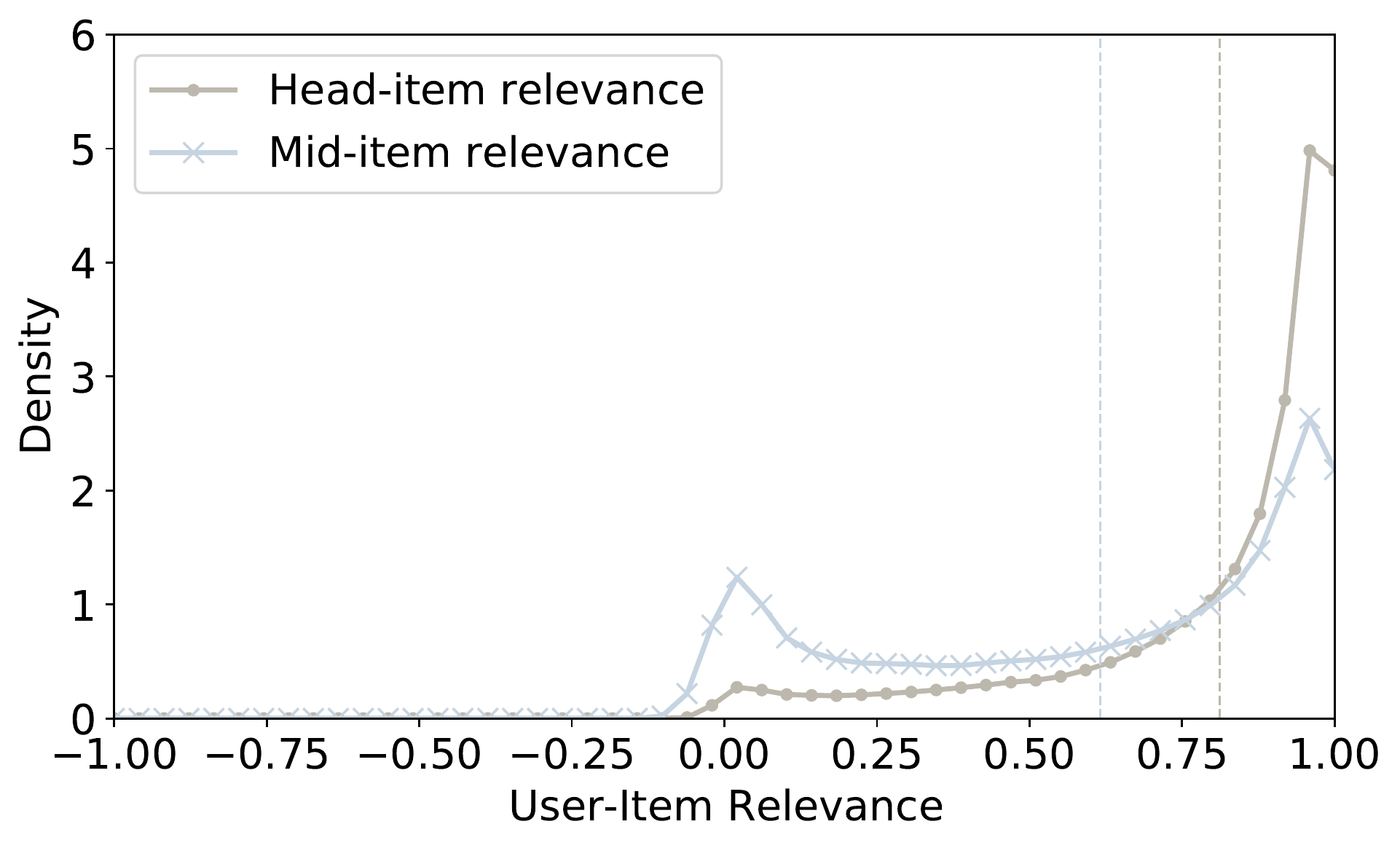}
    \caption{NeuMF on ML-1M.}
\end{subfigure}
\begin{subfigure}[t]{0.49\linewidth}
    \centering
    \includegraphics[width=1.0\linewidth]{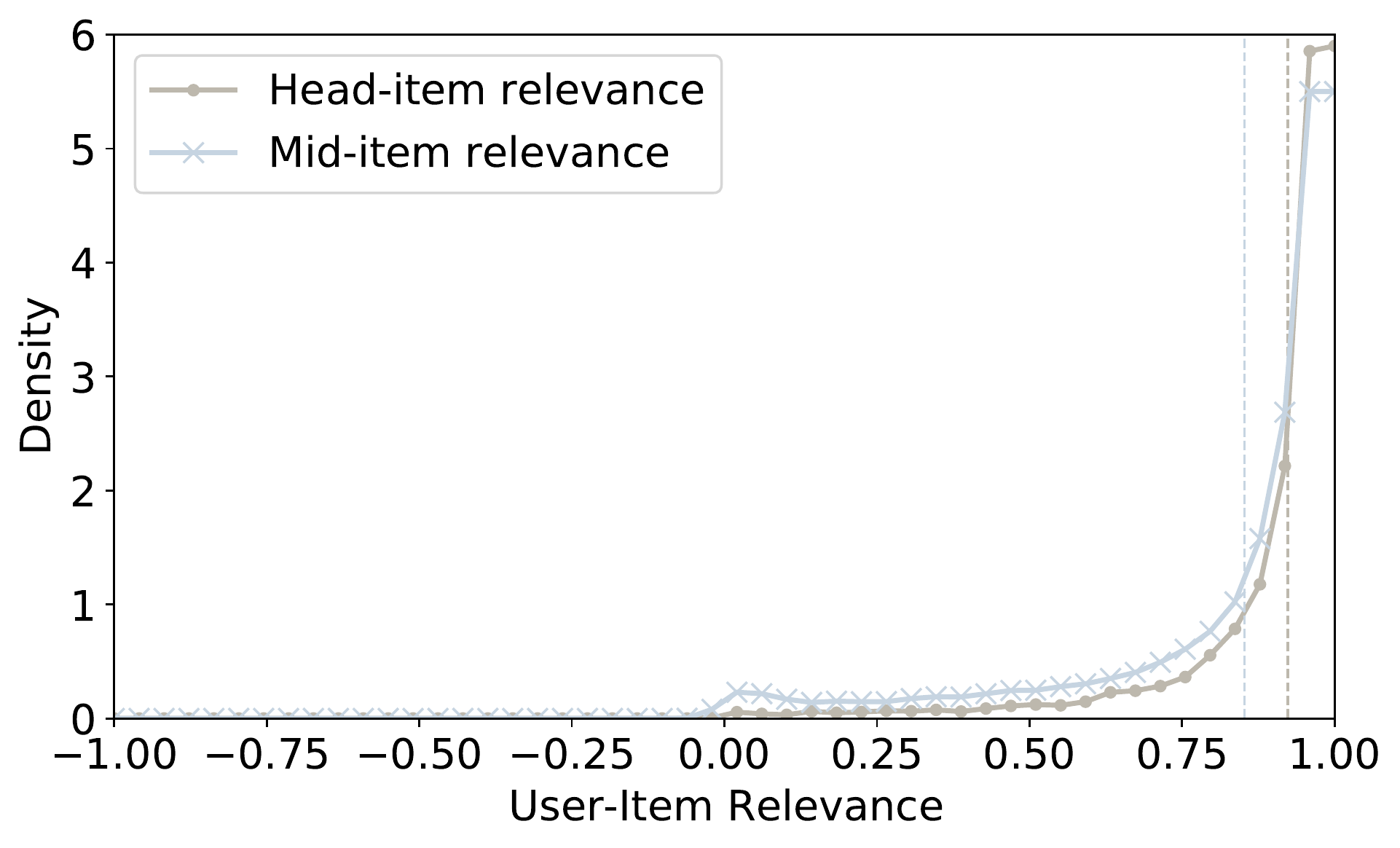}
    \caption{NeuMF on COCO.}
\end{subfigure}
\caption{\textbf{Relevance Score Distribution}. Head-item and mid-item relevance obtained for randomly-sampled pairs of users and observed items in ML1M and COCO, balancing the cases where the observed item belongs to the head or the mid part of the popularity tail.}
\label{fig:relevance-distr}
\end{figure} 

This result might depend on the fact that, in presence of a popularity bias, the biased differences in relevance scores across items play a key role in pair-wise accuracy. Thus, Figure~\ref{fig:relevance-distr} depicts the distribution of the user-item relevance scores obtained for observed head items and observed mid items in the training data. For each user $u$, we randomly sampled pairs of items, each including a head item and a mid item that user $u$ interacted with in the training set. Then, we computed the head-item and the mid-item relevance for user $u$. The process was repeated over the users' population in order to build two probability distributions. It can be observed that the distributions are significantly different in all the setups, and that there is a tendency of observed mid items of getting lower relevance. This result uncovered a biased behavior of the models, which were under-considering observed mid items regardless of the real users' interests. 
 
\vspace{2mm} \noindent \colorbox{gray!15}{\parbox{0.98\textwidth}{\textbf{Observation 4}. \textit{Relevance distributions for observed head items and mid items are significantly different. Observed head items are more likely to obtain higher relevance than observed mid items, being over-represented in top-k lists.}}} \vspace{2mm}

\begin{figure}[!t] 
\centering
\includegraphics[width=0.5\linewidth]{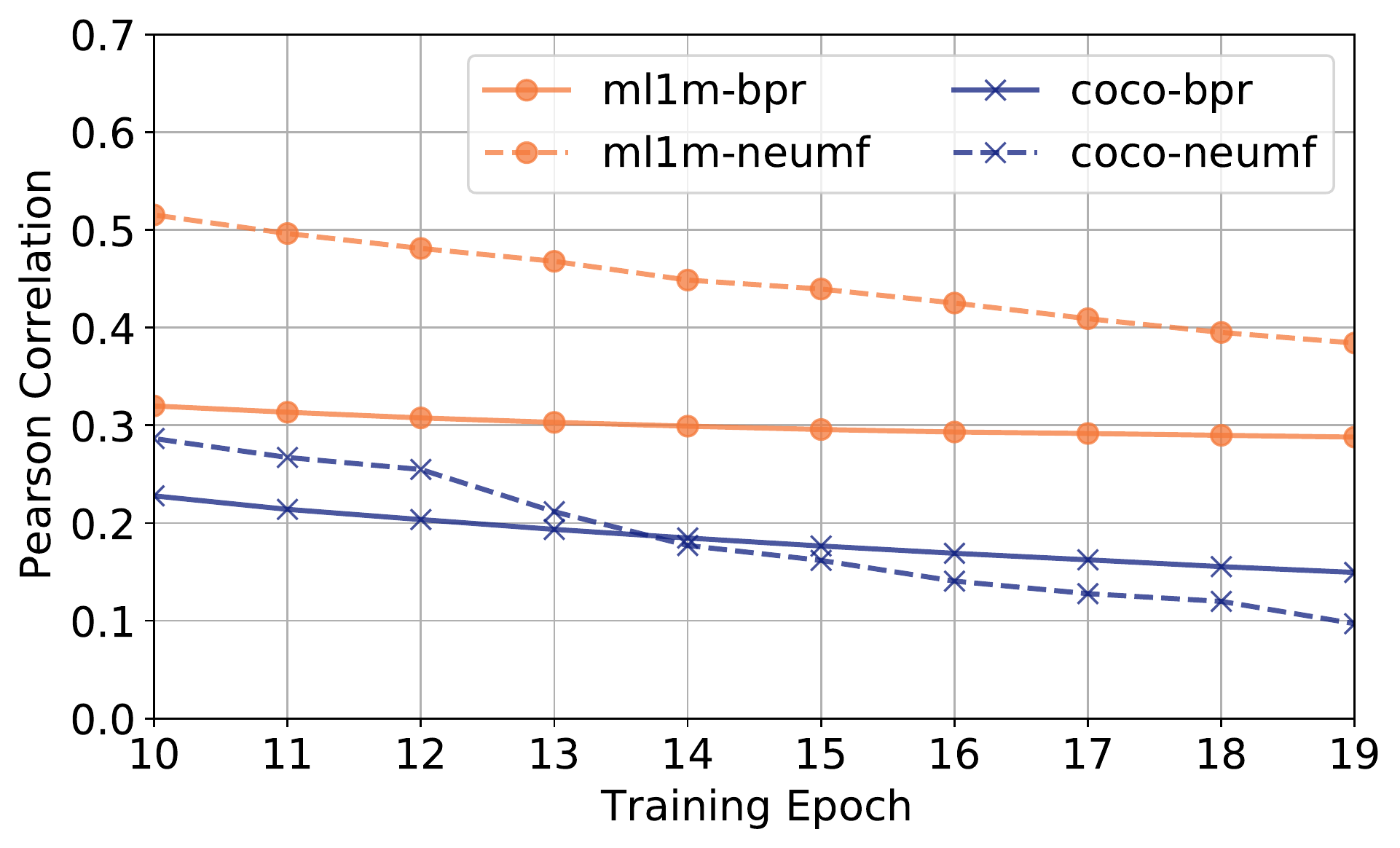}
\caption{\textbf{Correlation between Item Popularity and Predicted Item Relevance}. Mean Pearson correlation between the popularity of the observed item and the relevance predicted for that item, throughout batches included into each training epoch.}
\label{fig:corr-train}
\end{figure}

The observations seen so far are rooted in the fact that each recommendation algorithm emphasized a direct relation between item relevance and item popularity, emerged also throughout the training procedure (Figure \ref{fig:corr-train}). This effect makes the recommender system biasedly less accurate when mid items are considered, even when they are of interest. It is interesting to ask whether minimizing such a biased correlation might make a positive impact on recommended item popularity and beyond-accuracy objectives, retaining ranking accuracy.

\section{The Proposed Mitigation Procedure}
\label{popularity-reg}
With an understanding of the internal mechanics associated with point- and pair-wise optimization functions, we investigate how a recommender system can generate recommendations less biased against popularity. To this end, we propose a mitigation procedure that aims at minimizing both ($i$) the loss function targeted by the considered recommendation algorithm, either Eq.~\ref{eq:point-opt} or \ref{eq:point-pair}, and ($ii$) the biased correlation between the predicted relevance and the popularity of the observed item. Though we rely on BPR and NeuMF along our experiments, our procedure can be seamlessly applied to other algorithms from these families.

Correlation-based regularization approaches have been proved to be empirically effective in several domains \cite{DBLP:conf/kdd/BeutelCDQWWHZHC19,DBLP:conf/aies/BeutelCDQWLKBC19}. Differently from this prior work, the popularity debiasing task requires to relax the assumption of knowing group memberships of input samples, since we target individual items regardless of their head or mid membership. Our task inspects relative differences of relevance and popularity rather than differences of predicted labels and group labels. Further, we do not rely on any arbitrary split between head and mid items. Lastly, as we tackle a popularity debiasing perspective, the design choices we made lead to examine training processes and model facets so far under-explored.    

\begin{figure}[!t] 
\centering
\includegraphics[width=1.0\linewidth]{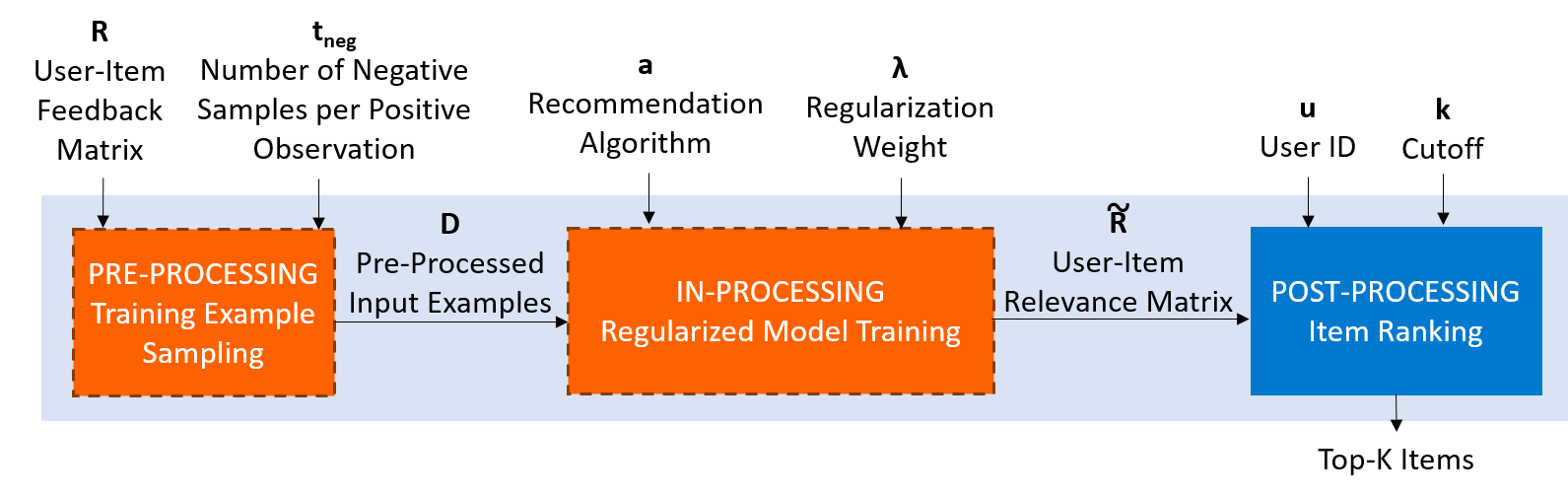}
\caption{\textbf{Popularity Bias Mitigation}. The proposed popularity-bias mitigation framework and its parameters. Orange dashed boxes indicate the components affected by our procedure.}
\label{fig:approach-schema}
\end{figure}

Our mitigation procedure combines both pre- and in-processing operations, which extend the common data preparation and model training procedures of the original recommendation algorithms (Figure \ref{fig:approach-schema}). Specifically, with a slight difference between point- and pair-wise settings, the proposed mitigation procedure is based on the following steps:

\vspace{2mm} \noindent \textbf{Training Examples Mining (sam)}. Under a point-wise optimization setting, $t$ unobserved-item pairs $((u,j),0)$ are created for each observed user-item interaction $((u,i),v)$. The observed interaction $((u,i),v)$ is replicated $t$ times to ensure that our correlation-based regularization will work. On the other hand, under a pair-wise optimization setting, for each user $u$, $t$ triplets $(u,i,j)$ per observed user-item interaction $(u,i)$ are generated. In both settings, the unobserved item $j$ is selected among the items less popular than $i$ for $t/2$ training examples, and among the items more popular than $i$ for the other $t/2$ examples. These operations enable our regularization, as the training examples equally represent both popularity sides associated with the popularity of the observed item, subjected to correlation computing. We denote training examples as $D$.  

\vspace{2mm} \noindent \textbf{Regularized Optimization (reg)}. The training examples in $D$ are fed into the original recommendation model $a$ in batches $D_{batch} \subset D$ of size $m$ to perform an iterated stochastic gradient descent. Regardless of the family of the algorithm, the optimization approach follows a regularized paradigm derived from the original point- and pair-wise optimization functions. Specifically, the regularized loss function we propose can be formalized as follows:     
\begin{equation}
\underset{\theta}{\operatorname{min}} \,\, (1 - \lambda) \; \mathcal{L}(D_{batch}|\theta) + \lambda \; \mathcal{C}(D_{batch}|\theta) 
\label{eq:correlation}
\end{equation}

where $\lambda \in [0,1]$ is a weight that expresses the trade-off between the accuracy loss and the regularization loss. With $\lambda=0$, we yield the accuracy loss, not taking the regularization loss into account. Conversely, with $\lambda=1$, the accuracy loss is discarded and only the regularization loss is minimized. 

The accuracy loss function $\mathcal{L}(\cdot)$ depends on the family of the involved recommendation algorithm. For instance, it could be either Eq. \ref{eq:point-opt} for point-wise optimization settings or Eq. \ref{eq:point-pair} for pair-wise optimization settings. This aspect will make our mitigation procedure easily applicable to other algorithms of the same families, with no changes on their original implementation. Lastly, $\mathcal{C}(.)$ is introduced in this paper to regularize the biased correlation between ($i$) the predicted residuals and ($ii$) the observed-item popularity, as:

\begin{equation}
\mathcal{C}(D_{batch}|\theta) = |Correlation(A_1, A_2)| 
\label{eq:reg-corr-type}
\end{equation}

\noindent where $Correlation$ is a function that computes the correlation across two distributions $A_1$ (predicted residuals) and $A_2$ (observed-item popularities):

\begin{equation}
A_1(b) = \mathcal{L}(D_{batch}(b)|\theta) \, \, \, \text{with} \, b \in \{0, \dots, |D_{batch}|\}
\label{eq:problem-definition}
\end{equation}

\noindent and:

\begin{equation}
A_2(b) = \frac{1}{|U|} \sum_{u \in U} min(R(u,s), 1) \, \, \, \text{with} \, b \in \{0, \dots, |D_{batch}|\}
\label{eq:problem-definition}
\end{equation}

\vspace{3mm}

\noindent where $s$ is the observed item $i$ of the example $b$ in the current batch $D_{batch}$, and $R(u,s)$ represents the feedback of user $u$ for the item $s$ in the training dataset. In other words, $A_1(b)$ and $A_2(b)$ indicate the predicted residual and the popularity of the observed item in the example $b$ of $D_{batch}$, respectively. Similar considerations are also valid for the point-wise optimization setting, where the predicted residuals in $A_1$ and the popularity estimates in $A_2$ are computed for both observed-item and unobserved-item pairs. The model is thus penalized if its ability to predict a higher relevance for the item directly depends on the popularity of the item. The proposed regularization is defined in a way that it can be applied to a wide range of learning-to-rank optimization procedures.

\vspace{2mm} 

In this paper, we refer to a model trained on data organized through the proposed sampling strategy as \texttt{sam} and to a model optimized through the proposed regularized loss function as \texttt{reg}. The operations included in the training example mining strategy are performed after each epoch, while the regularized optimization is computed for each batch, until convergence.

\section{Experimental Evaluation} \label{emp-eval}
In this section, we empirically evaluate the proposed mitigation procedure to assess its impact on accuracy, beyond-accuracy, and popularity debiasing objectives. We arranged the same experimental setting described for the exploratory analysis, including the same datasets (Section \ref{sec:data}), train-test protocols (Section \ref{sec:algo}), and metrics (Section \ref{sec:exp}). We aim to answer four key research questions: 

\begin{itemize}[leftmargin=*]
\item RQ1. What are the effects of our mitigation elements, separately and jointly?
\item RQ2. What is the impact of our mitigation on internal mechanics?
\item RQ3. To what degree of popularity debiasing can a model achieve the highest overall recommendation quality?
\item RQ4. How does our mitigation procedure perform compared with other state-of-the-art countermeasures against a popularity bias? 
\end{itemize}

\subsection{Effects of Mitigation Elements (RQ1)}
In this subsection, we show an ablation study that aims to assess ($i$) the influence of the new training examples sampling and the new regularized loss function on the model performance, and ($ii$) whether combining these two treatments can improve the trade-off between ranking accuracy and popularity bias.

\begin{figure}[!b] 
\begin{subfigure}[t]{0.49\linewidth}
    \centering
    \includegraphics[width=1.0\linewidth]{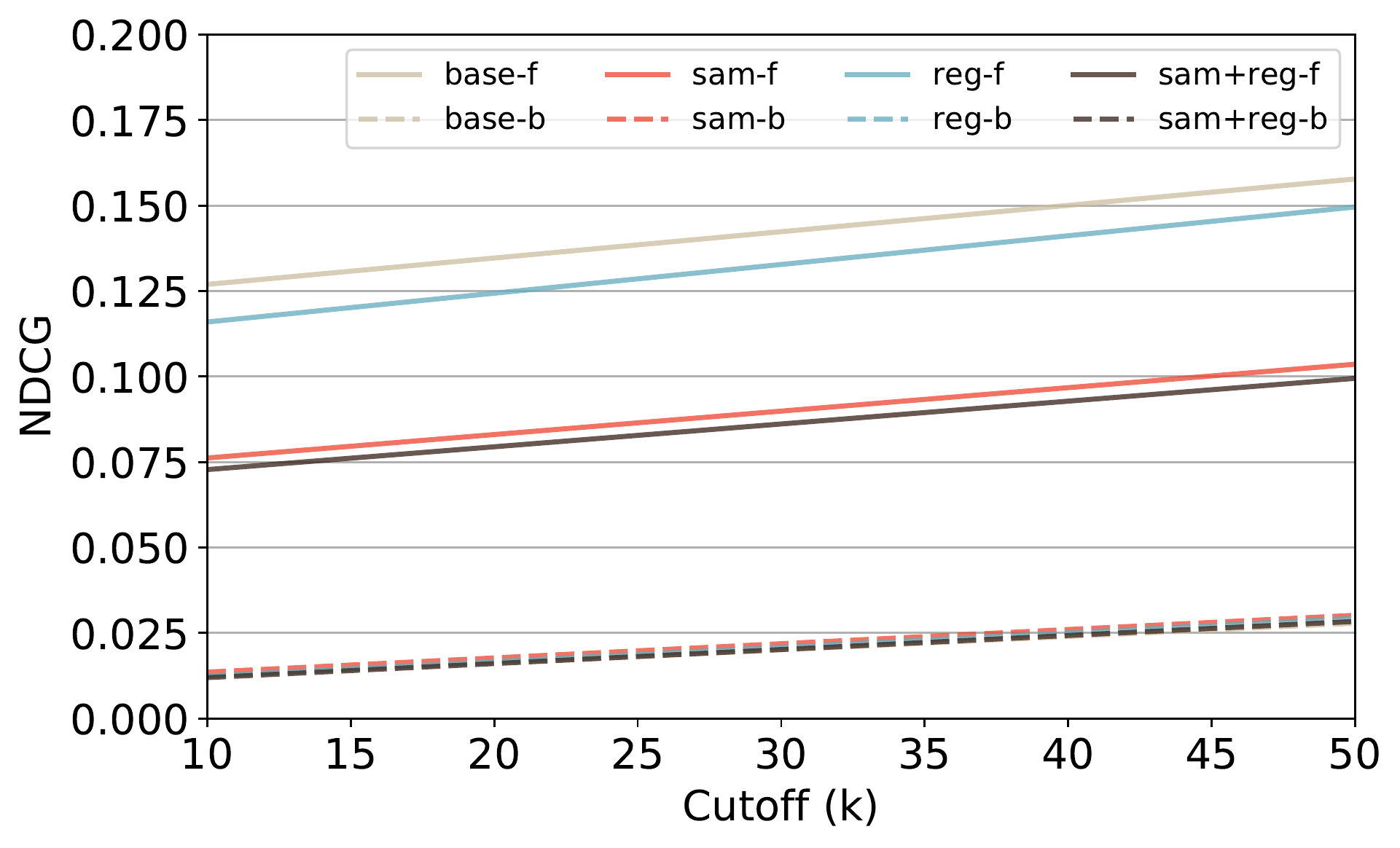}
    \caption{NDCG of BPR on ML1M.}
\end{subfigure}
\begin{subfigure}[t]{0.49\linewidth}
    \centering
    \includegraphics[width=1.0\linewidth]{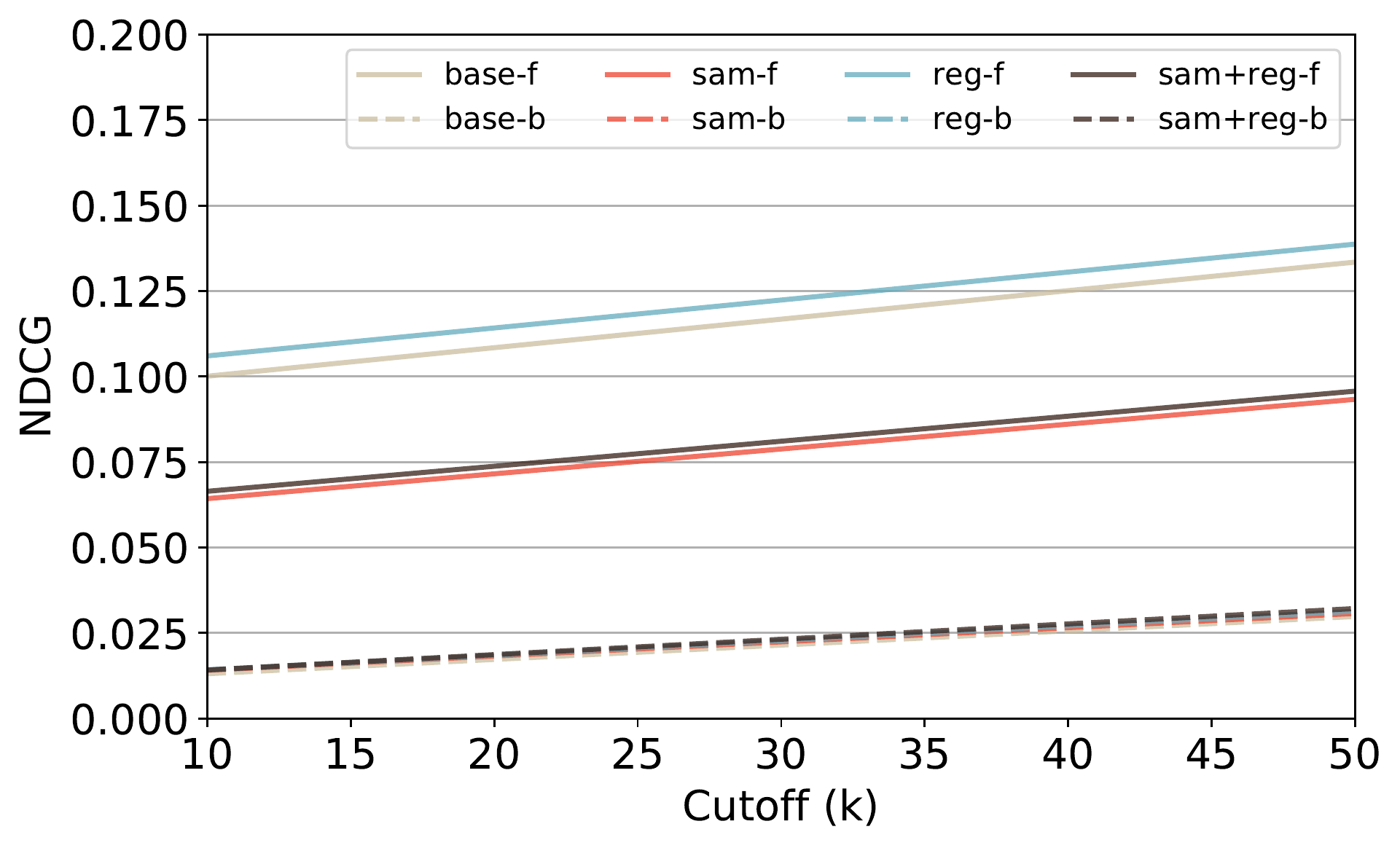}
    \caption{NDCG of NeuMF on ML1M.}
\end{subfigure}
\begin{subfigure}[t]{0.49\linewidth}
    \centering
    \includegraphics[width=1.0\linewidth]{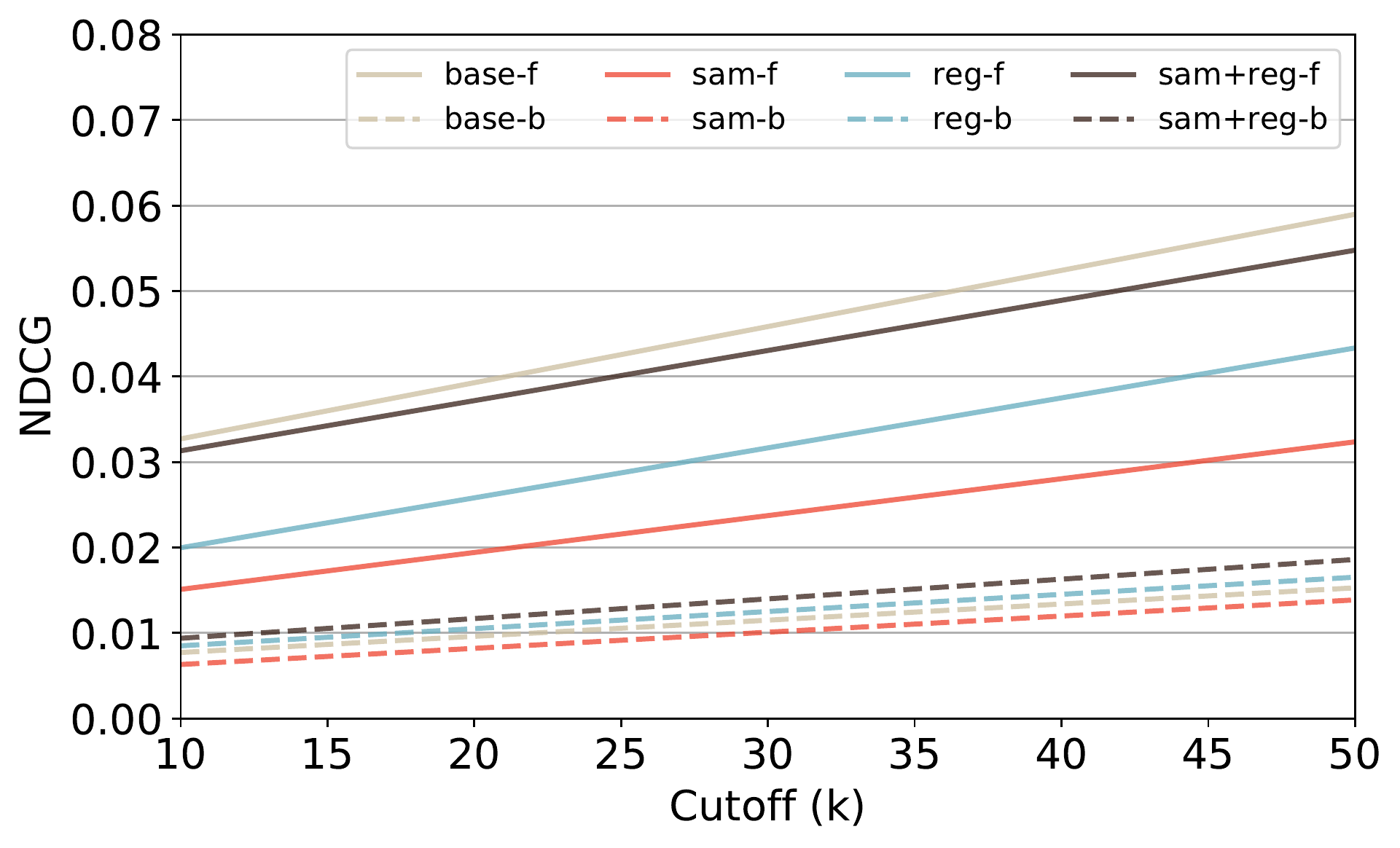}
    \caption{NDCG of BPR on COCO.}
\end{subfigure}
\begin{subfigure}[t]{0.49\linewidth}
    \centering
    \includegraphics[width=1.0\linewidth]{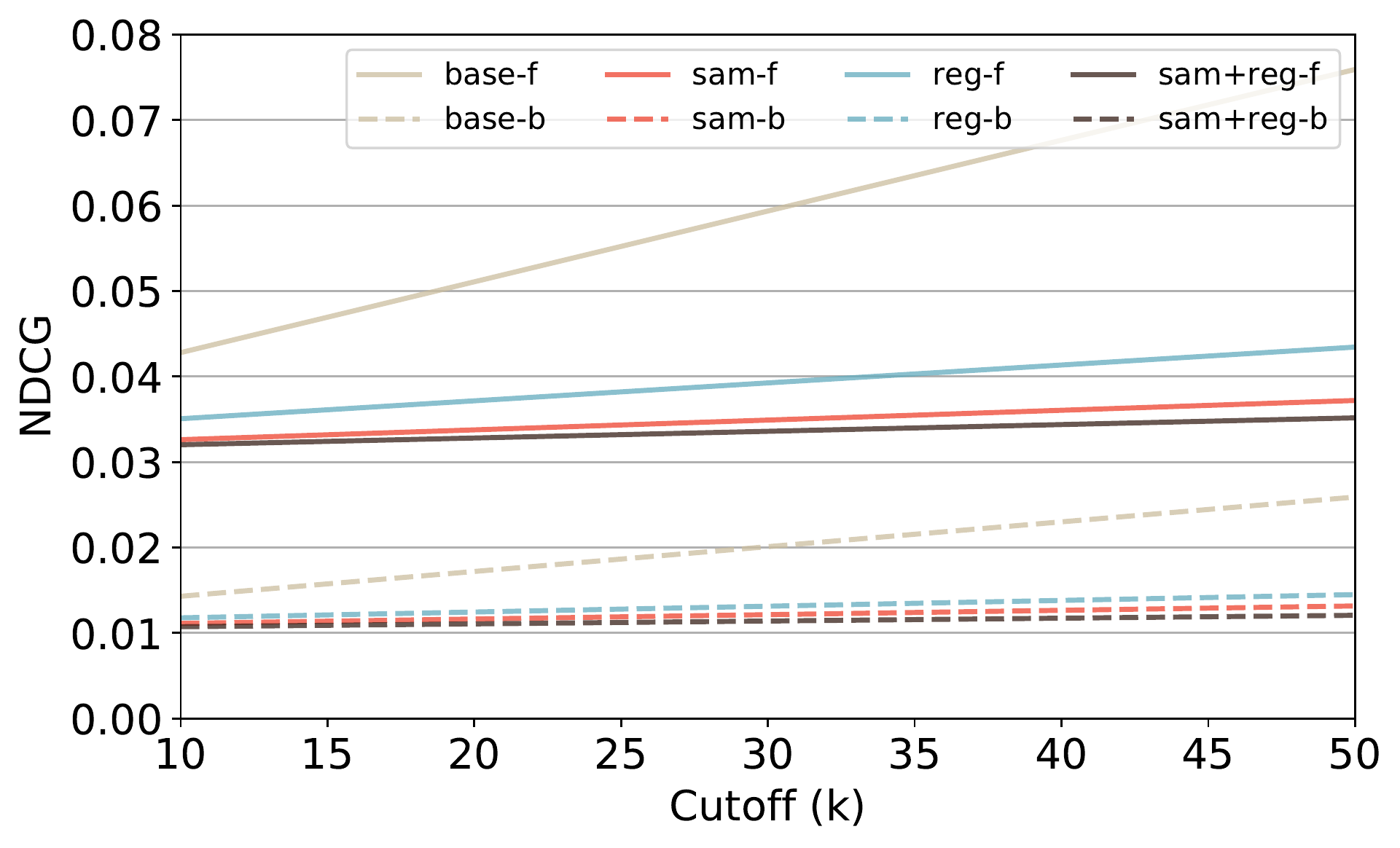}
    \caption{NDCG of NeuMF on COCO.}
\end{subfigure}
\caption{\textbf{Impact on Accuracy}. Normalized  Discounted  Cumulative  Gain  (NDCG) achieved by the considered recommender systems. The base label indicates the original recommender, sam and reg indicate the application of our treatments individually, and sam+reg combines both treatments. Straight lines indicate results on the full test set, while dashed lines refer to the results on a subset of the test set where all items have the same amount of observations. }
\label{fig:res-ndcg-01}
\end{figure} 

\begin{figure} 
\begin{subfigure}[t]{0.49\linewidth}
    \centering
    \includegraphics[width=1.0\linewidth]{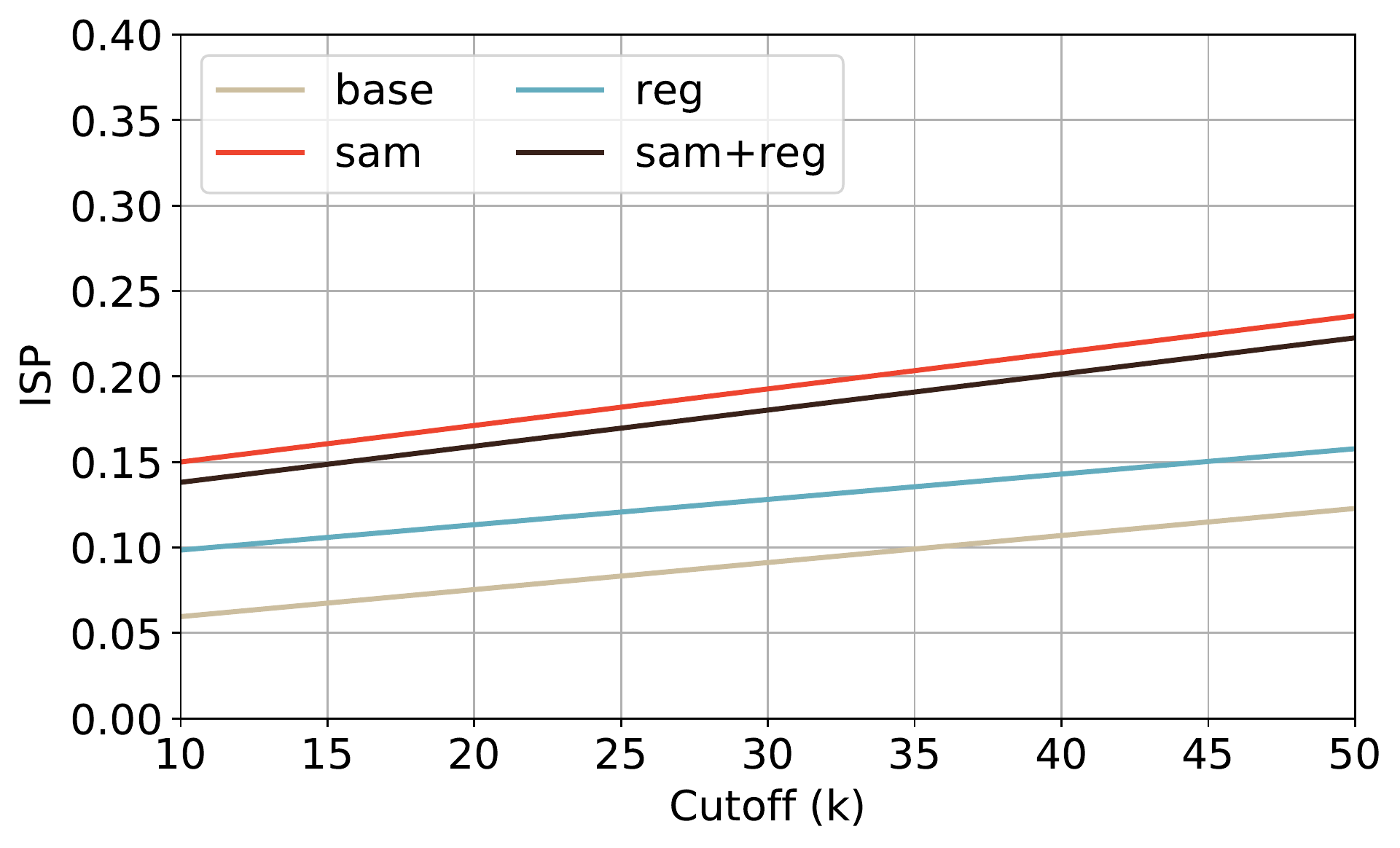}
    \caption{ISP of BPR on ML1M.}
\end{subfigure}
\begin{subfigure}[t]{0.49\linewidth}
    \centering
    \includegraphics[width=1.0\linewidth]{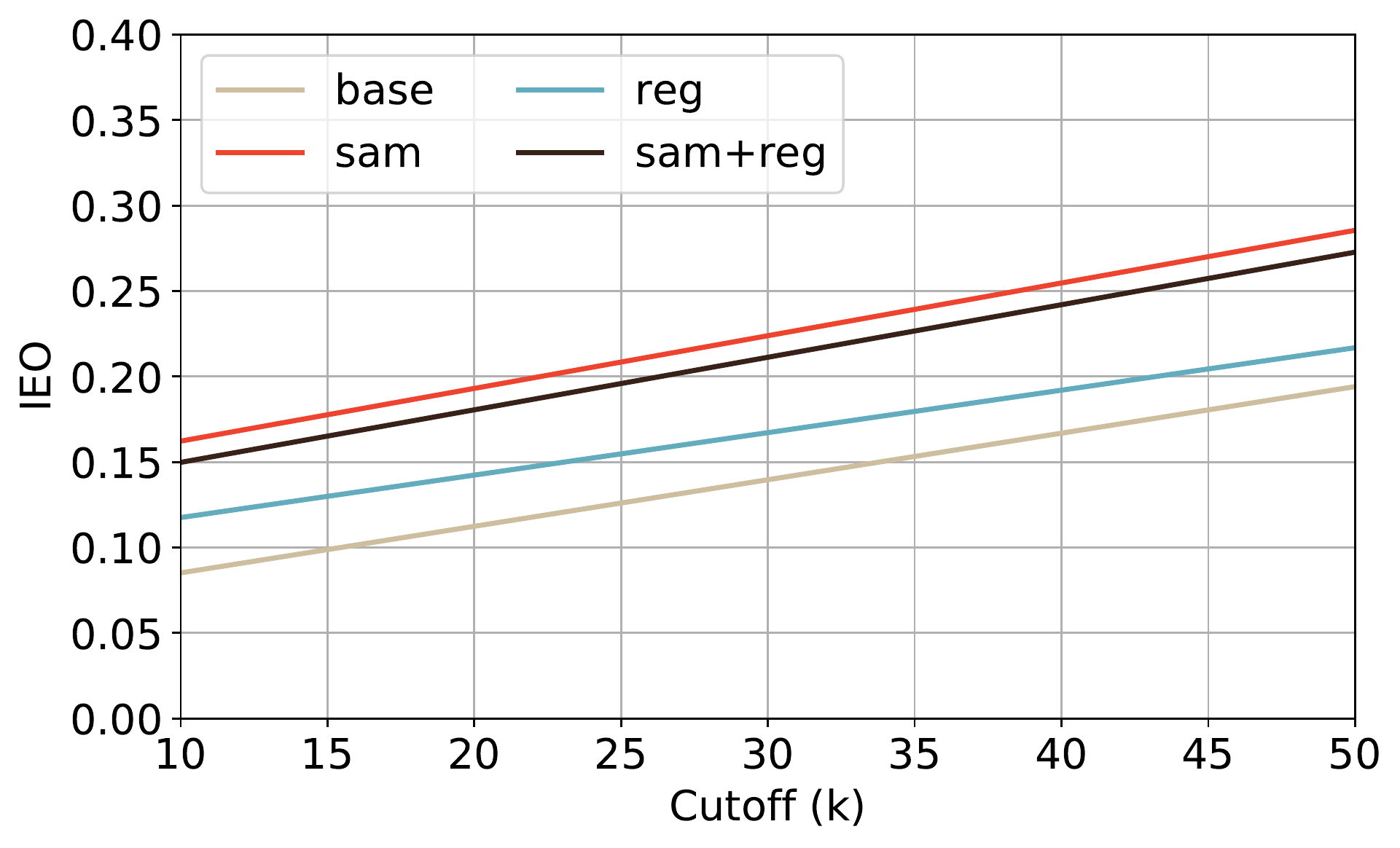}
    \caption{IEO of BPR on ML1M.}
\end{subfigure}
\begin{subfigure}[t]{0.49\linewidth}
    \centering
    \includegraphics[width=1.0\linewidth]{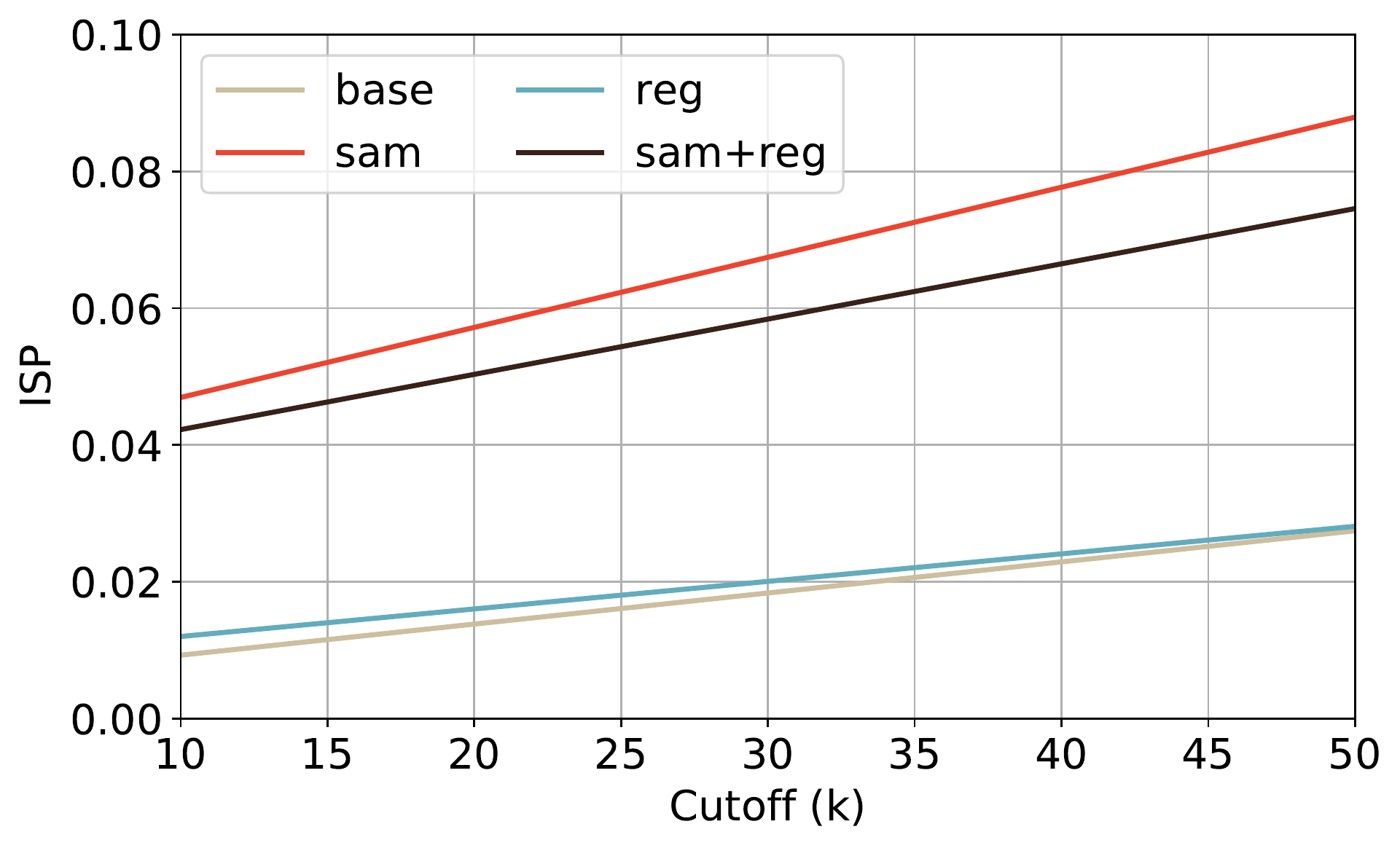}
    \caption{ISP of BPR on COCO.}
\end{subfigure}
\begin{subfigure}[t]{0.49\linewidth}
    \centering
    \includegraphics[width=1.0\linewidth]{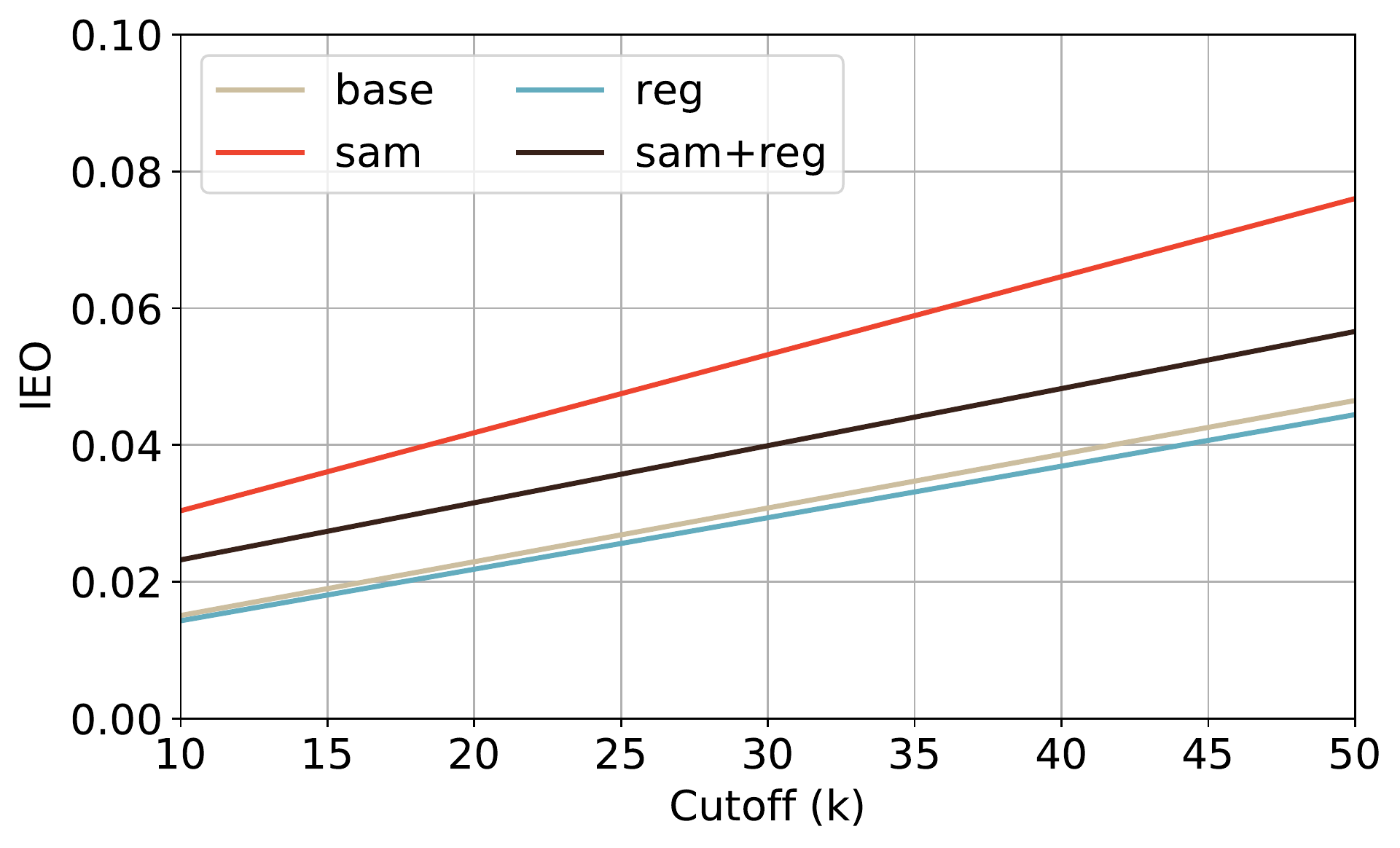}
    \caption{IEO of BPR on COCO.}
\end{subfigure}
\begin{subfigure}[t]{0.49\linewidth}
    \centering
    \includegraphics[width=1.0\linewidth]{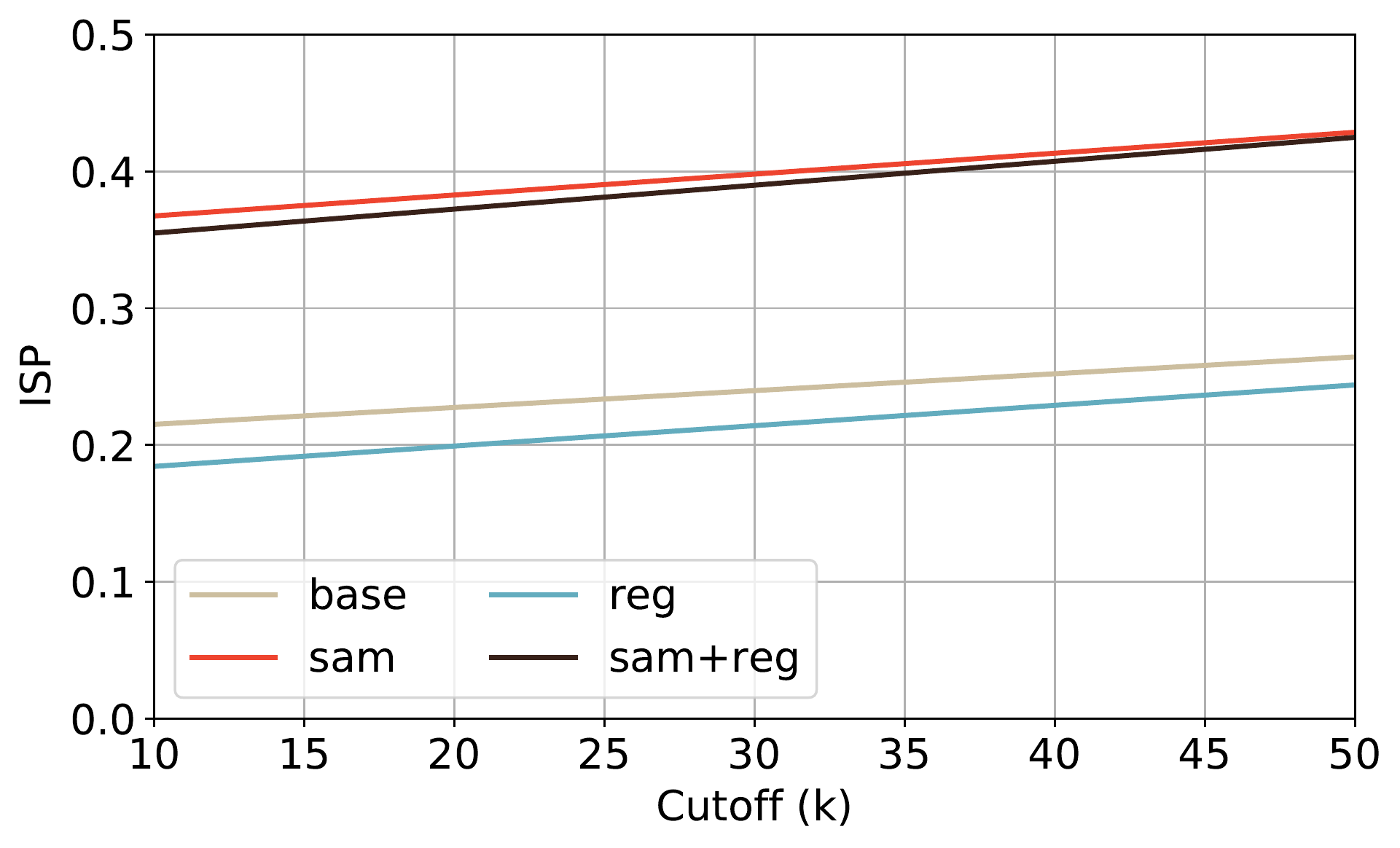}
    \caption{ISP of NeuMF on ML1M.}
\end{subfigure}
\begin{subfigure}[t]{0.49\linewidth}
    \centering
    \includegraphics[width=1.0\linewidth]{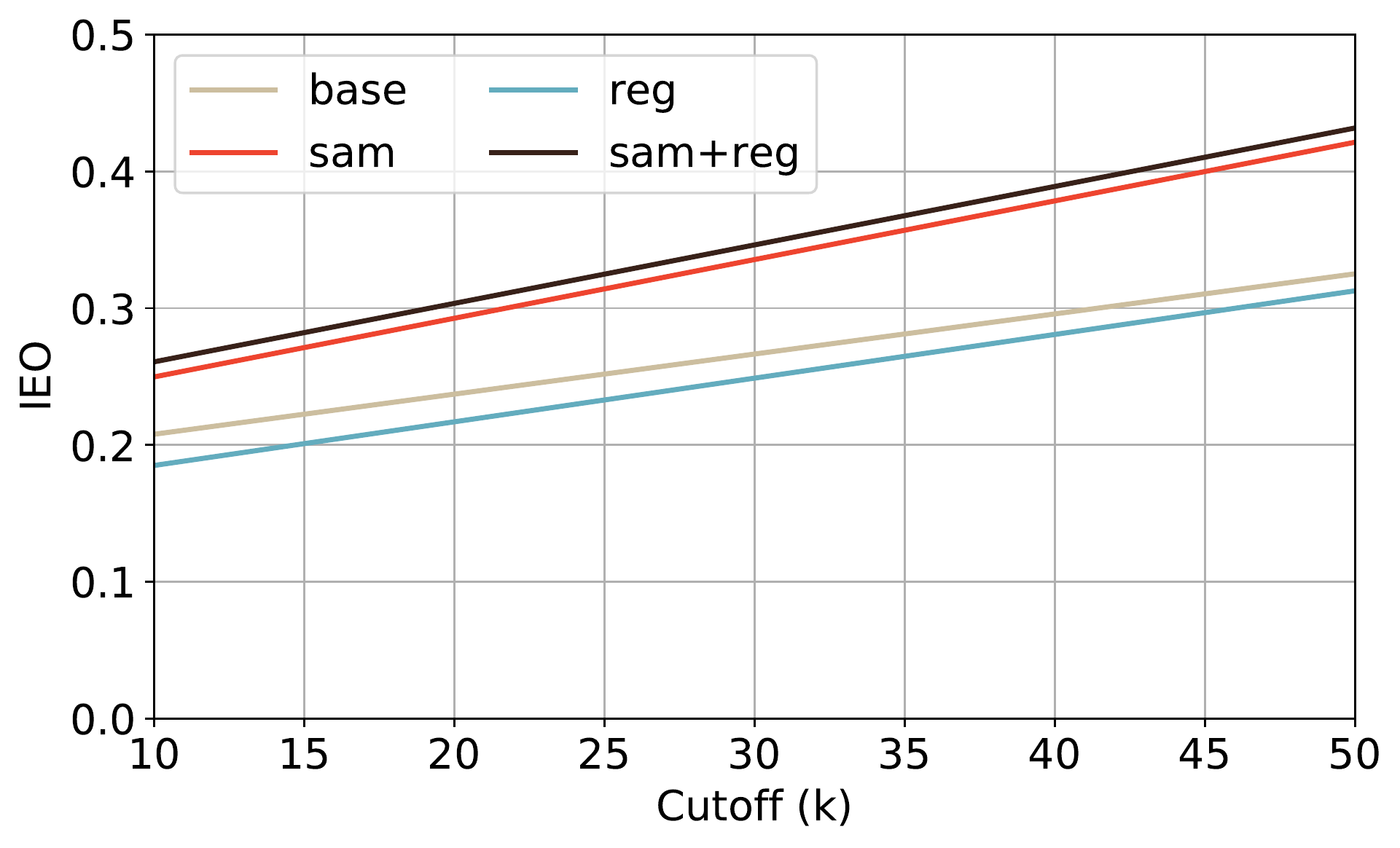}
    \caption{IEO of NeuMF on ML1M.}
\end{subfigure}
\begin{subfigure}[t]{0.49\linewidth}
    \centering
    \includegraphics[width=1.0\linewidth]{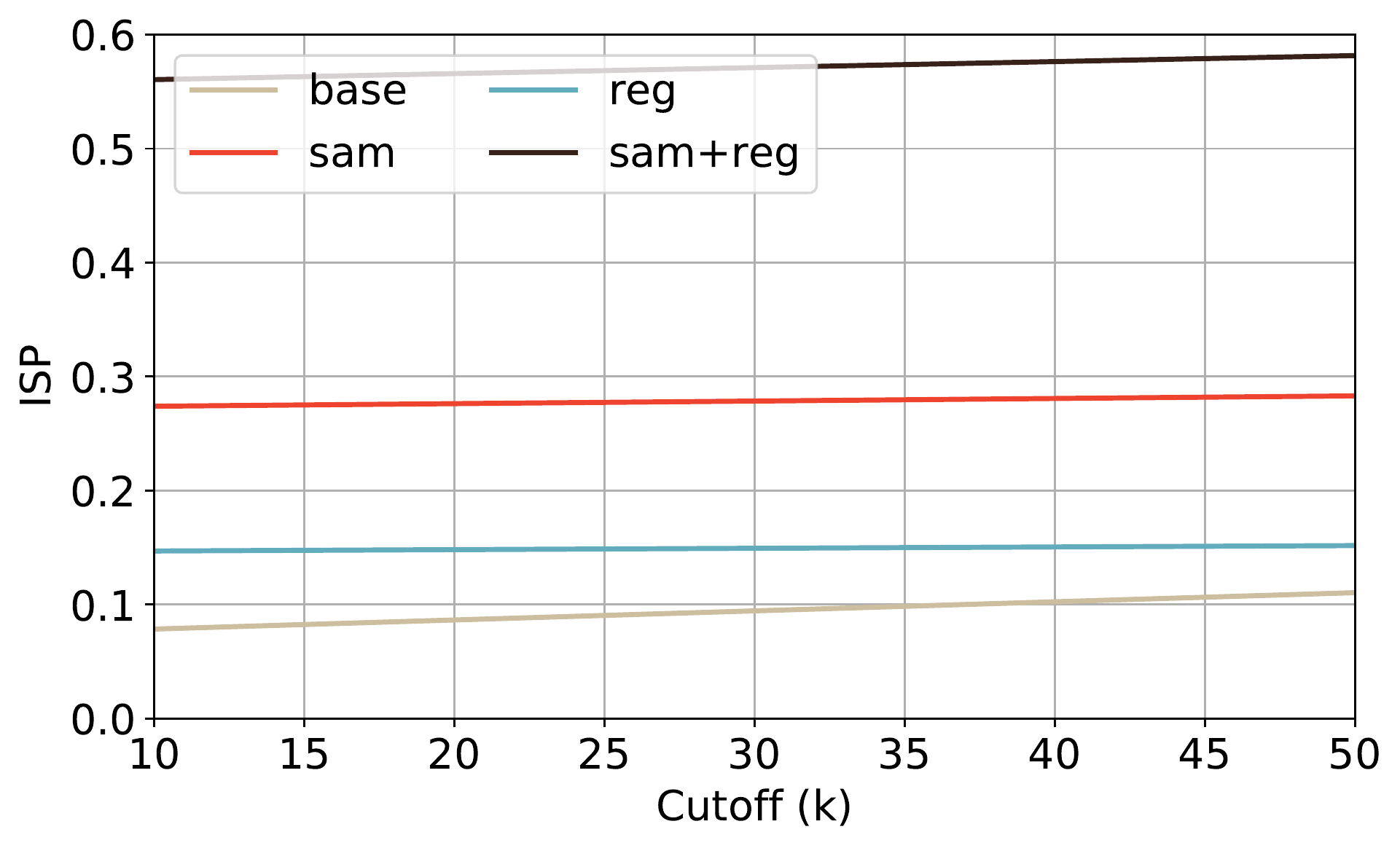}
    \caption{ISP of NeuMF on COCO.}
\end{subfigure}
\begin{subfigure}[t]{0.49\linewidth}
    \centering
    \includegraphics[width=1.0\linewidth]{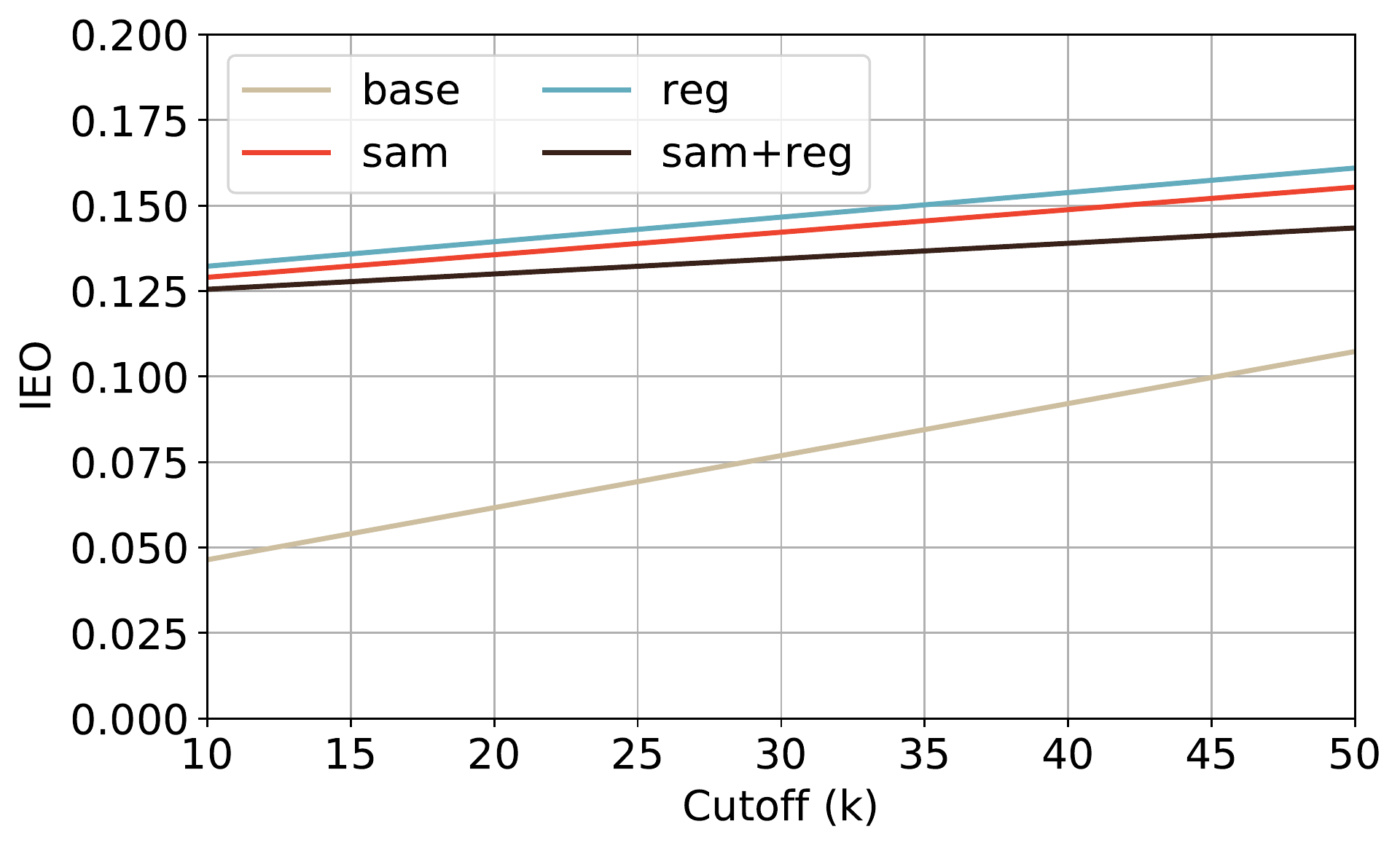}
    \caption{IEO of NeuMF on COCO.}
\end{subfigure}
\caption{\textbf{Impact on Popularity Bias}. Item Statistical Parity (ISP) and Item Equal Opportunity (IEO) achieved by the considered recommender systems. The base label refers to the original recommender, sam and reg apply our treatments individually, and sam+reg combines both treatments. Straight lines indicate results on the full test set. Dashed lines show results on a subset of the test set where all items have the same amount of observations.}
\label{fig:res-popularity-01}
\end{figure} 

To answer these questions, we compared the original model (base) against an instance of the model trained on data organized through the proposed sampling strategy only (\texttt{sam}), the model optimized through the proposed regularized loss function only (\texttt{reg}), and the model obtained after combining both our treatments (\texttt{sam+reg}). The regularized optimization for the last two setups was configured with $\lambda=0.2$, which gave us the best trade-off during experiments in Section \ref{sec:linking}. The results are presented and discussed below.

From Figure \ref{fig:res-ndcg-01}, it can be observed that all the newly introduced configurations (red, blue, and brown lines) have a loss in accuracy with respect to the original model (amber line), if we considered the full test set (straight lines). However, the gap in accuracy among the original and the regularized models is negligible, when we consider the same number of test observations for all the items (dashed lines). We argue that, as large gaps of recommendation accuracy in the full test set reflect only a spurious bias in the metric and the underlying test set (see \cite{DBLP:journals/ir/BelloginCC17} for a demonstration), the real impact of our treatments on accuracy should be considered on the balanced test set. In the latter evaluation setting, there is a small gap in accuracy across models. NDCG seems to vary across datasets. For the sake of conciseness, we do not report results on precision and recall, which showed similar patterns. On ML1M, with NeuMF, combining our data sampling and regularized loss function slightly improves accuracy. When the treatment are applied separately, there is no significant difference with respect to the original model. On COCO, the loss in accuracy is smaller, with reg outperforming \texttt{sam+reg} for NeuMF. 

Figure \ref{fig:res-popularity-01} shows that our data sampling (\texttt{sam}: red line) and our combination of data sampling and regularized loss function (\texttt{sam+reg}: brown line) made a positive impact on ISP and IEO, while our regularized loss function alone (\texttt{reg}: blue line) still emphasizes a bias comparable with the original model (base: amber line). No statistically significant difference was observed between \texttt{sam} (red line) and \texttt{sam+reg} (brown line) on ISP. It follows that the regularized loss function does not improve ISP, directly. Further, \texttt{sam+reg} led to an increase of IEO with respect to \texttt{sam}, better equalizing opportunities across items.   
 
\vspace{2mm} \noindent \colorbox{gray!15}{\parbox{0.98\textwidth}{\textbf{Observation 5}. \textit{Combining our sampling and regularization leads to higher ISP and IEO w.r.t. applying them separately. The loss in ranking accuracy is negligible with respect to the original model, if a balanced test set is considered.}}} \vspace{2mm}

\subsection{Impact on Internal Mechanics (RQ2)}
In this subsection, we investigate whether our bias-mitigation approach can reduce ($i$) the biased gap between head- and mid-item relevance distributions, and ($ii$) the biased gap in pair-wise accuracy among head and mid items.

\begin{figure}[!b] 
\begin{subfigure}[t]{0.49\linewidth}
    \centering
    \includegraphics[width=1.0\linewidth]{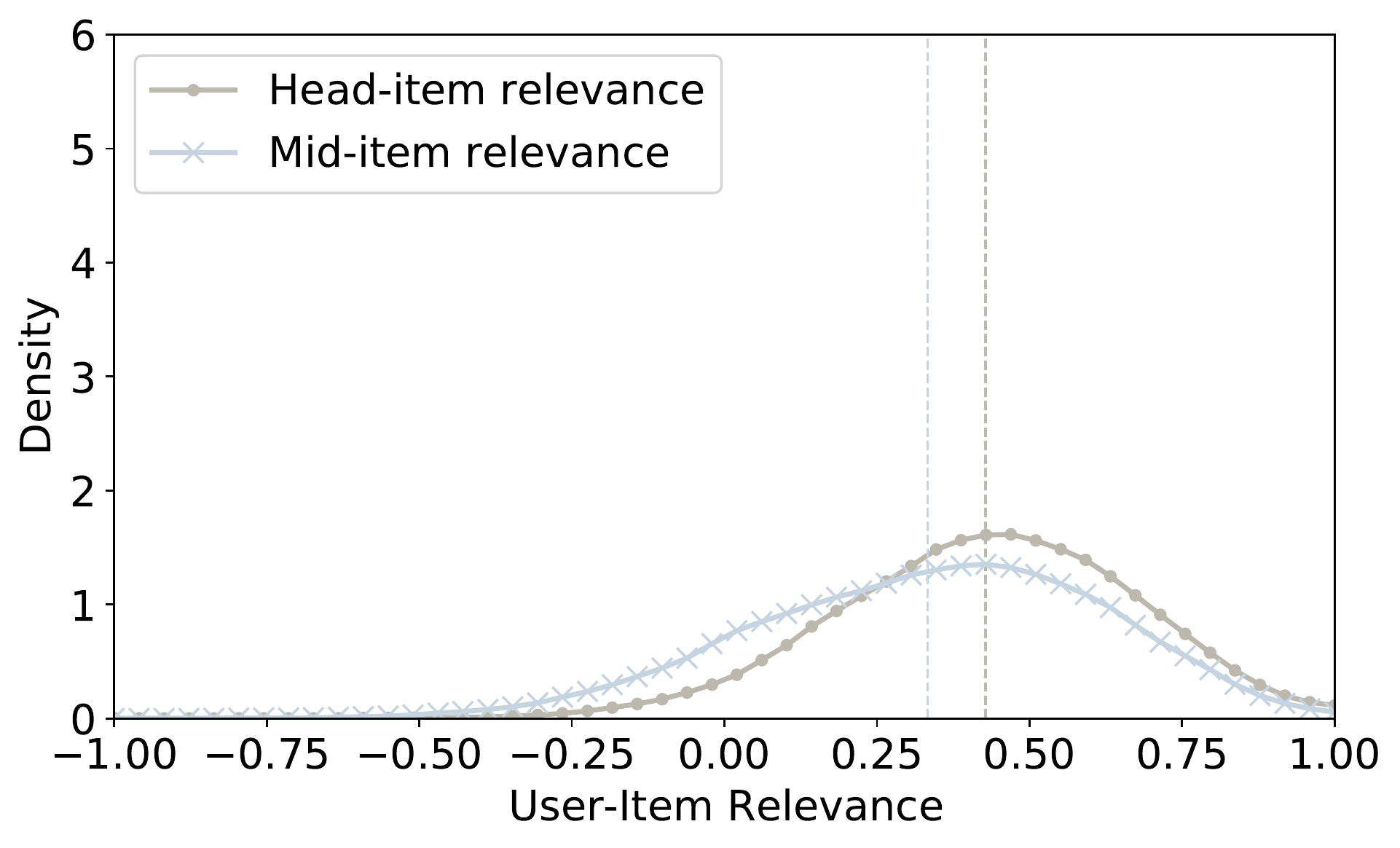}
    \caption{BPR on ML1M.}
\end{subfigure}
\begin{subfigure}[t]{0.49\linewidth}
    \centering
    \includegraphics[width=1.0\linewidth]{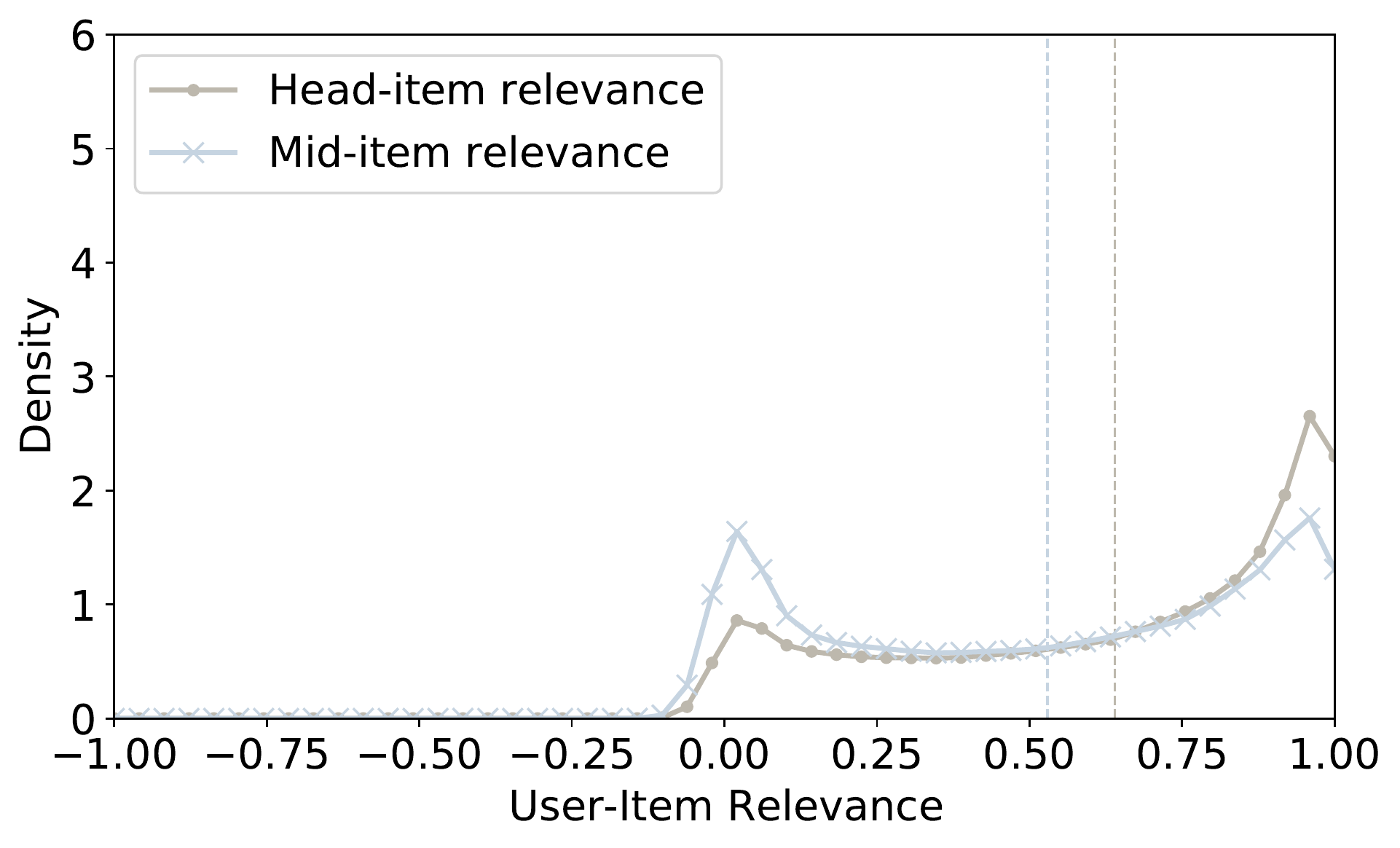}
    \caption{NeuMF on ML1M.}
\end{subfigure}
\begin{subfigure}[t]{0.49\linewidth}
    \centering
    \includegraphics[width=1.0\linewidth]{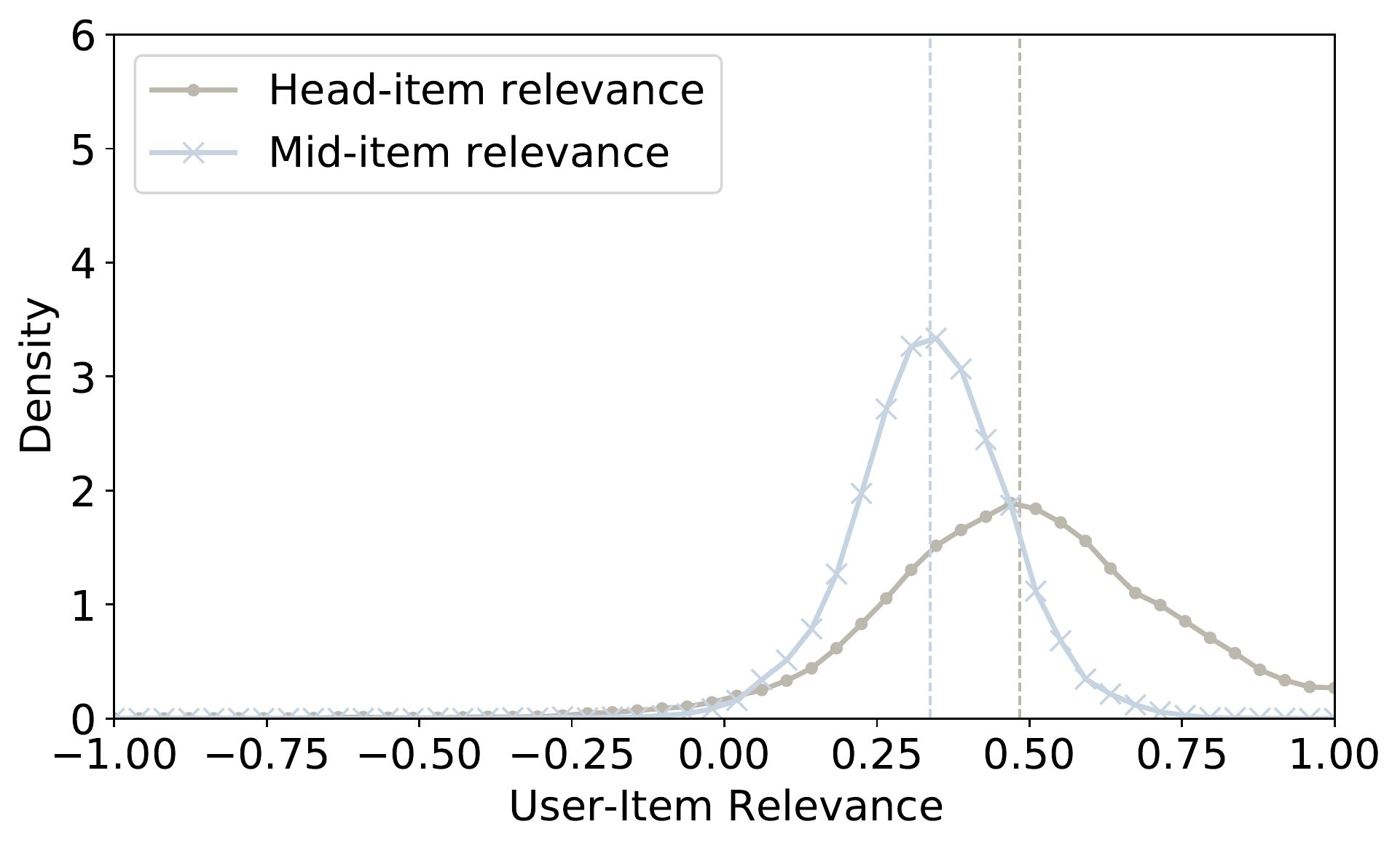}
    \caption{BPR on COCO.}
\end{subfigure}
\begin{subfigure}[t]{0.49\linewidth}
    \centering
    \includegraphics[width=1.0\linewidth]{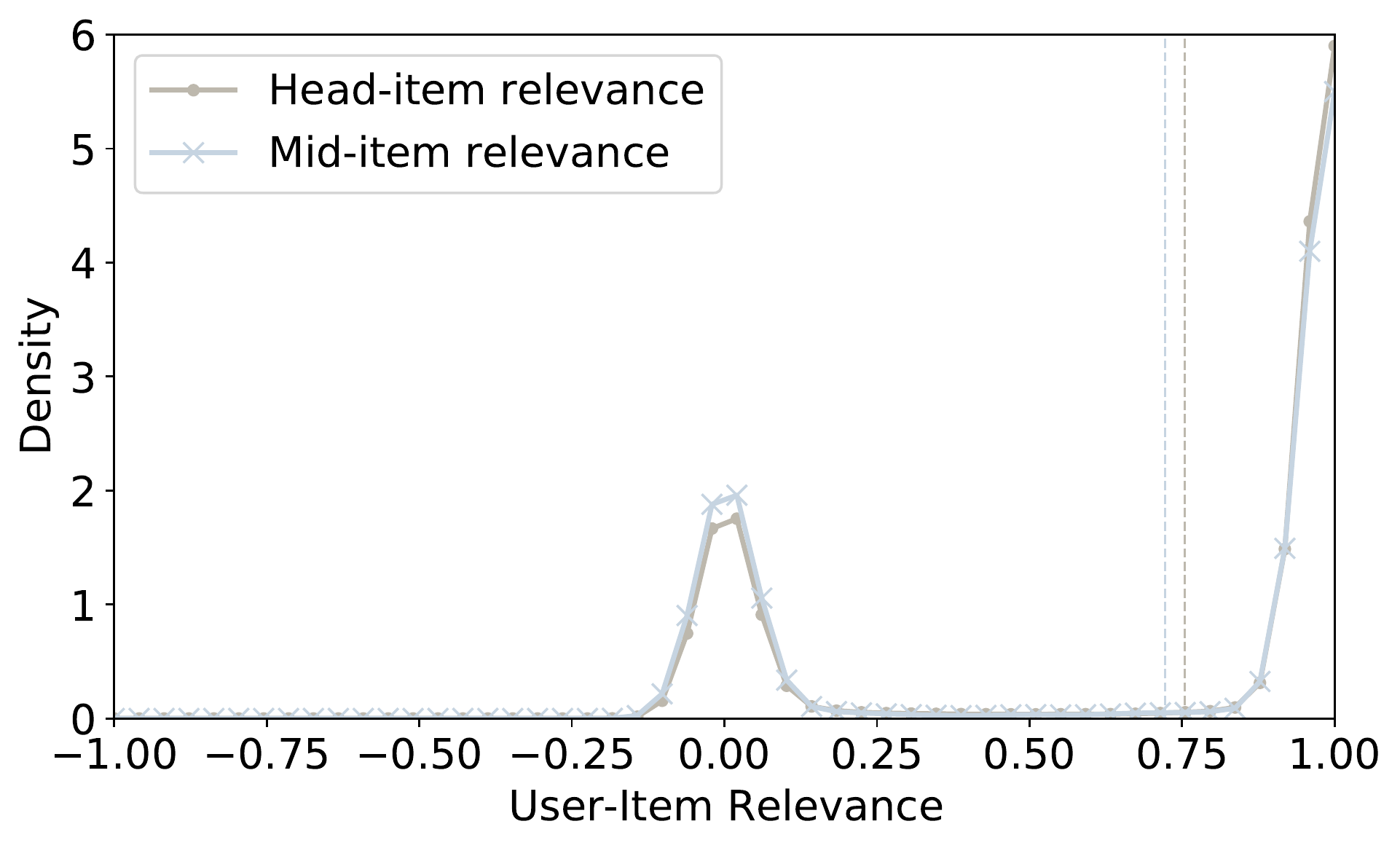}
    \caption{NeuMF on COCO.}
\end{subfigure}
\caption{\textbf{Impact on Internal Mechanics}. Head- and mid-item relevance for users in ML1M and COCO, after applying our sam-reg approach. The distributions are based on relevance scores obtained from randomly-sampled pairs of items, including a head item and a mid item the current user interacted with in the training set. }
\label{fig:res-popularity-03}
\end{figure}

To answer the first question, we computed the relevance score distributions for observed head and mid items in \figurename~\ref{fig:res-popularity-03}. Gray lines refer to the user-(head-item) pairs, and the blue lines are associated with the user-(mid-item) pairs. We followed the same methodology described for our exploratory analysis in Section \ref{sec:exp}. The proposed mitigation can reduce the gap between the relevance score distributions, when compared with the gaps in Fig. \ref{fig:relevance-distr}. It follows that our intuition and our mitigation procedure were proved to be valid. 

For the second question, we computed the pair-wise accuracy for observed head and mid items in Table \ref{tab:pair-acc-03}. The (mid, head) setup experienced a statistically significant improvement in pair-wise accuracy. Conversely, as far as mid items end up to be well-performing, pair-wise accuracy on the setups involving observed head items slightly decreased. The improvement is generally higher under a pair-wise optimization (BPR) and less sparse datasets (ML1M). To assess the impact of our mitigation in cases where the model does not show any biased behavior across head and mid items, we included NeuMF trained on COCO into our evaluation (last column). In this situation, our mitigation led to a decrease in performance over all the observed/unobserved items setups. Therefore, it should be applied only when the gap is considerable.  

\begin{table}[!b]
   
\resizebox{\textwidth}{!}{
\begin{small}
\begin{tabular}{cc|ll|ll}
\hline
\multirow{2}{*}{\textbf{Observed Item}} & \multirow{2}{*}{\textbf{Unobserved Item}} & \multicolumn{2}{c}{\textbf{ML-1M}} & \multicolumn{2}{c}{\textbf{COCO}} \\
 &  & \textbf{BPR} & \textbf{NeuMF} & \textbf{BPR} & \textbf{NeuMF} \\
\hline
Head & Any & 0.88 (-- 0.05) & 0.91 (-- 0.04) & 0.92 (-- 0.02) & 0.84 (-- 0.14) \\
Mid & Any & 0.78 (+0.04) & 0.85 (+0.01) & 0.89 (+0.06) & 0.82 (-- 0.14) \\
\hline
Head & Head & 0.77 (-- 0.11) & 0.87 (-- 0.05) & 0.89 (+0.00) & 0.85 (-- 0.11) \\
Head & Mid & 0.93 (-- 0.06) & 0.95 (-- 0.04) & 0.95 (-- 0.04) & 0.83 (-- 0.16) \\
Mid & Head & 0.68 (+0.10) & 0.80 (+0.06) & 0.82 (+0.13) & 0.82 (-- 0.10) \\
Mid & Mid & 0.89 (-- 0.02) & 0.90 (-- 0.04) & 0.94 (-- 0.04) & 0.81 (-- 0.18) \\
\hline 
\end{tabular} \end{small}}
\caption{\textbf{Impact on Pair-wise Accuracy}. Pair-wise accuracy across user-item pairs spanning different parts of the popularity spectrum, after applying our sam+reg approach. The more the cases where the relevance for the observed item is higher than the relevance for the unobserved item, the higher the pair-wise accuracy. Numbers between brackets indicate the difference in percentage-points with respect to the accuracy of the original model.}
\label{tab:pair-acc-03}
\end{table}
 
\vspace{2mm} \noindent \colorbox{gray!15}{\parbox{0.98\textwidth}{\textbf{Observation 6}. \textit{Our correlation-based regularization, jointly with the tailored sampling, leads to a reduction of the gap in relevance score of items along the popularity tail. This is stronger for pair-wise approaches and sparsed datasets.}}} \vspace{2mm}

\subsection{Linking Regularization Weight and Recommendation Quality (RQ3)} \label{sec:linking}
We investigate how the recommender system performs when we vary the regularization weight $\lambda$ in the proposed loss function. With this experiment, we seek to inspect to what degree the influence of popularity may be debiased to achieve the best quality of recommendation, according to ranking accuracy and beyond-accuracy objectives. For the sake of conciseness, we only report experimental results on ML1M, but the results on COCO showed similar patterns.

\begin{figure}[!b] 
\begin{subfigure}[t]{0.49\linewidth}
    \centering
    \includegraphics[width=1.0\linewidth]{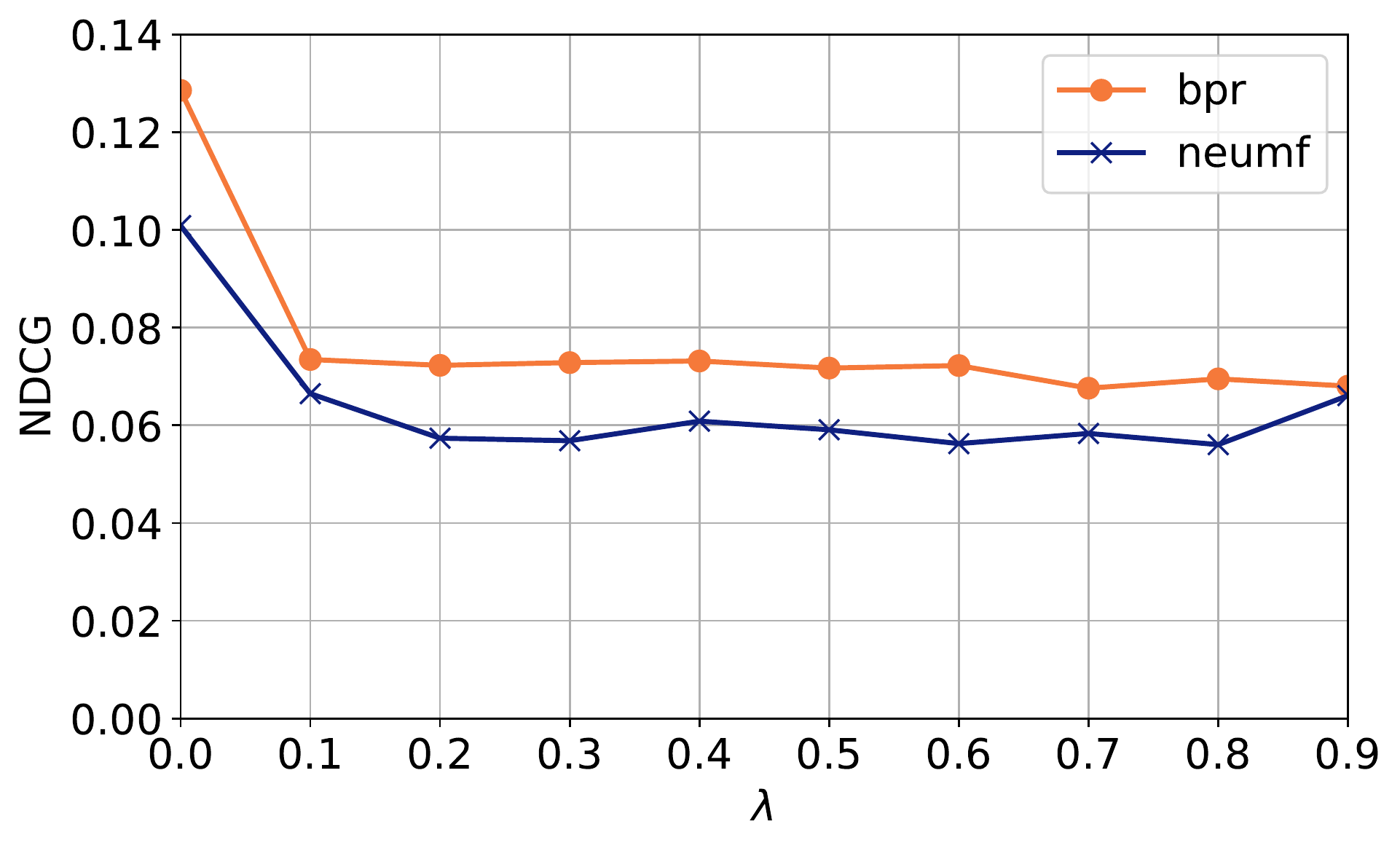}
    \caption{NDCG on the Full Test Set.}
\end{subfigure}
\begin{subfigure}[t]{0.49\linewidth}
    \centering
    \includegraphics[width=1.0\linewidth]{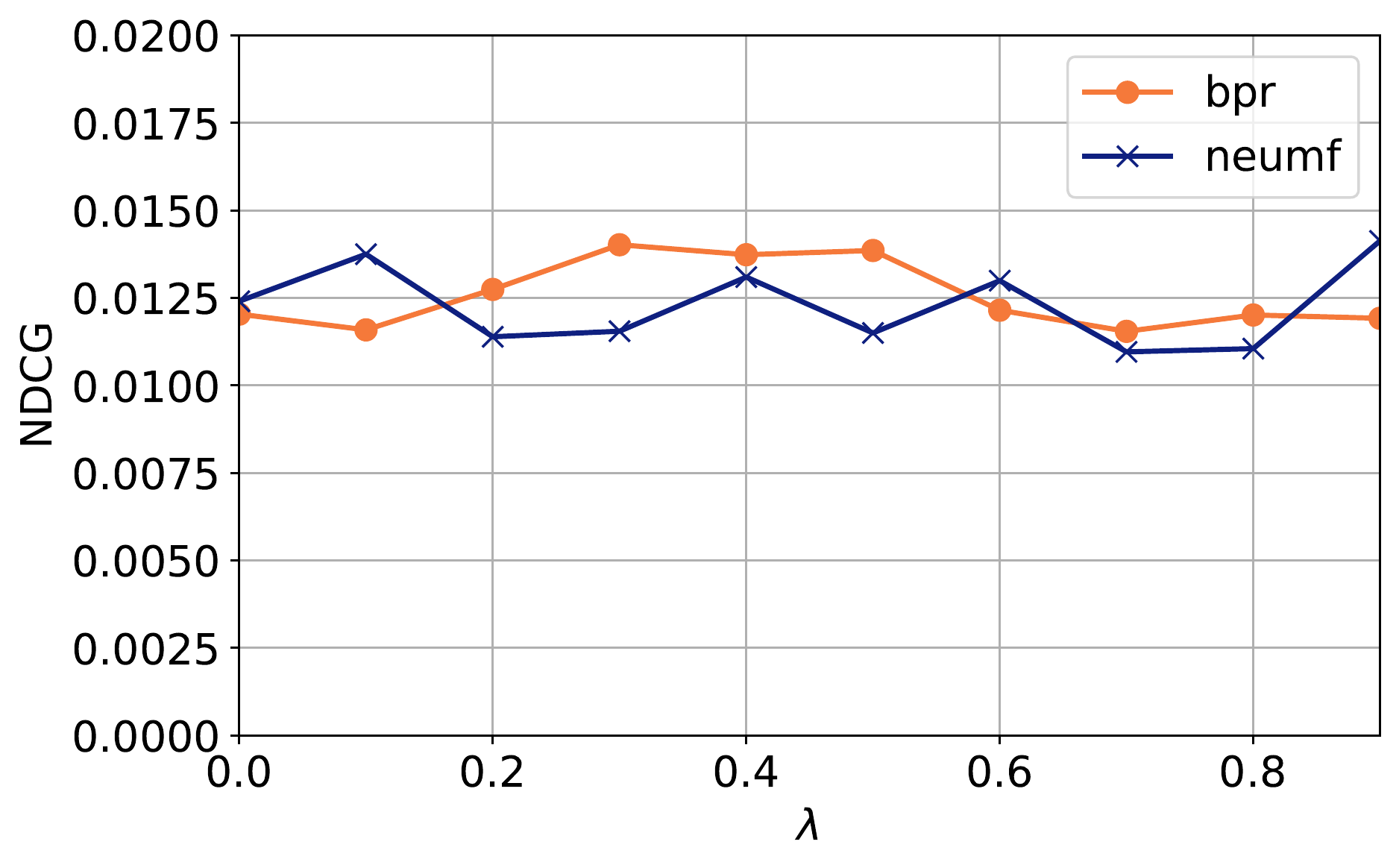}
    \caption{NDCG on the Balanced Test Set.}
\end{subfigure}
\begin{subfigure}[t]{0.49\linewidth}
    \centering
    \includegraphics[width=1.0\linewidth]{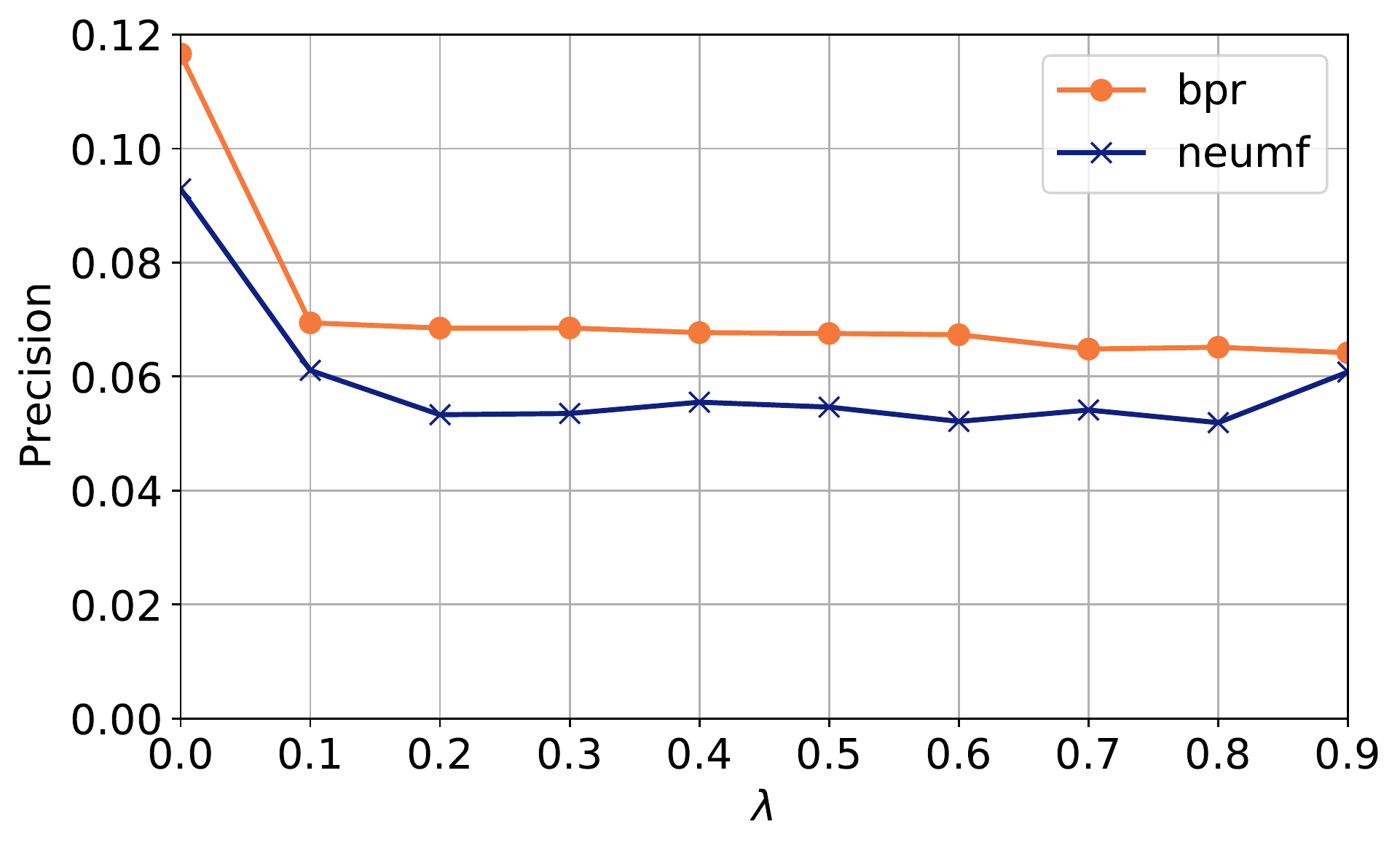}
    \caption{Precision on the Full Test Set.}
\end{subfigure}
\begin{subfigure}[t]{0.49\linewidth}
    \centering
    \includegraphics[width=1.0\linewidth]{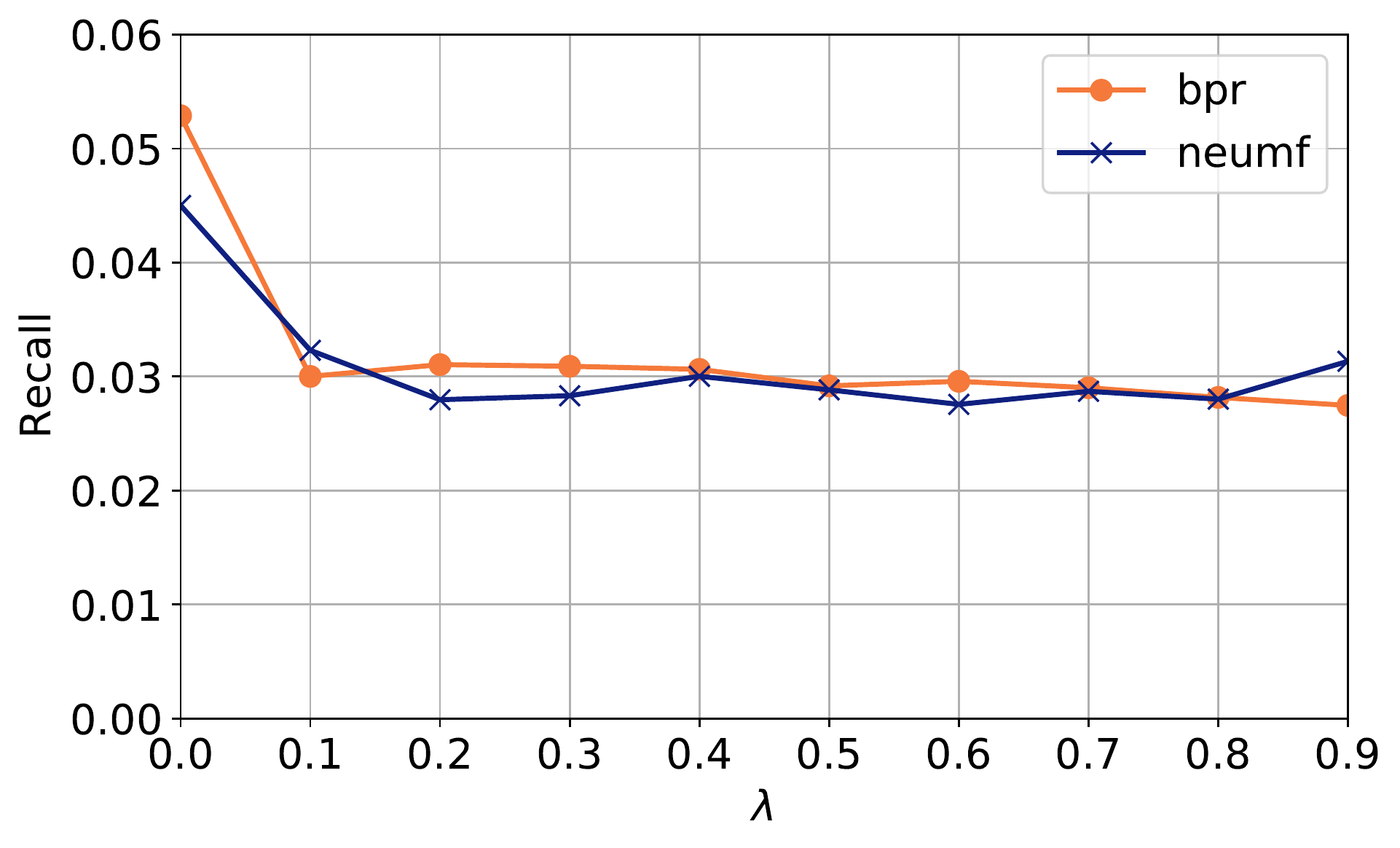}
    \caption{Recall on the Full Test Set}
\end{subfigure}
\begin{subfigure}[t]{0.49\linewidth}
    \centering
    \includegraphics[width=1.0\linewidth]{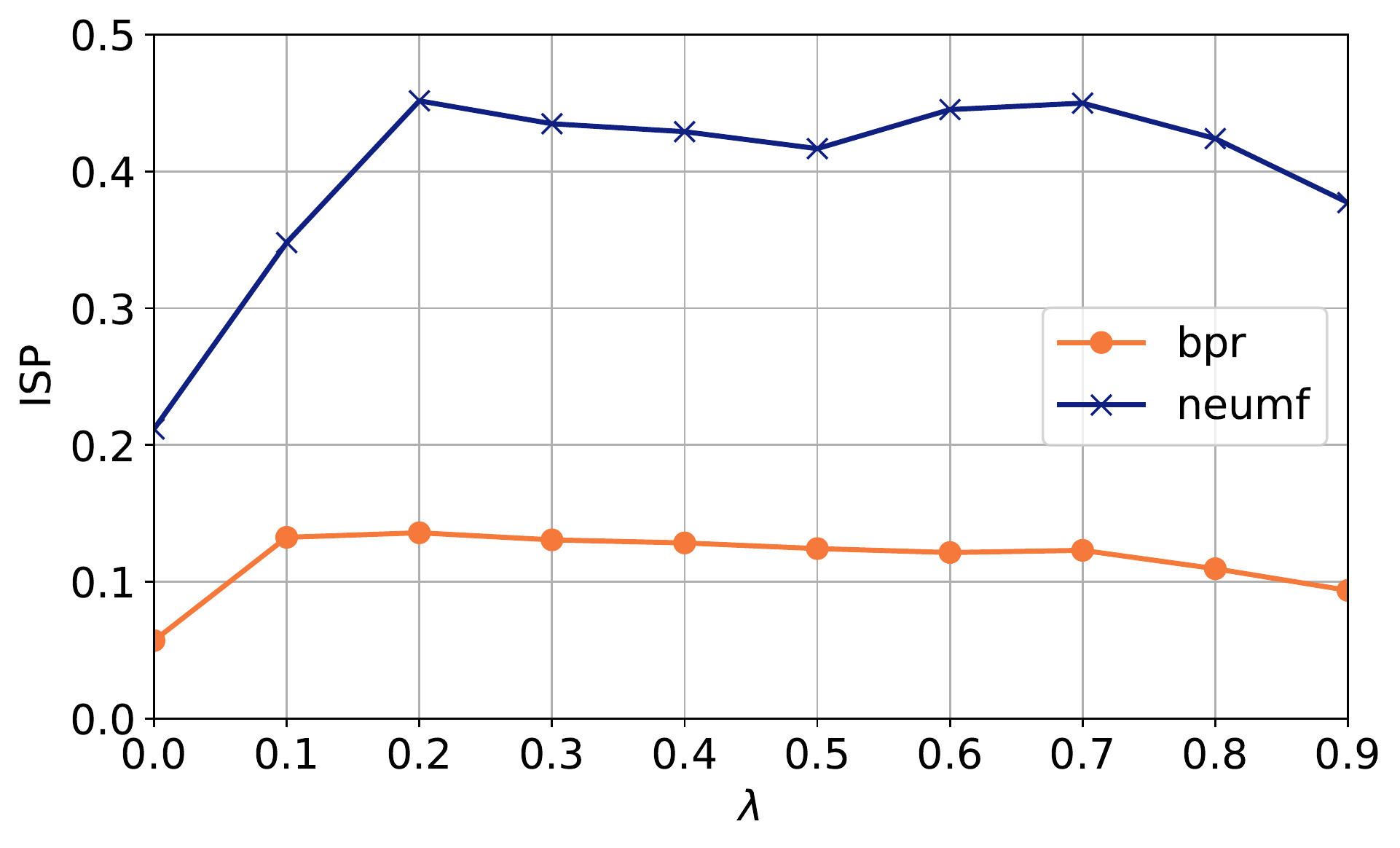}
    \caption{ISP.}
\end{subfigure}
\begin{subfigure}[t]{0.49\linewidth}
    \centering
    \includegraphics[width=1.0\linewidth]{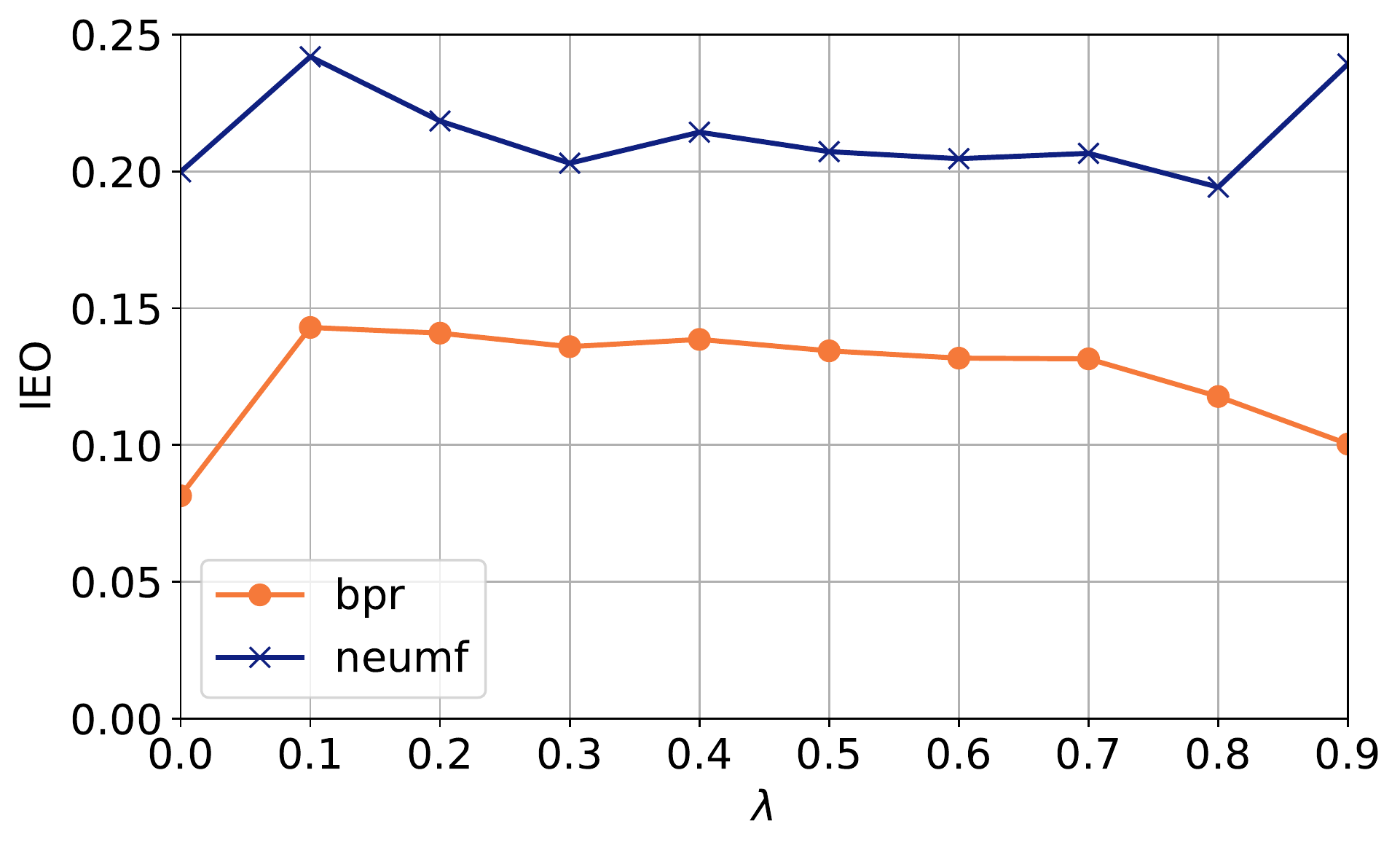}
    \caption{IEO}
\end{subfigure}
\begin{subfigure}[t]{0.49\linewidth}
    \centering
    \includegraphics[width=1.0\linewidth]{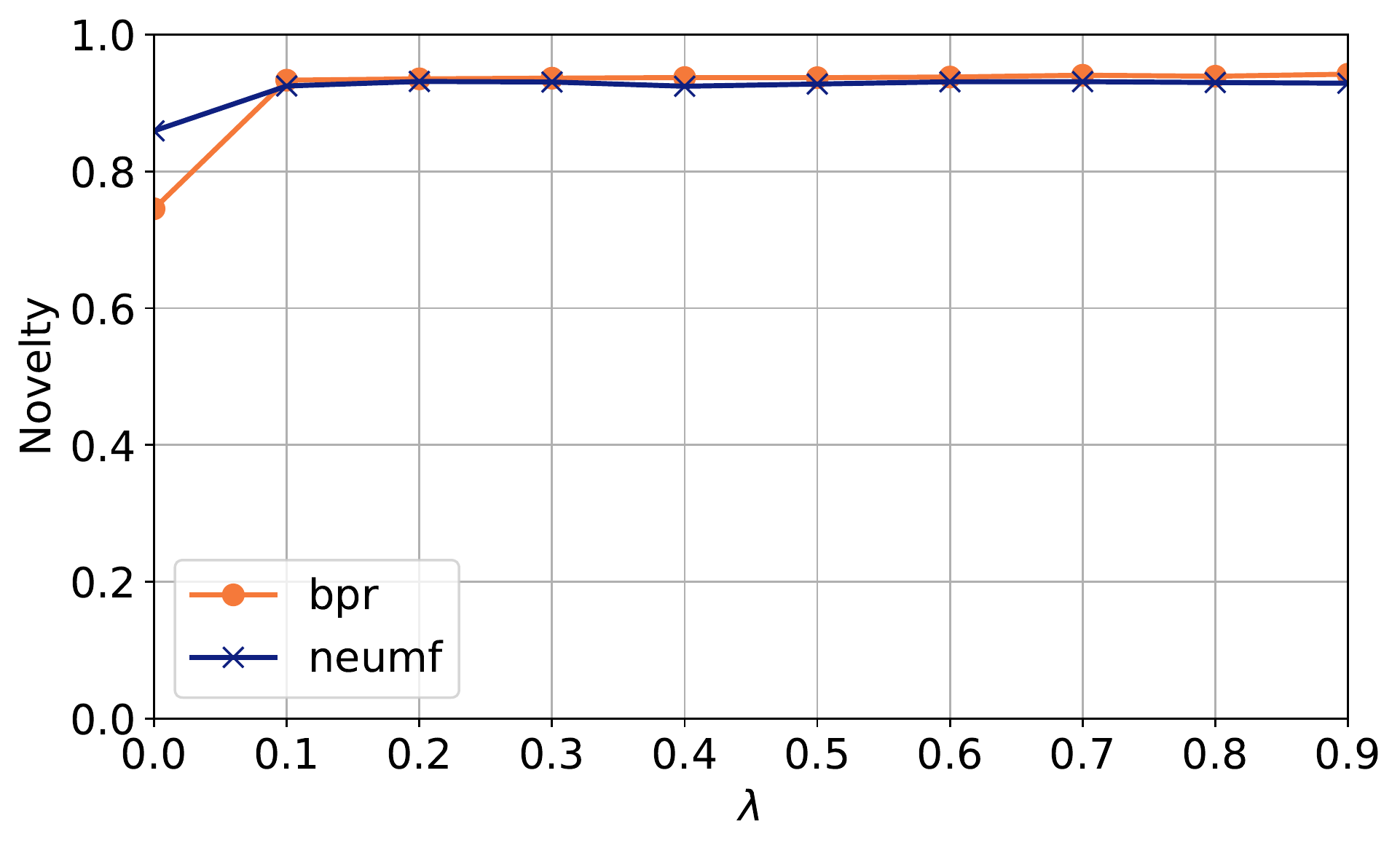}
    \caption{Novelty.}
\end{subfigure}
\begin{subfigure}[t]{0.49\linewidth}
    \centering
    \includegraphics[width=1.0\linewidth]{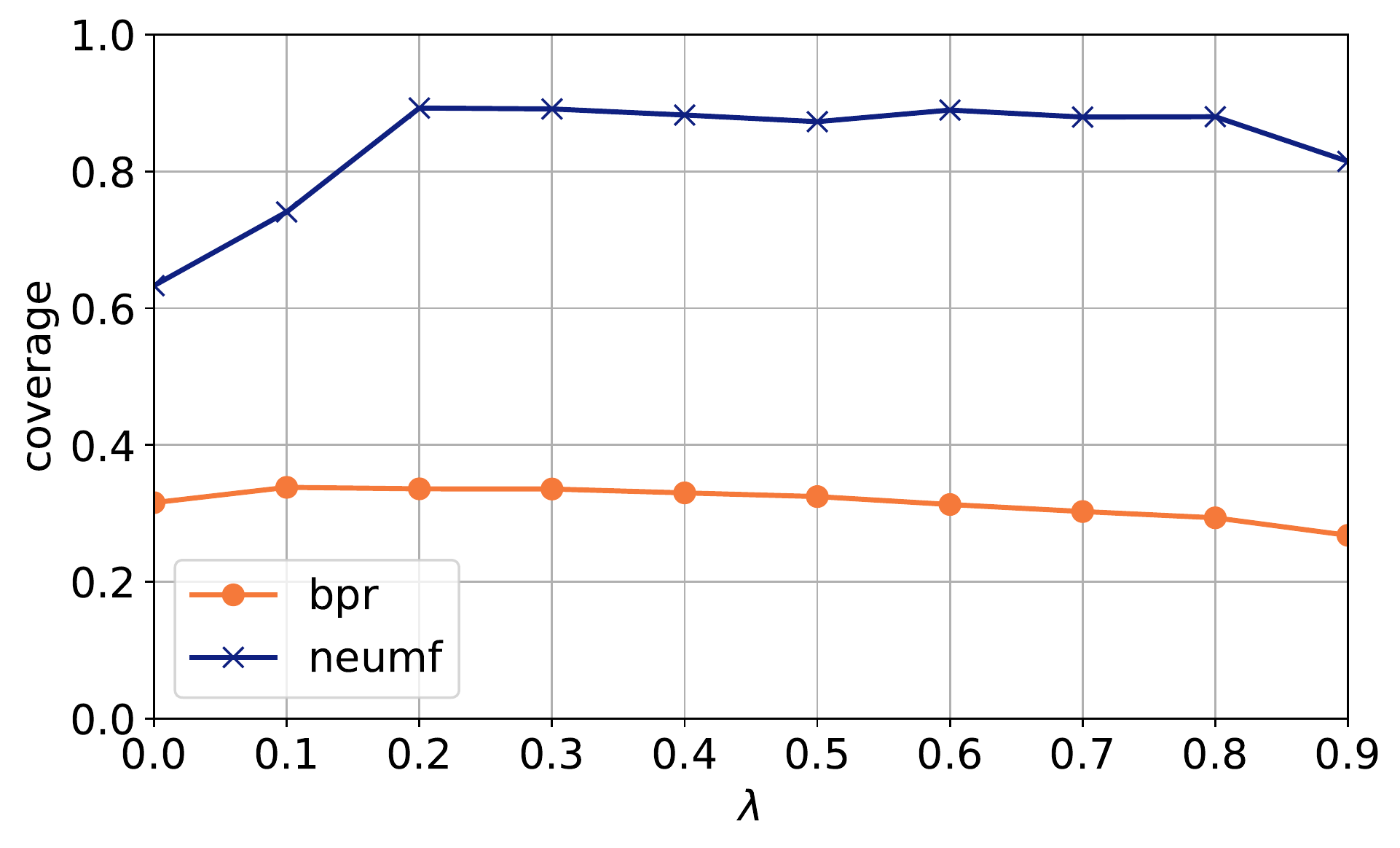}
    \caption{Catalog Coverage}
\end{subfigure}
\caption{\textbf{Influence of the Regularization Weight}. Normalized Discounted Cumulative Gain (NDCG), Precision, Recall, Item Statistical Parity (ISP), Item Equal Opportunity (IEO), Novelty, and Coverage achieved by BPR and NeuMF in ML1M, by varying $\lambda$.}
\label{fig:res-02}
\end{figure} 

We varied the regularizer weight $\lambda$ and computed the accuracy, popularity bias, and beyond-accuracy metrics (\figurename~\ref{fig:res-02}). The x-axis indicates the value of $\lambda$, while the y-axis shows the value measured for the corresponding metric at that value of $\lambda$. It can be observed that the regularization procedure experienced a quite stable performance at various $\lambda \ge 2$. Specifically, at the cost of a loss in NDCG on the full test set, our mitigation ensured comparable or even better NDCG values on the balanced test set, large gains in ISP and IEO, higher novelty and a more wider coverage of the catalog. Lower gains on coverage were experienced by BPR. To balance accuracy and other metrics, setting $\alpha = 0.2$ is a reasonable choice.
 
\vspace{2mm} \noindent \colorbox{gray!15}{\parbox{0.98\textwidth}{\textbf{Observation 7}. \textit{Mitigating popularity with our procedure makes a positive impact on recommendation quality. Lower ISP and IEO, higher novelty, and a wider coverage are achieved at the cost of a small loss in NDCG, if the model is evaluated on a balanced test set.}}} \vspace{2mm}

\subsection{Comparison with Other Mitigation Procedures (RQ4)} \label{sec:baselines}
We next compare the proposed \texttt{sam+reg} mitigation procedure with representative state-of-the-art alternatives to assess ($i$) how the proposed procedure performs in comparison with other countermeasures, and ($ii$) how they manage the trade-off between popularity bias and recommendation quality. We highlight the fact that we do not aim to show that an in-processing procedure beats a post-processing procedure (or vice versa), also because they could be jointly combined. Our goal here is to assess how far an in-processing strategy is from a post-processing strategy to reach good trade-offs. We leave the joint employment of both pre- and post-processing as a future work, to focus on the validation of our mitigation procedure. We compare the trade-off achieved by the proposed regularized models \texttt{sam+reg} against the one obtained by:
\begin{itemize}[leftmargin=*]
\item \textit{Pop-Weighted}~\cite{abdollahpouri2018popularity} re-ranks the output of the original model according to a weighted-based strategy. The relevance returned by the original model for a given item is multiplied by a weight inversely proportional to the popularity of that item, before re-ranking.  
\item \textit{Binary-xQuad}~\cite{DBLP:conf/flairs/AbdollahpouriBM19}, for each user, iteratively builds the re-ranked list by balancing the contribution of the relevance score returned by original model and of the diversity level related to head- and mid-item sets. It includes only the best mid item it can. The split between head and mid items was performed based on the percentiles shown in Figure \ref{fig:pop-dist}.
\item \textit{Smooth-xQuad}~\cite{DBLP:conf/flairs/AbdollahpouriBM19} follows the same strategy of Binary-xQuad, but it takes into account the likelihood that an item should be selected based on the ratio of items belonging to the head- and mid-item sets in the users' profile.
\end{itemize}

To answer these questions, we report accuracy, popularity bias, and beyond-accuracy metrics for all the considered procedures in Table \ref{tab:baseline-res}. The best performer approach per metric and algorithm is marked with a bold style. The same value of $\lambda$ is used for all the approaches to favor comparability. 

\begin{table}[!b]
\resizebox{\textwidth}{!}{
\begin{small}
\begin{tabular}{cl|rr|rr}
\multirow{2}{*}{\textbf{Metric}} & \multirow{2}{*}{\textbf{Approach}} & \multicolumn{2}{c}{\textbf{ML1M}} & \multicolumn{2}{c}{\textbf{COCO}} \\
 &  & \multicolumn{1}{c}{\textbf{BPR}} & \multicolumn{1}{c}{\textbf{NeuMF}} & \multicolumn{1}{c}{\textbf{BPR}} & \multicolumn{1}{c}{\textbf{NeuMF}} \\
 \hline
\multirow{5}{*}{\textbf{\begin{tabular}[c]{@{}c@{}}NDCG \\ F-T\end{tabular}}} 
 & Pop-Weighted & \bftab{0.128} & \bftab{0.098} & \bftab{0.031} & 0.019 \\
 & Binary-xQuad & 0.117 & 0.094 & 0.027 & \bftab{0.038} \\
 & Smooth-xQuad & 0.112 & 0.092 & 0.024 & 0.036 \\
 & Sam+Reg (\texttt{ours}) & 0.073 & 0.067 & 0.030 & 0.031 \\
 \hline
\multirow{5}{*}{\textbf{\begin{tabular}[c]{@{}c@{}}NDCG \\ B-T\end{tabular}}} 
 & Pop-Weighted & \bftab{0.012} & \bftab{0.014} & 0.007 & 0.007 \\
 & Binary-xQuad & \bftab{0.012} & 0.012 & 0.007 & \bftab{0.013} \\
 & Smooth-xQuad & 0.011 & 0.013 & 0.007 & \bftab{0.013}  \\
 & Sam+Reg (\texttt{ours}) & \bftab{0.012} & \bftab{0.014} & \bftab{0.009} & 0.011 \\
 \hline
\multirow{5}{*}{\textbf{\begin{tabular}[c]{@{}c@{}}Precision \\ F-T\end{tabular}}} 
 & Pop-Weighted & \bftab{0.116} & \bftab{0.091} & \bftab{0.011} & 0.007 \\
 & Binary-xQuad & 0.106 & 0.085 & 0.010 & \bftab{0.013} \\
 & Smooth-xQuad & 0.098 & 0.083 & 0.008 & 0.012 \\
 & Sam+Reg (\texttt{ours}) & 0.085 & 0.079 & 0.010 & 0.009 \\
 \hline
\multirow{5}{*}{\textbf{\begin{tabular}[c]{@{}c@{}}Recall \\ F-T\end{tabular}}} 
 & Pop-Weighted & \bftab{0.052} & \bftab{0.043} & \bftab{0.050} & 0.034 \\
 & Binary-xQuad & 0.048 & 0.041 & 0.044 & \bftab{0.059} \\
 & Smooth-xQuad & 0.046 & 0.041 & 0.039 & 0.057 \\
 & Sam+Reg (\texttt{ours}) & 0.037 & 0.038 & 0.047 & 0.048 \\
 \hline
\multirow{5}{*}{\textbf{ISP}} 
 & Pop-Weighted & 0.057 & 0.221 & 0.008 & 0.174 \\
 & Binary-xQuad & 0.073 & 0.213 & 0.010 & 0.069 \\
 & Smooth-xQuad & 0.093 & 0.215 & 0.016 & 0.041 \\
 & Sam+Reg (\texttt{ours}) & \bftab{0.132} & \bftab{0.347} & \bftab{0.047} & \bftab{0.556} \\
 \hline
\multirow{5}{*}{\textbf{IEO}} 
 & Pop-Weighted & 0.081 & 0.203 & 0.014 & 0.041 \\
 & Binary-xQuad & 0.086 & 0.193 & 0.014 & 0.040 \\
 & Smooth-xQuad & 0.109 & 0.195 & 0.014 & 0.075 \\
 & Sam+Reg (\texttt{ours}) & \bftab{0.142} & \bftab{0.241} & \bftab{0.022} & \bftab{0.125} \\
 \hline
\multirow{5}{*}{\textbf{Novelty}} 
 & Pop-Weighted & 0.746 & 0.864 & 0.977 & 0.987 \\
 & Binary-xQuad & 0.776 & 0.869 & \bftab{0.981} & 0.982 \\
 & Smooth-xQuad & 0.806 & 0.870 & 0.914 & 0.983 \\
 & Sam+Reg (\texttt{ours}) & \bftab{0.933} & \bftab{0.924} & \bftab{0.981} & \bftab{0.995} \\
 \hline
\multirow{5}{*}{\textbf{\begin{tabular}[c]{@{}c@{}}Catalog \\ Coverage\end{tabular}}} 
 & Pop-Weighted & 0.316 & 0.644 & 0.105 & \bftab{0.550} \\
 & Binary-xQuad & 0.357 & 0.646 & 0.104 & 0.438 \\
 & Smooth-xQuad & \bftab{0.377} & 0.653 & \bftab{0.120} & 0.400 \\
 & Sam+Reg (\texttt{ours}) & 0.368 & \bftab{0.740} & 0.078 & 0.964 \\
 \hline 
\end{tabular} \end{small}}
\vspace{-3mm}
\caption{\textbf{Comparison among Mitigation Procedures}. Ranking accuracy on the full test set (NDCG F-T) and on the balanced test set (NDCG B-T), Precision and Recall on the full test set, item statistical parity (ISP), item equal opportunity (IEO), novelty, and catalog coverage achieved by the bias-mitigation procedures.}
\label{tab:baseline-res}
\end{table}

From the top rows of the table, it can be observed that the proposed \texttt{sam+reg} mitigation procedure experienced the larger loss in NDCG, when the full test set was considered. Conversely, it achieved comparable NDCG with respect to Pop-Weighted on the balanced test set for both BPR and NeuMF. In highly sparsed datasets such as COCO, the gap on accuracy between \texttt{sam+reg} and the other strategies was lower. This different behavior might be also caused by the skewed popularity tail in COCO, which makes it harder to find input pairs/triplets where the observed item is less popular than the unobserved item. 

Going in depth with the popularity bias estimates, it can be observed that our \texttt{sam+reg} procedure largely improved ISP on both datasets. On ML1M, Pop-Weighted exhibited the highest bias on statistical parity. Binary- and Smooth-xQuad achieved comparable scores between each other, but lower than \texttt{sam+reg}. On COCO, ISP improved for NeuMF, but not for BPR. Smaller improvements of our proposal were achieved on ISP on both datasets. Similar patterns were observed for IEO, with \texttt{sam+reg} achieving higher values. Our proposal appeared highly competitive on both novelty and catalog coverage, especially on ML1M. This gain came at the cost of a higher NDCG loss with the full test set. Conversely, it achieved a comparable NDCG on the balanced test set. 

\subsection{Discussion} 
In this section, we connect the main findings coming from the individual experiments and present the implications and limitations of our proposal. 

The increasing adoption of recommender systems is requiring platform owners to address issues of bias in the system internal mechanics, given that providing biased recommendations may prevent the overlying platform from being successful. The outcomes of our exploratory analysis in Section \ref{sec:exp} highlighted that two widely-adopted families of algorithms, based on point- and pair-wise optimization functions, emphasize a bias against unpopular items; thus, the latter ones end up being under-recommended even when they are of interest, reducing the novelty and the coverage of items in recommendations. Our results on two fundamental state-of-the-art recommender systems might have implications on other systems optimized on the same or similar objective functions. 

Our results provide more evidence on the influence of item popularity in recommendation, which has been primarily investigated on rating prediction algorithms in the contexts of movies, music, books, social network, hotels, games, and papers \cite{DBLP:conf/ecir/BorattoFM19,DBLP:journals/jucs/PampinJO15,DBLP:journals/umuai/JannachLKJ15,DBLP:conf/um/JannachKB16,DBLP:conf/iconference/CollinsTAB18}. We extended the existing knowledge by linking observations on the model internal mechanics to the level of popularity bias experienced by the recommender system, quantifying a biased correlation between item relevance and item popularity (Section \ref{sec:diagnosis}). The methodology and lessons learned throughout our exploration represent one of the first attempts of linking internal mechanics to popularity bias and beyond-accuracy objectives.

Combining our training examples sampling with the correlation-based loss resulted in lower popularity bias at the cost of a decrease in ranking accuracy, which confirms the trade-off experienced by other debiasing procedures \cite{DBLP:conf/flairs/AbdollahpouriBM19,DBLP:journals/umuai/JannachLKJ15}. However, trading a small degree of accuracy for popularity bias mitigation has been proved to improve the overall recommendation quality (Section \ref{sec:linking}). Our study additionally brings forth the discussion about the impact of popularity debiasing on beyond-accuracy objectives, going beyond ranking accuracy and recommended item popularity estimates. Finally, as the models analyzed in Section \ref{sec:baselines} showed very different trade-off patterns, our comparison among debiasing procedures might support stakeholders when they need to choose the most suitable bias-mitigation procedure for their recommendation scenario. 

Throughout our study, a range of limitations emerged at different steps of the pipeline. The most representative limitations relate to the following aspects:

\begin{itemize}[leftmargin=*]
\item \textbf{Limitations of data}. Our study was conducted on observations extracted from the ratings provided by users. This assumption does not account for the implicit behaviour of users who interacted with items, without necessarily providing ratings. However, as we target a learning-to-rank task, our mitigation procedure can be applied even to matrices which do not include ratings (e.g., binary data or frequencies). Learning to rank requires a definition of what it is of interest or not for the user (e.g., by applying thresholds to ratings or frequencies). Our procedure does not make any assumption on the type of users' feedback, which depends on the scenario under consideration. 
\item \textbf{Limitations of recommendation algorithms}. While we validated algorithms that use two key families of optimization function, there are other algorithms we did not consider. However, our study makes it easy to run our analyses on additional algorithms. Our experiments highlighted that the proposed mitigation procedure works well under a pair-wise optimization setting, while it leads to lower gains in recommendation quality under a point-wise optimization setting. As we target a learning-to-rank task, it should be noted that our mitigation procedure can be also applied to algorithms originally designed for rating prediction, when they are optimized for ranking accuracy.  
\item \textbf{Limitations of experimental protocol}. Our analysis does not allow us to assess whether the gains/losses experienced by the considered metrics are caused by (i) the actual differences of the recommender system’s performance or (ii) the differences of the evaluation protocol effectiveness at monitoring the overall recommendation quality. Furthermore, there is no concrete evidence on the real-world impact of the bias-mitigated recommendations on the users' satisfaction, which requires subsequent online evaluation studies.
\item \textbf{Limitations of evaluation metrics}. There are several metrics capable of monitoring a certain aspect associated with the quality of recommendations. Our study focused on item statistical parity and equal opportunity, novelty, and coverage. We also monitored NDCG, Precision, and Recall as proxies of recommendation effectiveness. However, the metrics and the target trade-offs depend on the scenario and are often selected by the platform owners.
\end{itemize}

Our study incorporates elements of beyond-accuracy importance, which can be shaped by adjusting the item popularity of the recommendations. As recommender systems move further into online platforms, it becomes more and more essential that they embrace and consider aspects similar to ours.

\section{Conclusions}
\label{conclusions}
In this paper, we proposed two new metrics aimed at monitoring popularity bias in a ranking-based recommendation task. Then, we empirically showed that representative learning-to-rank algorithms based on point- and pair-wise optimization functions are vulnerable to imbalanced data across items, and tend to generate biased recommendations with respect to the proposed metrics. To mitigate this bias, we proposed a countermeasure that incorporates a new training example sampling strategy and a new regularized loss function. Finally, we performed extensive experiments to monitor the trade-off among popularity bias, accuracy, and beyond-accuracy goals. Based on the results, we conclude that:

\begin{enumerate}[leftmargin=*]
\item Predicted users' relevance distributions for observed head and mid items are biasedly different; the first one includes higher relevance scores, on average. 
\item The pair-wise accuracy for observed mid items is lower than the one for observed head items; mid items are under-ranked regardless of users' interests. 
\item The combination of our sampling strategy and our regularized loss function leads to a lower gap in pair-wise accuracy between observed head and mid items; higher statistical parity, equal opportunity, and beyond-accuracy estimates can be achieved by the models treated with our mitigation procedure. 
\item The treated models exhibit comparable accuracy against the original model, when the same number of test observations are used for each item, which was proved to be a proper testing setup when popularity bias is considered \cite{DBLP:journals/ir/BelloginCC17}.
\item Compared to state-of-the-art alternatives, our treated models nearly reduce popularity bias while achieving competing ranking accuracy and beyond-accuracy estimates, generalizing well across users' populations and domains.
\end{enumerate}

In our next steps, we are interested in investigating temporal- and relevance-aware metrics, which respectively take the item popularity and relevance at a given time into account, for monitoring a popularity bias. We will explore the possibility of defining post-processing countermeasures which fine-tune pre-trained user/item vectors to reduce a popularity bias. Furthermore, we plan to inspect the inter-play between the system-level and the user-level tendency of preferring more (less) popular items. Finally, we will link the resulting observations to beyond-accuracy objectives associated with multiple stakeholders (e.g., users and providers), subsequently deriving countermeasures tailored for them.  

\section*{Acknowledgments}
This work has been partially supported by the Sardinian Regional Government, POR FESR 2014-2020 - Axis 1, Action 1.1.3, under the project ``SPRINT'' (D.D. n. 2017 REA, 26/11/2018, CUP F21G18000240009), and by the Ag\`encia per a la Competivitat de l'Empresa, ACCI\'O, under the project ``Fair and Explainable Artificial Intelligence (FX-AI)''.

\bibliographystyle{elsarticle-num}
\bibliography{bibliography}

\end{document}